\newcommand{\beq}{\begin{equation}}
\newcommand{\eeq}{\end{equation}}
\newcommand{\bed}{\begin{displaymath}}
\newcommand{\eed}{\end{displaymath}}
\newcommand{\beqa}{\begin{eqnarray}}
\newcommand{\eeqa}{\end{eqnarray}}
\newcommand{\beqan}{\begin{eqnarray*}}
\newcommand{\eeqan}{\end{eqnarray*}}
\newcommand{\bfig}{\begin{figure}}
\newcommand{\efig}{\end{figure}}
\newcommand{\imp}{\Rightarrow}
\newcommand{\goto}{\rightarrow}

\newcommand{\ovl}{\overline}
\newcommand{\fn}{\footnote}
\newcommand{\lae}{\mathrel{<\kern-1.0em\lower0.9ex\hbox{$\sim$}}}
\newcommand{\gae}{\mathrel{>\kern-1.0em\lower0.9ex\hbox{$\sim$}}}
\newcommand{\noi}{\noindent}
\newcommand{\sun}{\hbox{$\odot$}}
\newcommand{\non}{\nonumber}
\newcommand{\ogam}{\ovl{\gamma}}

\newcommand{\rch}{r_{\rm ch}}
\newcommand{\lb}{\langle}
\newcommand{\rb}{\rangle}
\newcommand{\gamc}{\gamma_{\rm core}}    
\newcommand{\game}{\gamma_{\rm env}}    
\newcommand{\cc}{c_{\rm core}}    
\newcommand{\ce}{c_{\rm env}}    
\newcommand{\mc}{m_{\rm core}}

\documentstyle[emulateapj,apjfonts,psfig]{article}

\begin{document}
\title{COMPOSITE POLYTROPE MODELS OF MOLECULAR CLOUDS.\\ 
I.\ THEORY}
\author{CHARLES L.\ CURRY\altaffilmark{1,2}}
\affil{Astronomy Department, University of California, Berkeley, 
CA  94720}
\author{and}
\author{CHRISTOPHER F.\ MCKEE}
\affil{Departments of Physics and Astronomy, University of California, 
Berkeley, CA  94720}
\altaffiltext{1}{Present address: Department of Physics and Astronomy, 
University of Western Ontario, London, ON  N6A 3K7}
\altaffiltext{2}{Email: curry@astro.uwo.ca} 

\begin{abstract} 
We construct spherical, hydrostatic models of dense molecular cores 
and Bok globules consisting of two distinct, spatially separate 
gas components: a central, isothermal region surrounded by a 
negative-index, polytropic envelope.  The clouds are supported against 
their own self-gravity by a combination of thermal, mean magnetic, and 
turbulent wave pressure.  The latter two are included by allowing for 
locally adiabatic, non-isentropic pressure components.  Such models 
are meant to represent, in a schematic manner, the velocity and density 
structure of cores and globules, as inferred from molecular line and 
dust continuum observations.  In addition, our picture reflects the 
theoretical expectation that MHD wave motions, which are important 
at scales $\gae 0.1$ pc in typical low-mass star-forming regions,
are damped at smaller scales, 
giving rise to a finite-sized, thermally-dominated core region.   
We show that if the pressure components are isentropic,
then the pressure drop from the center to the edge of the composite
polytropes we consider is limited to 197, the square of the value
for the Bonnor-Ebert sphere.  If the pressure components are
non-isentropic, it is possible to have arbitrarily large
pressure drops, in agreement with the results of McKee \& Holliman
(1999).  However, we find that even 
for non-isentropic pressure components, the ratio of the {\it mean} 
to surface pressure in the composite polytropes we consider is less 
than 4.  We show by explicit construction that it is possible to have
dense cores comparable to the Jeans mass embedded in stable clouds of 
much larger mass.  In a subsequent paper, we show that composite 
polytropes on the verge of gravitational instability can reproduce 
the observed velocity and density structure of cores and globules 
under a variety of physical conditions.  
\end{abstract} 

\section{INTRODUCTION}
\label{sec-intro}
Observations of cold, dense molecular clouds in the interstellar medium 
present a unique challenge to theorists.  Isolated Bok globules and dense 
molecular cores in molecular cloud complexes, in particular, are 
definitive testbeds for theories of molecular cloud structure and 
low-mass star formation.  As ``the simplest unit structures in the dense 
interstellar medium'' (Dickman \& Clemens 1983), an adequate description 
of these objects should contribute significantly to our understanding 
of the molecular component of the Galaxy as a whole. 

\subsection{Observed Properties of Dense Cores and Bok Globules}
\label{sec-obs}

Dense cores are compact regions within molecular clouds whose number 
density exceeds about $10^4$ cm$^{-3}$.  The connection of these 
regions to star 
formation in the complexes has been explored in some detail (Loren 
1989; Onishi et al 1998); in many cases 
there is close spatial agreement between young stars and cores (Beichman 
et al 1986).  In fact, IRAS sources have 
been detected in approximately half of the known ($\sim 100$) nearby cores 
(Benson \& Myers 1989).  The relatively isolated cores in Taurus, which 
have been extensively surveyed in NH$_3$ (Myers \& Benson 1983; Benson 
\& Myers 1989), have sizes on the order of 0.1 pc, masses $\sim 
1-10\; M_{\sun}$, kinetic temperatures $\sim 10$ K, and NH$_3$ 
linewidths $\sim 0.3$ km s$^{-1}$.  The narrow linewidths 
suggest that these objects are relatively quiescent; thermal 
motions are responsible for at least half of the observed line broadening.
This is in 
contrast to the well-known increase of nonthermal linewidth with size 
scale in the regions immediately surrounding the ammonia cores  
(Fuller \& Myers 1992; Goodman et al 1998).  The properties of cores 
in more crowded regions, 
such as Ophiuchus and Orion, differ from those in Taurus mainly insofar   
as they tend to be more massive and compact, and have higher temperatures 
and linewidths (Caselli \& Myers 1995; Motte, Andr{\'e}, \& Neri 1998).

    A catalog of 248 optically selected Bok globules was 
compiled by Clemens \& Barvainis (1988).  A comparison of these objects 
with the Benson \& Myers (1989) sample of dense cores in molecular cloud 
complexes reveals that 
the two populations are quite similar, with a trend toward higher masses 
and densities (by about a factor of two) in the dense cores (Bourke et 
al.\ 1995).  Associated IRAS point sources have been detected in about 
one-quarter of the Bok globules, 
down to a detection limit of 0.7 $M_{\sun}$ 
(Yun \& Clemens 1990).  The degree of similarity is perhaps surprising, 
given the different environments in which the two types of object reside.  
For example, the two populations should differ with respect to their 
exposure to the interstellar radiation field.  Whereas cores in 
complexes are often shielded from such radiation by their parent 
clouds (Benson \& Myers 1989; Onishi et al 1996), photoelectric heating 
from the interstellar radiation field is expected to produce an
outward temperature gradient in Bok globules 
lacking embedded young stars (Boland \& de Jong 1984).  
This effect has been observed in dark clouds and cores  
in close proximity to star-forming regions (Young et al 1982; Bachiller, 
Guilloteau, \& Kahane 1987), but direct evidence for isolated Bok 
globules is more scarce.  According to Clemens, Yun, \& Heyer 
(1991), the spectral energy distributions of globules 
in the Clemens \& Barvainis sample preclude single-temperature models 
for all objects.  On the other hand, 
molecular line observations of objects in the same sample display 
remarkably uniform temperatures over their projected areas (Lemme et 
al 1996).  The latter might be explained, however, by the fact that 
the dense molecular tracers used in such studies do not sample the 
outer, more exposed regions of the globules.

\subsection{Virial Parameters: Implications for Stability}
\label{sec-virial}

A key question relating to cores and Bok globules is: how near are they to 
gravitational collapse?  A stable interstellar cloud need not be highly 
centrally condensed and self-gravitating; it can also be nearly uniform 
and confined by an external pressure.  In an analysis of clump data
\fn{We shall follow the terminology of Williams, Blitz,
\& McKee (1999) in describing the structure of molecular clouds. 
``Clumps" are coherent regions in $l-b-v$ space, generally
identified from spectral line maps of molecular emission.
Star-forming clumps are the massive clumps out of which
stellar clusters form; cores (sometimes referred to as
``dense cores") are the regions out of which individual
stars or stellar systems form.  The clumps studied by
Bertoldi \& McKee were identified based on
$^{13}$CO line maps, whereas dense cores in low-mass
star-forming regions are generally within star-forming
clumps and are identified from their NH$_3$ emission.}
from several active star-forming regions, 
Bertoldi \& McKee (1992) noted a correlation between the 
``observed virial parameter'' for a clump, 
$\alpha = 5 \ovl{R} \lb\sigma^2\rb/GM$, and the clump mass, 
$M$.  Here, $\lb\sigma^2\rb$ is the mean square velocity 
along the line of sight and 
$\ovl{R}$ an average cloud radius, both determined from intensity 
contours in the $J = 1-0$ line of $^{13}$CO.\fn{Assigning a ``radius'' 
to a non-spherical clump or core is common practice.  The authors showed 
that, if $\alpha$ is interpreted as as average over all orientations of 
an ellipsoidal cloud, then this approximation is likely to be in error 
by no more than 30\%.}  Stable clumps with 
$\alpha \gg 1$, they argued, are likely to be confined by the pressure 
of the external medium, and not very centrally concentrated, while those 
with $\alpha \sim 1$ would be expected to be strongly condensed and 
self-gravitating (no clumps with $\alpha \ll 1$ were observed; such
clumps are unstable and would collapse unless supported
by a strong magnetic field). The authors' analysis revealed that only the 
most massive clumps ($\gae 100\; M_{\sun}$) in the complexes had $\alpha 
\sim 1$, and thus that the majority of the clumps seen were likely to be 
pressure-confined.  For example, representative values for
a $\sim 10\; M_{\sun}$ clump in Ophiuchus
from Loren (1989) are $\sigma \simeq 0.46$ km s$^{-1}$ and  
$\ovl{R} \simeq 0.43$ pc, implying $\alpha \simeq 10.5$.  

	It is important to emphasize that 
estimates of $\alpha$ from molecular line studies are highly dependent 
on the observed molecule.  Dense molecular tracers, such as NH$_3$ and 
CS, are generally only excited in 
small regions where the mean density of hydrogen exceeds $\sim 10^4$ 
cm$^{-3}$.  The mean density of the Ophiuchus $^{13}$CO clumps, on the 
other hand, is only $\sim 3 \times 10^3$ cm$^{-3}$.  It is therefore 
desirable to recalculate $\alpha$ for these denser molecular tracers, 
in order to ensure that one is truly probing the densest gas.  For this 
purpose, we turn to the NH$_3$ study of dense cores by Myers \& Benson  
(1983).  Mean values from their sample are: $\ovl{\sigma} \simeq 0.23$ km 
s$^{-1},\; \ovl{R} \simeq 0.06$ pc, and $M \simeq 4.1\; M_{\sun}$.  These 
yield
$\alpha \simeq 0.9$, indicating that the cores in their sample, which 
are of very low mass compared to those in the $^{13}$CO studies, are 
nevertheless nearer to a state of marginal stability.  Further support 
is provided by dust continuum studies in Ophiuchus, which also give 
$\alpha \sim 1$ for cores of this size and smaller (Motte et al 1998).  

In this paper, we adopt the view that both low- and high-mass cores 
that have been detected in ammonia and other {\it dense} molecular 
tracers (e.g.\ C$^{18}$O and CS) are near a state of marginal 
stability, or, as we shall refer to it, near the {\it critical point}.  
The latter is more precisely defined as the unique state 
such that, upon suffering a small perturbation at its surface, 
the cloud must collapse.  
These cores may themselves be embedded in molecular gas
that is gravitationally bound to the core; however, we
assume that this external gas acts primarily to set
the pressure at the surface of the core, not to
determine the cloud stability.

\subsection{The Initial Conditions for Star Formation} 
\subsubsection{Pre-Protostellar Cores}

	The high percentage of dense cores and Bok globules with 
embedded IRAS point 
sources can, in principle, put restrictions on theories of star 
formation.  Furthermore, the roughly equal samples of ``starless'' cores 
and Bok globules should place constraints on the initial conditions for 
protostellar  
collapse, assuming that these clouds still have a reasonable chance of 
forming stars ({\S}\ref{sec-virial}).  Hence the recent usage of the term 
``pre-protostellar core'' to denote these objects (Ward-Thompson et al 
1994; Andr{\'e}, Ward-Thompson, \& Motte 1996).  We shall 
henceforth refer to both starless dense cores found within GMCs and 
starless Bok globules by the latter term, for brevity 
abbreviated to PPC.  
In the quasi-static picture of low-mass star formation,
PPCs evolve from low density and possibly low mass along a
near-equilibrium sequence of denser, possibly more massive states, until 
a state of marginal stability is reached.  This evolution may be achieved 
in a number of ways: ambipolar diffusion is likely to play a key role,  
and possibly ongoing accretion from the surrounding medium 
(Mestel \& Spitzer 1956; Boland \& 
deJong 1984; Bonnell, Bate, \& Price 1996).  As defined, PPCs generally 
only form one or at most a small group of stars.  

     Historically, theories of cloud equilibria have 
been intimately connected 
with calculations of gravitational collapse, inasmuch as the former 
fill the important role of initial conditions for the latter.  
Models with singular density profiles are often adopted for
computational simplicity.  In particular, the singular isothermal 
sphere (hereafter SIS), which has a static density 
structure given by $\rho(r) \propto r^{-2}$, has been extensively 
studied.  This model is unstable to collapse, and has a corresponding 
time-dependent solution first studied by Shu (1977).  

      As a model for PPCs, there are three principal caveats to the 
SIS, two of which 
stem directly from observation.  The first is the existence of a 
nonthermal component of the molecular linewidth in PPCs.  
The nonthermal line width
increases with size scale as $r^{0.5}$ in low-mass cores and as $r^{0.2}$ 
in high-mass cores (Caselli \& Myers 1995). 
As mentioned in {\S}\ref{sec-obs} above, thermal motions prevail within  
$\sim 0.1$ pc of the intensity maximum; but even along the line of sight 
toward the intensity maximum, 
a finite nonthermal component is detected (Goodman et al 1998).
Second, most objects that have been mapped with sufficient resolution 
exhibit non-singular---indeed, nearly flat---column density profiles 
as one approaches the intensity maximum.  This is 
true of both stable Bok globules  
(Dickman \& Clemens 1983; Turner, Xu, \& Rickard 1992; Lehtinen et al 
1995) and dense cores (Gaida, Ungerechts, \& Winnewisser 1984) mapped in 
molecular lines,  
as well as PPCs mapped in sub-mm dust continuum (Casali 1986; 
Ward-Thompson et al 1994; Motte et al 1998).  In contrast, radial 
profiles of cores {\it with} embedded stars lack such an extended, 
flat inner region, 
and tend to be more compact than PPCs (Motte et al 1998).  

    An additional, theoretical, 
argument against the SIS model is that
there is no known way for a cloud to evolve to a
{\it static} SIS, since isothermal spheres are unstable
to gravitational collapse for center-to-surface density
ratios greater than 14 (Ebert 1955; Bonnor 1956).  Models of the 
collapse of isothermal clouds (Hunter 1977; Foster \& Chevalier 
1993) and of magnetized clouds (Safier, McKee, \& Stahler 1997)
show that a $1/r^2$ structure does develop during the collapse,
but it is not static; as a result, the accretion rate
at early times is greater than that predicted by the SIS model.
However, at late times, both solutions behave similarly to
the SIS.

These considerations argue both against a singular density profile toward 
the centers of PPCs, and against the existence of a purely thermal 
region filling most of the PPC.  On these grounds, 
we conclude that 
PPCs {\it as a whole} are unlikely to bear any close relation to 
(singular or non-singular) isothermal spheres.  

\subsection{Previous Models of Self-Gravitating Cloud Equilibria}
\label{sec-prev}

Any complete model of a molecular cloud must include a detailed treatment 
of thermal balance throughout the gas, which entails a full description 
of the physical and chemical properties at every point.  While progress 
has been made along these lines (de Jong, Dalgarno, \& Boland 1980; 
Falgarone \& Puget 1985), polytropic models, which postulate a
power-law relation between pressure and density
at each point, 
\beq
P=K_p\rho^{\gamma_p}=K_p\rho^{{1+\frac{1}{n}}},
\eeq
are often adopted as a simplifying 
assumption.  Here, $K_p$ is constant and $\gamma_p$ and $n$ are
different versions of the polytropic index.
Yet, spherical, non-singular, polytropic models of dense 
cores and Bok globules have met with only limited success.  
Somewhat surprisingly, 
the dominant problem lies not with details of chemistry, but rather in 
simultaneously reproducing the bulk properties---mass, radius, and 
density contrast---of the objects.  
      
      Positive-index polytropes ($n>0$, $\gamma_p>1$)
cannot account for the observed variation of line width with radius in
molecular clouds
since their temperatures decrease with radius (Shu et al 
1972).  Gas temperatures in cores and globules are found to 
be either uniform, at $T \simeq 10$ K (perhaps a factor of 2--3 higher 
in more massive objects and in regions with recent nearby star formation), 
or outwardly increasing (Bachiller et al 1987).  Furthermore, on nearly 
all spatial scales, the nonthermal linewidth increases with size 
(Larson 1981).  Negative-index polytropes ($n<-1$, $0<\gamma_p<1$;
hereafter, NIPs), which have a temperature $T\propto P/\rho
\propto \rho^{1/n}$
that increases with linear scale, can qualitatively reproduce this 
behavior if the polytropic temperature is taken to represent the 
{\it total} velocity dispersion (Maloney 1988); i.e., 
$T_{NIP} (r) \propto P(r)/\rho(r) \propto \sigma^2 (r)$.  
The limiting case of the NIP as
$\gamma_p\rightarrow 0$ is the ``logatrope", introduced by
Lizano \& Shu (1989) as a phenomenological model
for turbulent pressure and studied further by
McLaughlin \& Pudritz (1996) and Gehman et al (1996).

	Molecular clouds and Bok globules are supported by
several different pressure components---in addition to
the thermal and turbulent pressures discussed above, there
is also the static magnetic field.  Mouschovias (1976) and 
Tomisaka, Ikeuchi, \& Nakamura (1988) developed models of 
axisymmetric
clouds supported by static magnetic fields and isothermal
gas.  Lizano and Shu (1989) extended these 
calculations to include turbulent pressure, modeled as a logatrope. 
More recently, McKee \& Holliman (1999; hereafter MH) developed
{\it multi-pressure polytrope} models for clouds, in which
each of the important pressure components 
is represented by an appropriate polytrope.  Comparison
of these models with observations was carried
out by Holliman (1995).  A phenomenological alternative
to multi-pressure polytropes is the ``TNT" model
developed by Myers \& Fuller (1992) and Caselli \& Myers (1995).
In this model, the density is assumed to be the sum of two terms,
\bed
n\propto\left(\frac{r_0}{r}\right)^2 + \left(\frac{r_0}{r}\right)^p,
\eed
where the first represents the effect of thermal pressure and
the second the effect of nonthermal pressure.  By inserting
this expression into the equation of hydrostatic equilibrium,
it is possible to infer both the density at $r_0$ and the
variation of the line width with $r$; reasonably good fits
are obtained for most cores by an appropriate choice of
the parameters $r_0$ and $p$.

	    Fits of the envelopes of globules in the 
Clemens \& Barvainis sample to a volume density law of the form 
$n(r) \propto r^{-k_\rho}$ indicate a rather narrow range: 
$1 \lae k_\rho \lae 2$ (Yun \& Clemens 1991).  This further supports 
the use of NIPs, whose behavior asymptotically 
approaches a power law with precisely the same range of indices.
[The singular polytropic sphere has $n(r)\propto r^{-2/(2-\gamma_p)}
=r^{-2n/(n-1)}$.] 
Similar fits to dense core data produce the same range of $k_\rho$ 
(Cernicharo, Bachiller, \& Duvert 1985; Bachiller et al 1987; 
St{\"u}we 1990).  The use of NIPs in 
the thermal modeling of molecular clouds has been validated by
de Jong et al (1980), Boland \& de Jong (1984), and Falgarone \& 
Puget (1985), each  
of whom attempted to determine the equation of state relevant to dense 
molecular gas using detailed thermal and chemical balance calculations.
While the representation of turbulence as a spatially 
varying isotropic pressure is clearly simplistic, Alfv{\'e}n
waves, a significant contributor to MHD turbulence,
have a polytropic index $\gamma_p=1/2$, corresponding
to $n=-2$ (Wal{\'e}n 1944; McKee \& Zweibel 1995).

    Non-singular NIPs that are truncated by a medium of constant 
pressure have well-defined stability properties.  
In keeping with the above discussion of virial equilibrium, it is 
of interest to compare the theoretical properties of critically 
stable NIPs with the observed masses, radii, and density contrasts 
of PPCs.  Under the usual assumptions that the NIP is isentropic 
(see {\S}2.1), the mass, radius, and density contrast 
cannot exceed the values (Viala \& Horedt 1974), 
\bed 
R_{BE} = 0.485 \frac{c_s^2}{(G P_s)^{1/2}} \;\;\;\;\; M_{BE} = 
1.18 \frac{c_s^4}{(G^3 P_s)^{1/2}}, \;\;\;\;\; \left( 
\frac{\rho_c}{\rho_s} \right)_{BE} = 14.04, 
\eed
where $M_{BE}, \; R_{BE}$, and $(\rho_c/\rho_s)_{BE}$ are the 
mass, radius and density contrast of the critically
stable, bounded isothermal sphere, also known as the Bonnor-Ebert 
sphere (Ebert 1955; Bonnor 1956).  That is, the Bonnor-Ebert sphere 
(which has $n=\pm \infty$, $\gamma_p=1$) sets the upper limits of the 
physical parameters for any NIP, provided that the sound speed, 
$c\equiv (P/\rho)^{1/2}$, in 
the isothermal sphere is the same as the {\it surface} value, $c_s$, 
in the NIP.  The same limits apply if $c_s$ is replaced by the rms sound
speed in the cloud, $\lb c^2\rb^{1/2}$,
which is often more easily observed than $c_s$ (MH). 
The fact that $\alpha \sim 1$ for observed PPCs ensures that their masses 
and radii are in reasonable accord with the above, provided that $c$ 
includes both thermal and nonthermal motions.  However, 
the inferred density contrast between the intensity peak of many cores 
and the boundaries of their surrounding clumps (as determined from 
$^{13}$CO) is typically $\gae 30$ (Stutzki \& Guesten 1990; Williams, 
Blitz, \& Stark 1995; Andr{\'e} et al 1996), which significantly exceeds 
$(\rho_c/\rho_s)_{BE} = 14.04$.  The problem may be even more pronounced 
for Bok globules, which are expected to have a lower boundary pressure 
than cores 
since they are not embedded in GMCs.  As stated succinctly by Hasegawa 
(1988), who attempted to model the Bok globules L134 and L183, ``\ldots 
a single hydrostatic gas sphere cannot explain the high density contrast 
between the environment and the center of globules.''

Attempts have been made to elude this difficulty.  In their model of a 
quiescent Bok globule, Dickman \& Clemens (1983) proposed a new 
stability criterion for NIPs; namely, that the central temperature,  
not the entropy, remains constant during a perturbation (see Appendix).  
This criterion was later adopted by Maloney (1988), who explicitly 
showed that it renders all bounded NIPs unconditionally stable.  Of 
course, this 
removes the upper bound on the density contrast given above, allowing 
it to be fit to observations.  Predictably, however, this can result 
in an overestimate of the mass, radius, and visual extinction of the 
object (Hasegawa 1988; Turner et al 1992).  An important exacerbating 
factor in the modelling difficulties faced by both Dickman \& Clemens 
and Hasegawa was their assumption that the total boundary pressure was 
equal to the thermal pressure ($P_{\rm th} \lae 3700$ K cm$^{-3}$) 
{\it only}.  In contrast, current estimates including the interstellar 
magnetic field (but not cosmic rays, which pervade both the cloud and 
the ambient medium) give $P \simeq 2 \times 10^4$ K cm$^{-3}$ (Boulares 
\& Cox 1990).  While this lessens somewhat the discrepancies they 
uncovered, it does not solve the problem.  Nor does it change the 
fact that NIPs of any radius are unable to fit the detailed 
shape of the line width--size relation (Paper II).  

Implicit in the stability analyses described so far is the 
assumption that the gas sphere responds {\it isentropically}, i.e.\ 
with spatially and temporally constant entropy, to perturbations applied 
at its surface.  While this is appropriate for an isothermal gas alone,   
observations suggest a more complex situation, in which several 
distinct pressure components---primarily thermal, magnetic, and 
turbulent---coexist in the gas (Myers \& Goodman 1988).  As shown 
recently by MH, this necessitates a {\it non-isentropic} description 
of cloud structure (the significance of this will be discussed in {\S}2 
below).  Non-isentropic polytropes that accurately reflect the relative 
pressures of these distinct components tend to be more stable than their 
isentropic counterparts.  This improved treatment obviates the perceived 
need for the ad hoc, constant central-temperature restriction introduced 
by the above authors (see Appendix).  

\subsection{Motivation and Evidence for Core-Envelope Structure}

In this paper, we construct a model for PPCs that explicitly 
reflects their observed velocity structure; i.e., thermal and subsonic 
nonthermal motions on small scales ($\lae 0.1$ pc), and supersonic 
nonthermal motions on larger scales.  Theoretically, the transition 
between the two regimes is believed to occur via the damping of Alfv{\'e}n 
waves in regions of low ionization; i.e.\ near the density maximum where 
the optical depth to incoming FUV photons is largest.  The critical 
wavelength for damping is of order 0.1 pc, under typical PPC conditions 
({\S}\ref{sec-core}).  This suggests, in the apt description of Goodman et 
al.\ (1998), a view of PPCs as ``islands of calm in a turbulent sea.''  

Another theoretical argument for such a division comes from the theory 
of magnetized molecular clouds in the absence of wave motions.  Central 
to the quasi-static picture is the notion of a magnetically-supported 
cloud, slowly contracting via ambipolar diffusion 
(Mouschovias 1991).  Over time, magnetic support in the central region 
decreases as the mass and density increase there, while in the envelope, 
little 
changes.  This leads naturally to a division between core and envelope, 
in that the latter remains ``magnetically subcritical,'' while the 
core becomes ``magnetically supercritical'' (McKee et al 1993). 
Eventually, the 
core alone collapses, leaving behind the magnetically subcritical
envelope that accretes more slowly (Safier et al 1997). 
Inclusion of the effects of hydromagnetic waves should moderate the 
inward flow of mass via ambipolar diffusion, but the decay of the waves 
in the central region should allow it to operate as predicted, 
preserving---and perhaps enhancing---the distinction between core and 
envelope (see also Mouschovias 1991).  

	 Studies of the thermal structure of clouds exposed to the 
interstellar radiation field also suggest a core-envelope structure.
Falgarone \& Puget (1985) studied the effect of external 
infrared and UV radiation on thermally-supported self-gravitating 
condensations of gas and dust.  For the conditions they assumed, 
they found equilibria featuring isothermal, 
molecular cores (of typical radius $0.1$ pc and mass $2\; M_{\sun}$) 
surrounded by warmer and more diffuse, partially atomic, envelopes 
(with $r \sim 1$ pc, $M \sim 10\; M_{\sun}$, depending on the UV field 
intensity and external pressure).  More recently, 
detailed numerical calculations by Nelson \& Langer (1997) demonstrated 
that the thermal evolution of a $100 - 400\; M_{\sun}$, uniform density 
cloud subjected to the interstellar radiation field leads to a 
core-envelope structure, in which 
the core region has roughly the same properties as those observed for PPCs 
in C$^{18}$O, while containing only one-fifth of the total mass.  Elmegreen 
(1989) calculated the relative mass and radius fractions of molecular 
cores to atomic envelopes for spherical polytropes exposed to the 
interstellar radiation field.  
While this molecular/atomic core--envelope structure is generally not 
relevant to PPCs, which are shielded by the ambient molecular gas,
it may be relevant to the more massive Bok globules, which have 
bulk properties quite similar to those 
predicted by the models of Nelson \& Langer (1997), 
except for the large nonthermal linewidths in their envelopes. 

     On the observational side, Bachiller \& Cernicharo (1984) noted marked 
differences in the density and velocity structure of the large dark cloud 
B1 depending on the 
molecular tracer used, prompting them to divide the cloud into an NH$_3$ 
core ($r \lae 0.3$ pc, $M \lae 60\; M_{\sun}$), a C$^{18}$O and $^{13}$CO 
envelope ($r \lae 1$ pc, $M \lae 350\; M_{\sun}$), and a $^{12}$CO halo 
($r \sim 2$ 
pc, $M \sim 400\; M_{\sun}$).  Interestingly, despite these differences, 
the three regions showed negligible variations in kinetic temperature 
($T = 12 \pm 2$ K).  While this layering could 
simply reflect the density threshold at which different molecular species 
are excited, 
a later study of similar clouds in the Taurus-Auriga-Perseus complex using 
dust extinctions to derive H$_2$ column densities found similar results 
(Cernicharo et al 1985).  On smaller scales, Fuller (1989) proposed the 
same division on the basis of higher-resolution NH$_3$ and C$^{18}$O 
observations.

As a particular example of a cloud one might keep in mind for our models, 
consider R65, a well-studied, starless clump in the $\rho$ Oph complex 
(Loren 1989).  $^{13}$CO 
maps of this clump give the following estimates: $T \approx 20$ K, FWHM 
size $\approx 0.77 \times 0.44$ pc, $\ovl{n}_{H_2} \approx 3.5 \times 10^3$ 
cm$^{-3}, \; \sigma \approx 0.39$ km s$^{-1}$, and $M \approx 21\; 
M_{\sun}$; no IRAS sources were detected within the FWHM contour.  Note 
that we have assumed a H$_2$ to $^{13}$CO abundance ratio of $4.8 \times 
10^5$, in accord with the downward revision in Loren's original estimate 
by Bertoldi \& McKee (1992).  Using 
these values, and taking for $R$ one-half of the geometric mean of the 
cloud dimensions, one finds $\alpha \approx 2.4$.  The dense core of R65, 
L1689B, has been detected in C$^{18}$O, NH$_3$ 
(Benson \& Myers 1989) and, more recently, 1.3 mm dust continuum 
(Andr{\'e} et al 1996).  The latter study gives: $T \approx 18$ K, FWHM 
geometric mean radius $\approx 0.046$ pc, 
$\ovl{n}_{H_2} \approx 4.1 \times 10^4$ cm$^{-3}$, and $M \approx 1\; 
M_{\sun}$.  According to Andr{\'e} et al, the derived density 
contrast between the intensity peak of L1689B and the edge of R65 is 
$\gae 30$, a figure much higher than simple equilibrium NIP models
can accomodate, even at their critical states (i.e.\ $\rho_c/\rho_s \leq 
14.04$). 

In {\S}2, we develop the theory of polytropes with core--envelope
structures in detail, including an analysis of their stability.  
After reviewing the properties of single-component polytropes in 
\S 3, we examine the physical properties of composite polytropes 
in {\S}4.  Some of the broader implications of these properties for 
star formation in PPCs are mentioned in {\S}5.  In a 
subsequent paper, we compare the models to available observations. 

\section{COMPOSITE POLYTROPES: GENERAL THEORY}
\label{sec-genth}
\subsection{Locally Adiabatic Pressure Components}
\label{sec-locad}

	Polytropic models have long been used to model molecular 
clouds.  The discussion of the data on molecular cloud cores and 
Bok globules above suggests that two generalizations of the
standard polytropic theory are needed in order to successfully
model these objects:  First, as mentioned above, it
is necessary to include several different pressure
components, each with pressure $P_j(r)$, so that the
total pressure is $P=\sum P_j$.  If $\gamma_{p,j}$ is
the polytropic index for the $j$th pressure component, then
the polytropic index for the resulting multi--pressure polytrope 
is given by (MH)
\beq
\gamma_p\equiv\frac{d\ln P}{d\ln\rho}
	=\sum_{j=1}^N\left(\frac{P_j}{P}\right)\gamma_{p,j},
\eeq
where $N$ is the number of distinct pressure components.
To simplify the analysis, we make the approximation in
this paper that the spatially varying polytropic index $\gamma_p(r)$
can be replaced by a suitable average, $\ogam_p$.

	Second, it is necessary to consider two (or more)
spatially distinct polytropic regions, each with its own
polytropic index---a {\it composite polytrope}.  
For a core and an envelope, we write
\bed
P_{\rm core} = K_{\rm core} \rho^{{\ogam}_{p,{\rm core}}} \;\;\;\;\;\;   
P_{\rm env} = K_{\rm env} \rho^{{\ogam}_{p,{\rm env}}},   
\eed
where $K_{\rm core}$ and $K_{\rm env}$ are constants.  
Composite polytropes have found wide application in stellar structure:
Milne (1930, 1932) constructed such models for stars with degenerate 
cores.  Henrich \& Chandrasekhar (1941) considered models of a star 
with an isothermal core and an $n = 3$ (radiative) 
envelope, while Sch{\"o}nberg \& Chandrasekhar (1942) investigated the 
effect of compositional gradients in such models.  
The radial stability of a composite polytrope with an isothermal core 
and an $n = 3/2$ ($\gamma_p=5/3$) envelope was studied by Yabushita 
(1975) in the context of neutron stars.  
More recently, Beech (1988) constructed an analytic 
model for low-mass stars, joining an approximate solution for a $n =
3$ ($\gamma_p=4/3$) polytrope onto a $n = 1$ ($\gamma_p=2$) envelope. 

       The specification of $\ogam_p$ in each of the two regions suffices  
for the determination of an entire equilibrium sequence; e.g., the 
total mass as a function of central density.  In order to decide the 
issue of stability for members of that sequence, however,  
one needs to know the response of each pressure component 
to small perturbations.  We shall assume in this work that: (1) the 
perturbations are adiabatic, and; (2) for each component, the time scale 
for internal heat transfer---a direct consequence of the 
perturbation---is long compared to the dynamical time scale.  Following 
MH, we refer to the latter as ``locally adiabatic'' pressure 
components\fn{MH also consider 
the case in which the timescale for internal heat transfer is short 
compared to the dynamical timescale; they term such components ``globally 
adiabatic.''}.  Such components satisfy the usual adiabatic relation 
(Ledoux \& Walraven 1958), 
\bed
\delta\; {\rm ln}\; P_j = \gamma_j \delta\; {\rm ln} \;\rho,
\eed
where the different pressure components
are assumed to be thermally insulated from each other.  
The adiabatic indices $\gamma_j$, then, describe {\it temporal} 
variations of the pressure components, while the polytropic indices, 
$\gamma_{p,j}$, describe {\it spatial} variations.  For locally 
adiabatic polytropes, MH show that the overall adiabatic index is
\beq
\gamma\equiv\frac{\delta \ln P}{\delta \ln \rho}
	=\sum_{j=1}^N \left(\frac{{P}_j}{{P}}\right)\gamma_j.
\label{eq-gamma}
\eeq
For the purposes of this discussion, we shall assume that, as in the 
case of the polytropic indexes, this spatially varying adiabatic index 
can be approximated by a mean adiabatic index in each region, $\ogam$.
Calculations using multi--pressure polytropes (MH) do not suggest a
simple prescription for determining $\ogam_p$ or $\ogam$ from the
properties of the constituent polytropes.  However, these calculations do
indicate that $\ogam_p$ and $\ogam$ are weighted means in the sense
that if there are two pressure components with adiabatic indexes
$\gamma_1$ and $\gamma_2$, for example, then $\gamma_1 < \ogam < \gamma_2$. 
In Paper II we shall infer values of $\ogam_p$ and $\ogam$ from
observations. 

	When $\ogam_p = \ogam$ in a region, the gas is isentropic 
there (i.e., it has a constant specific entropy), and as a result
no internal heat transfer is required 
to maintain the polytropic structure after a perturbation.  
Most studies of polytropic models for interstellar clouds
have assumed that the clouds are isentropic. More generally, 
however, $\ogam_p \neq \ogam$.  For example, small-amplitude, 
undamped Alfv{\'e}n waves have $\gamma_{p,w} 
= 1/2$ and $\gamma_w = 3/2$ (McKee \& Zweibel 1995). 
Note that when $\gamma< \gamma_p$, 
polytropic spheres are convectively unstable
according to the Schwarzschild criterion (Cowling 1941);
none of the known pressure components are in this regime, however.
For the pressure components in molecular clouds,
we have $\gamma_{\rm th}=\gamma_{p, \rm th}$ for the isothermal gas;
$\gamma_{B} \geq \gamma_{p, B}$ for the magnetic field in order
that the cloud be stable against the interchange instability;
and, as we have just seen, $\gamma_w > \gamma_{p,w}$ for the
Alfv{\'e}n waves.  
Since each pressure component has $\gamma\geq\gamma_p$, we expect
$\ogam \geq \ogam_p$ in each part of a composite polytrope model
for molecular clouds.
This is significant, since non-isentropic polytropes are
{\it more} stable than their isentropic counterparts (MH). 
This is essentially due to the fact that as $\ogam$ approaches 
$4/3$, single-component polytropes are unconditionally stable 
(Ledoux \& Walraven 1958).  

\subsection{Core Properties}
\label{sec-core}

All available evidence suggests that the dense, thermally-dominated 
core of a PPC is an unusually quiescent region, in comparison to 
its surroundings.  We presume, as have others, that there is a  
scale above which the cloud is supported against its own weight 
primarily by a mean magnetic field and MHD waves (Shu, Adams, \& 
Lizano 1987; McKee et al 1993); for typical low-mass star-forming 
regions in Galactic molecular clouds, this length scale is about 
0.1 pc.  In the 
core, however, the observed near-thermal velocity linewidths suggest 
that the waves important for maintaining equilibrium on large scales 
are damped at smaller scales; specific damping mechanisms for MHD waves 
were discussed by Zweibel \& Josafatsson (1983). 
The cutoff wavelength for damping along the direction of the field has 
been estimated as (Kulsrud \& Pearce 1969)
\beq
\lambda_{\rm cut} = 0.13 \;{\rm pc} \left(\frac{B}
{20 \mu {\rm G}}\right)  
\left(\frac{n_{{\rm H}_2}}{10^3 {\rm cm}^{-3}}\right)^{-1} 
\left(\frac{\zeta_{CR}}{10^{-17} {\rm s}^{-1}}\right)^{-1/2}, 
\label{eq-damp}
\eeq
where $B$ and $n_{{\rm H}_2}$ are characteristic values of the 
magnetic field strength and H$_2$ density in the relevant region, 
and $\zeta_{CR}$ is 
the cosmic-ray ionization rate. We have assumed the standard balance 
between ionization and recombination in order to compute the coefficient 
(McKee et al 1993).  However, $\zeta_{CR}$ can range between $10^{-18} 
\lae \zeta_{CR} \lae 10^{-16}$ depending on the degree of shielding 
(Caselli et al 1998), meaning that $\lambda_{\rm cut}$ is uncertain by as 
much as a 
factor of 3 in either direction.  Nevertheless, equation (\ref{eq-damp}) 
suggests an important constraint on the models; i.e.\ that the 
thermally dominated core region should 
have a radius $r_{\rm core} \simeq \lambda_{\rm cut}/2$. 

        We now turn to the polytropic and adiabatic indices.  By 
construction, $\gamma_{p,{\rm th}} = \gamma_{\rm th} = 1$.  The value of 
$\gamma_{p,B}$ is somewhat more uncertain; it depends both on the degree 
of isotropy in the contraction of the cloud from a more uniform initial 
state, and on the amount of ambipolar diffusion that has taken place.
For a PPC formed by compressing a uniform density cloud under strict 
flux conservation (i.e.\ no ambipolar diffusion), $\gamma_{p,B} = 4/3$, 
while for a PPC of ``infinite age'' (i.e.\ maximal ambipolar diffusion), 
$\gamma_{p,B} = 0$.  As an intermediate value, 
we shall assume $\gamma_{p,B} = 1$ in the core of our composite 
polytrope, indicating a moderate amount of ambipolar diffusion.  
Since this is identical to the value of $\gamma_p$ for the 
gas pressure, the other pressure component in the core, we adopt
$\ogam_{p,{\rm core}} = 1$; thus, the isothermal solution can be used 
for the core.  In a subsequent paper, we shall use 
observations to estimate a range of likely values for 
$\ogam_{p,{\rm core}}$, and show that this too is consistent with the 
use of the isothermal approximation. 
Since the redistribution of flux is negligible on the time scale of the 
perturbation, we take $\gamma_B = 4/3$, appropriate for adiabatic, 
flux-conserving compression.  
We expect the adiabatic index for the core to have a value
intermediate between that of the thermal gas and the magnetic field:
$1\lae \ogam_{\rm core} \lae 4/3$.

\subsection{Envelope Properties}
\label{sec-envp}
The observed increase of nonthermal linewidth with size to supersonic
values in molecular clouds strongly suggests the presence of MHD turbulence 
at scales larger than about 0.1 pc.  We therefore include a third, 
Alfv{\'e}n-wave pressure component, $\ovl{P}_w$, in the total mean 
pressure of the envelope.

        The assumption of a polytropic relation for Alfv{\'e}n waves 
was examined by McKee \& Zweibel (1995). They showed that as 
long as the waves are linear in amplitude, undamped, and of 
wavelengths shorter than variations in the background density, 
they exert an isotropic pressure obeying $P \propto \rho^{\gamma_p}$ 
with $\gamma_p = 1/2$ (this relation was originally derived by 
Wal{\'e}n 1944).  Further, the adiabatic index of such waves is 
$\gamma_w = 3/2$.  Now, were Alfv{\'e}n waves treated as a locally 
adiabatic pressure component, they would be unconditionally stable, 
since $\gamma_w > 4/3$.  However, as argued by 
MH, undamped Alfv{\'e}n waves are more likely to behave as a globally 
adiabatic pressure component, since wave energy tends to flow from 
high to low density regions in a centrally-concentrated cloud.  These 
authors demonstrated that single-component, globally adiabatic spheres  
are generally less stable than their locally adiabatic cousins.  In 
particular, a critically stable cloud with $\gamma_p = 1/2$ and 
$\gamma_w = 3/2$ has a finite center-to-surface pressure contrast 
of 4.15, and a density contrast of 17.26.  

        In our description, in which Alfv{\'e}n waves are treated in 
an averaged sense, we make the approximation that the waves,
like the mean magnetic field, are locally adiabatic.  
As shown by MH, the location of the critical point for
a multi--pressure polytrope including Alfv{\'e}n waves
is the same as that for a similar multi--pressure polytrope
in which the Alfv{\'e}n waves are replaced by a locally
adiabatic pressure component with $\gamma_{p,j}=1/2$
and with a suitable value of $\gamma_j$.
For a cloud supported entirely by Alfv{\'e}n waves, the
value of $\gamma$ for the equivalent locally adiabatic
component is $\gamma_{w,{\rm eq}}=0.6056$.  However, as other pressure 
components become important, the value of $\gamma_{w,{\rm eq}}$ 
increases, which complicates the choice a suitable value for $\ogam$.
In any case, since $\ogam$ is some weighted mean of the 
$\gamma_j$'s of the individual components, we expect $\ogam<4/3$.

     We now proceed to the actual construction of composite polytropes, 
starting with the illustrative example of the composite isothermal 
sphere.  Note that in what follows, we omit the overline indicating 
the averaged character of $\gamma_p$ and $\gamma$; for the theory, 
it matters only that they are constants. 

\subsection{Isothermal Cores and Envelopes}
\label{sec-iso}

In order to gain a feel for composite equilibria, we begin with the 
illustrative example of the composite isothermal sphere.  Consider a 
core of isothermal gas at temperature $T_{\rm core}$ surrounded 
by an isothermal envelope at temperature $T_{\rm env} > T_{\rm core}$.  
Let $c_{\rm core,~env}$ denote the constant isothermal
sound speed in the core and envelope, respectively.  
Define the jump in conditions at the interface by
\beq
\tau_i\equiv \left(\frac{\ce}{\cc}\right)^2
	     = \frac{T_{\rm env}}{T_{\rm core}}>1,
\label{eq-tau}
\eeq
where in the second step
we have assumed both regions to be of the same composition.

	Introducing the customary dimensionless 
variables $\xi$ and $\psi (\xi)$ 
(Chandrasekhar 1939), the density, radius, and mass within the core  
are given by
\beqa
\rho & = & \rho_c {\rm exp} (-\psi), \label{eq-rho1} \\
r & = & \rch \xi \equiv \frac{\cc}{(4\pi G\rho_c)^{1/2}} 
\xi, \label{eq-rc} \\
M(r) & = & 4\pi \rho_c \rch^3 \xi^2 \frac{d\psi}{d\xi}, 
\label{eq-Mc}
\eeqa
where $\rho_c$ is the density at the center ($r = 0$)
and $\rch$ is a characteristic radius proportional to 
the Jeans length in the core.  In the
interest of notational simplicity, it is desirable to retain $\xi$ and
$\psi$ in the envelope as well as in the core.  In the envelope then,
\beqa
\rho & = & \rho_e {\rm exp} (-\psi), \label{eq-rho2} \\
r & = & \rch \xi \equiv \frac{\ce}{(4\pi G\rho_e)^{1/2}} \xi, 
\non
\\
M(r) & = & 4\pi \rho_e \rch^3 \xi^2 \frac{d\psi}{d\xi}, \non
\eeqa
where $\rho_e$ is an undetermined constant.  

At the interface of core and envelope, the radius, pressure, and mass 
must be continuous (but {\it not} the density; the present class of models 
differs from composite {\it stellar} models in precisely this sense).  The 
above expressions therefore give 
\beqa
\cc^2 \rho_c {\rm exp} (-\psi_-) & = & \ce^2 
\rho_e {\rm exp} (-\psi_+),  \non \\
\cc \rho_c^{-1/2} & = & \ce \rho_e^{-1/2}, 
\label{eq-iso} \\
\cc^3 \rho_c^{-1/2} \psi_-' & = & \ce^3 
\rho_e^{-1/2} \psi_+', \non
\eeqa
where a prime denotes the derivative of a function with respect to its 
lone argument, a subscript `$-$' indicates a quantity evaluated on the
core side of the interface, and a subscript `$+$' indicates a quantity 
evaluated on the envelope side of the interface.  

        The equilibrium in both regions is governed by the Lane-Emden 
equation, 
\beq
\frac{1}{\xi^2} (\xi^2 \psi')' = {\rm exp}(-\psi). \label{eq-isole}
\eeq
The unique solution of equation (\ref{eq-isole}), subject to the
standard (i.e.\ regular) boundary conditions, 
\bed
\psi = 0, \;\;\;\;\;\; \psi' = 0 \;\;\; {\rm at} \;\;\; 
\xi = 0,
\eed
is the isothermal function tabulated by Chandrasekhar \& Wares (1949).
Thus, for a given dimensionless core radius, $\xi_i = r_i/\rch, \;
\psi_-$, and $\psi_-'$ are known (a subscript $i$ denotes quantities 
that are continuous across the interface).  Equations (\ref{eq-iso}) 
are then three equations in three unknowns: $\psi_+, \;\psi_+'$, and
$\rho_e/\rho_c$.  Solving these, one obtains the following relations: 
\beq
\psi_+ = \psi_- + 2 \;{\rm ln}~\tau_i, \;\;\;\;\;\; \psi_+' = \psi_-'/\tau_i, 
\;\;\;\;\;\; \rho_e/\rho_c = \tau_i.
\label{eq-jiso}
\eeq

	Equations (\ref{eq-jiso}) give starting values for the integration of 
equation (\ref{eq-isole}) out to some radius $r_s \equiv \rch \xi_s$, 
where an outer boundary condition is imposed.  We imagine that the 
composite cloud is ultimately surrounded by a tenuous (compared to 
the envelope) intercloud medium, 
whose entire dynamical effect is to exert a finite, constant pressure 
everywhere on the outer boundary.  This situation is effectively the 
same as is assumed for the well-known bounded isothermal sphere (Ebert 
1955; Bonnor 1956) except that we suppose $P_s$ to be comprised of 
both thermal and nonthermal components.  Using equations (\ref{eq-tau}), 
(\ref{eq-rho1}), (\ref{eq-rho2}) and (\ref{eq-jiso}), we find that this 
pressure is related to the central pressure, $P_c = \cc^2 \rho_c$, via
\beq
\frac{P_c}{P_s} = \tau_i^{-2} \;{\rm exp} (\psi_s). \label{eq-bciso}
\eeq
Note that the corresponding density contrast is simply $\tau_i (P_c/P_s)$.  
Alternatively, one may specify a given configuration by
its ratio of core-to-total radius and/or mass.  The definitions of $M$ 
and $r$ along with equations (\ref{eq-jiso}) yield
\beqa
\frac{r_i}{r_s} & = & \frac{\xi_i}{\xi_s}, \label{eq-rrat} \\
\frac{M_i}{M_s} & = & \tau_i^{-1} \frac{(\xi^2 \psi')_-} 
{(\xi^2 \psi')_s}. \non
\eeqa
Note that when 
expressing products of quantities, any one of which is not continuous 
across the interface, we retain the `$-$' subscripts for the entire product.

\subsection{Isothermal Cores with Polytropic Envelopes}
\label{sec-poly}

	We next consider composite polytropes with isothermal
cores and polytropic envelopes as an approximate model
for molecular clouds.  (In principle, the core of a
composite polytrope can have an arbitrary polytropic index, but
we restrict consideration here to the special case in which
it is isothermal.)  As remarked above, composite polytropes
have long been used in studies of stellar structure.
The general matching conditions for two-component spheres of differing 
polytropic indices $n$ were presented by Chandrasekhar (1939).  

	  The Lane-Emden equation governing the envelope is 
\beq
\frac{1}{\xi^2} (\xi^2 \theta')' = \Delta \theta^n, \;\;\;\;\;\; 
\Delta \equiv -\frac{|n+1|}{n+1} = \pm 1, \label{eq-ptle}
\eeq
where the suitably non-dimensionalized variables for the envelope are 
\beqan
\rho & = & \rho_e \theta^n, \;\;\;\;\;\; P = K_{\rm env} \rho^{1+1/n} = 
K_{\rm env} \rho_e^{1+1/n} \theta^{n+1}, \\
r & = & \left[ \frac{|n+1|}{4\pi G} K_{\rm env} \rho_e^{1/n-1}
\right]^{1/2} \xi, \\
M(r) & = & 4\pi \Delta\rho_e \rch^3 \xi^2 \theta', 
\eeqan
where we have used the continuity of $\xi$ to express the
mass in terms of characteristic radius for the isothermal
core, $\rch$.  Recall that $n$ and $\gamma_p$ are related via 
$\gamma_p = 1 + 1/n$.  The matching 
conditions at the core-envelope interface are  
\beqa
\cc^2 \rho_c {\rm exp} (-\psi_-) & = & K_{\rm env} \rho_e^{1+1/n} 
\theta_+^{n+1}, \non \\
\cc \rho_c^{-1/2} & = & (|n+1| K_{\rm env} \rho_e^{1/n-1})^{1/2}, 
\label{eq-poly} \\
\rho_c \psi_-' & = & \Delta \rho_e \theta_+'.
\non
\eeqa
Generalizing the definition of $\tau_i$ (eq. \ref{eq-tau}) to the case
of a polytropic envelope, we have
\beq
\tau_i\equiv\frac{c_+^2}{\cc^2}
	=\frac{\rho_-}{\rho_+}=\frac{\rho_c {\rm exp} (-\psi_-)}{
	\rho_e \theta_+^n}.
\eeq
The first two of equations (\ref{eq-poly}) can be combined to give
\beq
\left(\frac{\rho_e}{\rho_c}\right)^2 {\rm exp} (\psi_-) \theta_+^{n+1}
= |n+1|. 
\label{eq-thetai}
\eeq
Equations (\ref{eq-poly})--(\ref{eq-thetai}) then yield the jump 
conditions:
\beqa
\theta_+ & = & [\tau_i^2 \;|n+1|\;{\rm exp} (\psi_-)]^{1/(1-n)}, \non \\
\theta_+' & = & \Delta \tau_i \psi_-' {\rm exp} (\psi_-) \theta_+^n,
\label{eq-jpoly} \\
\rho_e/\rho_c & = & \tau_i^{-1} {\rm exp} (-\psi_-) \theta_+^{-n}. \non
\eeqa
These are the equivalents of equations (\ref{eq-jiso}) in the case of a 
polytropic envelope.

The ratio of core-to-total radius is again given by equation 
(\ref{eq-rrat}), while the mass ratio is 
\bed
\frac{M_i}{M_s} = \Delta \tau_i \;{\rm exp} (\psi_-) \theta_+^n 
\frac{(\xi^2 \psi')_-}{(\xi^2 \theta')_s}.
\eed
Finally, the center-to-surface pressure and density contrasts are
given by 
\beqa
\frac{P_c}{P_s} & = & |n+1| \;[\tau_i \;
{\rm exp}(\psi_-)\;\theta_+^n]^2 \theta_s^{-n-1} \label{eq-bcpoly1} \\
\frac{\rho_c}{\rho_s} & = & \tau_i \;{\rm exp}(\psi_-) (\theta_+/\theta_s)^n. 
\non
\eeqa

        As discussed in {\S}{\S}2.2-2.3 above, the adiabatic indexes 
for both core and envelope are expected to be less than 4/3.  As a 
result, the composite polytrope must be bounded by a medium of finite 
and constant pressure at its surface (MH).  Also note that the 
envelope temperature $T \propto P/\rho \propto \rho^{1/n}$
is an increasing function of $\xi$ in NIPs (Shu et al 
1972; Viala \& Horedt 1974).  As mentioned in {\S}\ref{sec-envp}, we 
interpret the departure from isothermality in the envelope as arising 
entirely from the presence of mean magnetic fields and Alfv{\'e}nic 
turbulence, approximated as isotropic pressure components.  The latter 
appear as nonthermal contributions to the total sound speed, given by 
(McKee \& Zweibel 1995)
\beq
c^2 (r) \equiv \frac{P (r)}{\rho (r)} = \sigma_{\rm th}^2 + \frac{3}{2} 
\sigma_w^2 (r) + \frac{1}{2} v_A^2 (r),
\label{eq-ss}
\eeq
where $\sigma_{\rm th}$ is the (constant) thermal speed, $\sigma_w (r)$ 
is the velocity dispersion associated with Alfv{\'e}n waves, and $v_A (r) 
\equiv B(r)/\sqrt{4\pi \rho (r)}$ is the Alfv{\'e}n speed.  
In our model, $\sigma_w$ is the same as the nonthermal
velocity dispersion $\sigma_{\rm nt}$.
Note that the total velocity dispersion is related to $c(r)$ by 
\beq
\sigma (r) = [\sigma_{\rm th}^2 +  \sigma_w^2 (r)]^{1/2}  = 
[\frac{2}{3} c^2 (r) + \frac{1}{3} \sigma_{\rm th}^2  
- \frac{1}{3} v_A^2 (r)]^{1/2}.
\label{eq-vel}
\eeq

\subsection{Stability of Composite Configurations}
\label{sec-stab}

While polytropic models of molecular clouds clearly represent a 
simplification of the real situation, they have a distinct 
advantage over more detailed models, in that the analysis of their 
stability is straightforward.  We note, however, that previous studies 
disagree as to whether such a cloud responds 
isothermally or adiabatically to radial perturbations applied at its 
surface.  We review this situation in the Appendix, and justify our 
choice of the adiabatic stability criterion adopted here. 

The equation of radial motion for small, adiabatic perturbations about 
equilibrium is, quite generally (Eddington 1926; Ledoux \& Walraven 1958)
\beqa
\frac{d^2 h}{dr^2} &+& \left[ \frac{4}{r} + \frac{1}{\gamma}\frac{d\gamma}
{dr} - \frac{G M(r)\rho}{r^2 P} \right] \frac{dh}{dr} \nonumber \\ 
&+& \left[ 
\frac{\omega^2 \rho}{\gamma P} + \frac{3}{\gamma r}\frac{d\gamma}{dr} - 
\left(3 - \frac{4}{\gamma} \right)\frac{G M(r)\rho}{r^3 P} \right] h = 0,
\label{eq-pert}
\eeqa
where $h \equiv \delta r/r$ is the relative displacement of a fluid 
element at a radius $r$ and $\omega$ is the 
frequency of the oscillations.  
We assume that the adiabatic index, $\gamma$, can take on any (positive 
and constant) 
value in either the core or the envelope.  Thus, in general, $\gamma$ 
has a discontinuity at the interface.  For the reasons outlined {\S}
\ref{sec-virial}, we confine our interest to critically stable 
clouds, setting $\omega = 0$ in what follows. 

Under these assumptions, equation (\ref{eq-pert}) becomes, in the 
isothermal core, 
\beq
\frac{d^2 h_{\rm core}}{d\xi^2} + \left( \frac{4}{\xi} - \psi' 
\right) \frac{dh_{\rm core}}{d\xi} - (3 - 4/\gamc) 
\frac{\psi'}{\xi} h_{\rm core} = 0 
\label{eq-isoh} 
\;\;\;\;\;\; {\rm (isothermal)}.
\eeq
When $\gamc = 1$, Yabushita (1974) showed that there exists 
an exact solution of equation (\ref{eq-isoh}); namely,
\bed
h_{\rm core} (\xi) = \epsilon \left[ \frac{\psi'}{\xi} {\rm exp} 
(\psi) - 1 \right], 
\eed
where $\epsilon$ is a small, dimensionless constant.

The solution of equation (\ref{eq-isoh}) must be matched onto the 
corresponding solution in the envelope, $h = h_{\rm env}$.  For an 
isothermal envelope, the governing equation is identical in form 
to equation (\ref{eq-isoh}); for a polytropic envelope, one has  
\beqa
\frac{d^2 h_{\rm env}}{d\xi^2} &+& \left[ \frac{4}{\xi} + (n+1) 
\frac{\theta'}{\theta} \right] \frac{dh_{\rm env}}{d\xi} \nonumber \\
&+& (3 - 4/\gamma_{\rm env})(n+1) \frac{\theta'}{\theta\xi} 
h_{\rm env} = 0 
\;\; {\rm (polytropic)}. 
\label{eq-polyh} 
\eeqa

By definition, $h$ must be continuous at the interface; i.e.\ 
$h_{\rm env}(\xi_i) = h_{\rm core}(\xi_i)$.  Its derivative must 
satisfy the condition (Ledoux \& Walraven 1958)
\bed
\left( \frac{dh_{\rm env}}{d\xi} \right)_+ =  
\frac{\gamc}{\game} \left(\frac{dh_{\rm core}}
{d\xi} \right)_- + 
\frac{3 h_{\rm core}}{\xi_i} \left(\frac{\gamc}
{\game} - 1 \right), 
\eed
which takes into account the possibility of a discontinuity in $\gamma$ 
across the interface.  

       The condition for the critical point is that the Lagrangian 
variation in the pressure at the boundary vanish,
$(\delta P/\delta r)_{{M_s}}=0$ (Ledoux \& Walraven 1958; MH). 
This places the following restriction on $h_{\rm env}$:
\beq
\left( \frac{dh_{\rm env}}{d\xi} + 3\frac{h_{\rm env}}{\xi} 
\right)_{\xi = \xi_s}  = 0.  \label{eq-bcgen} 
\eeq
This condition holds for both isothermal and polytropic envelopes, 
and differs from those expressed by equations (\ref{eq-bciso}) and 
(\ref{eq-bcpoly1}), in that it holds {\it only} for critically 
stable states.  Finally, note that although we shall take $n < -1$ 
($0<\gamma_p<1$)
in what follows, the results of this section hold for all $n \neq -1$. 
     
\subsection{Solution Procedure}
\label{sec-proc}
 
In practice, the solution procedure for critical states is as follows: 
\begin{quote}
(1) Choose a model by specifying five parameters: $n$ (or $\gamma_p),
\; \gamc, \; \game, \; \xi_i$, and $\tau_i$. 

(2) Solve the Lane-Emden equation for an isothermal sphere on the 
interval $[0, \xi_i]$; i.e., find $\psi (\xi)$ and $\psi'$. 

(3) Using $\psi_-, \; \psi_-', \; h_{\rm core}(\xi_i)$, and 
$(dh_{\rm core}/d\xi)_-$ as starting values, calculate $\theta_+,\; 
\theta_+',\; h_{\rm env} (\xi_i)$, and $(dh_{\rm env}/d\xi)_+$.  These are 
the starting values for the integration of the two ODES (\ref{eq-isole}) 
and (\ref{eq-isoh}) (isothermal envelope) or (\ref{eq-ptle}) and 
(\ref{eq-polyh}) (polytropic envelope). 

(4) Integrate the system outward in $\xi$ until the outer boundary 
condition (\ref{eq-bcgen}) is satisfied.  
\end{quote}
The resulting unique solution yields, as dimensionless output parameters: 
(i) the center-to-surface pressure and density contrasts; (ii) the 
core-to-total radius ratio; and (iii) the core-to-total mass ratio.  

\section{SINGLE-COMPONENT POLYTROPES: A REVIEW}
\label{sec-sing}

	Consider the variation of the mass $M$ of a 
single-component NIP ($-\infty < n < -1$;
$0<\gamma_p<1$) if the surface pressure and entropy are 
held fixed while the central density is increased.
For fixed surface conditions, the behavior of $M$ follows that
of the the dimensionless mass $\mu $ defined by MH,
\beq
\mu \equiv\frac{M}{c_s^4/(G^3P_s)^{1/2}}
	=\frac{M}{c_s^3/(G^3\rho_s)^{1/2}}.
\label{eq-surfnd}
\eeq
The sequence of $\mu $ versus $\rho_c/\rho_s$ 
ranges from zero at $\rho_c/\rho_s=1$ up to some maximum 
$\mu _{\rm max}$, beyond which it oscillates about the
value corresponding to a singular polytropic sphere
as $\rho_c/\rho_s\rightarrow\infty$
(Chandrasekhar 1939; MH), i.e.
\bed
\mu _{\infty} = \left( \frac{2}{\pi}\right)^{1/2} 
\frac{(4 - 3\gamma_p)^{1/2} \gamma_p^{3/2}}{(2 - \gamma_p)^2} =   
\left( \frac{2}{\pi}\right)^{1/2} \frac{|n - 3|^{1/2} 
|n + 1|^{3/2}}{(n - 1)^2}. 
\eed
The sequence is qualitatively similar for all values of 
$0<\gamma_p<6/5$: 
the maximum possible mass of a polytrope
in this range is determined by the conditions at its surface.
(The same conclusion holds if $c_s$ is replaced by the
rms value $\lb c^2\rb^{1/2}$.)

	The largest {\it stable} density drop
along this sequence occurs at the critical mass 
$\mu _{\rm cr}$, which is less than or equal to $\mu _{\rm max}$.  
In the isentropic case, $\gamma = \gamma_p$, one has $\mu _{\rm cr} 
= \mu _{\rm max}$ (Shu et al 1972).  In this case, all higher density, 
lower mass models are 
gravitationally unstable.  When $\gamma > \gamma_p,\; \mu _{\rm cr} 
< \mu _{\rm max}$, even though the corresponding 
$(\rho_c/\rho_s)_{\rm cr}$ is greater 
than in the isentropic case.  This behavior is illustrated in Figure 1,  
which displays the partial equilibrium sequences of both an $n = -3\; 
(\gamma_p = 2/3)$ polytrope 
and the isothermal sphere.  Recall that when $\gamma < \gamma_p$, 
both polytropic and isothermal spheres are convectively unstable, 
according to the Schwarzschild criterion (Cowling 1941).

	  Sometimes we know the conditions at the {\it center} of
the polytrope, and can ask how much mass can be supported.  In this  
case, the equilibrium mass can increase without limit as the surface 
pressure and density are lowered (Viala \& Horedt 1974); however, 
such equilibria are stable only if they are sufficiently non-isentropic.
Noting that $c_s^2=P_s/\rho_s\propto \rho_s^{1/n}$, we have 
$M\propto \mu \rho_s^{(3-n)/2n}$; as $\rho_s/\rho_c\rightarrow
0$ for fixed $\rho_c$, $\mu \rightarrow \mu _\infty$ and
$M\rightarrow\infty$ for $n>3$ and $n<-1 \;(0 \leq 
\gamma_p < 4/3$).  

\section{THE COMPOSITE ISOTHERMAL SPHERE}
\label{sec-isor}
\subsection{Pressure and Density Structure}

	In Figures 2$a$ and $b$, we plot, for various values of $\tau_i$, 
the critical density contrast of the composite isothermal sphere (CIS),   
$(\rho_c/\rho_s)_{CI, \rm cr}$, as a function of $r_i/r_s$ and $M_i/M_s$, 
respectively.  We first present results for the isentropic 
case only, i.e.\ $\gamc = \game = \gamma_p = 1$. 
Note that the corresponding pressure contrasts are simply a factor of 
$\tau_i$ lower than $(\rho_c/\rho_s)_{CI,\rm cr}$.  
The left-hand border of Fig.\ 2
corresponds to the case of an almost pure envelope with a core
at the origin, whereas the right-hand border corresponds
to the case of an almost pure core with an envelope at the surface.  
In each case, the limiting values of the curves at the borders are 
$(\rho_c/\rho_s)_{CI,\rm cr} \goto \tau_i (\rho_c/\rho_s)_{BE}$, 
indicating that $(P_c/P_s)_{CI,\rm cr} \goto (P_c/P_s)_{BE}$, as expected.  
Note that for values 
of $\tau_i \gae 5$, configurations with the {\it same} $r_i/r_s$ can have 
{\it different} $(\rho_c/\rho_s)_{CI,\rm cr}$.  These equilibria are still 
unique, however, as evidenced from the absence of this feature in 
Fig.\ 2$b$, which employs the Lagrangian variable $M_i/M_s$.  

      For each value of $\tau_i$, there is a maximum 
in $(\rho_c/\rho_s)_{CI,\rm cr}$, which increases with increasing 
$\tau_i$.  As $\tau_i \goto \infty$, the density contrast across the interface 
becomes infinite, and thus an infinite center-to-edge density contrast can 
be supported.  The maximum {\it pressure} contrast, however, is finite, and 
can be estimated as follows.  Since the core becomes less  
important with increasing $\tau_i$ (it is already only 5\% of the total radius 
and 1\% of the total mass at $\tau_i = 5$), one can view this region in the 
large-$\tau_i$ limit as a tiny, cold and dense ``pea'' in the midst of an 
essentially infinite, hot and tenuous background.  These are precisely the 
conditions pertaining to the bounded isothermal sphere.  Thus, on the scale at 
which pressure gradients in the core are significant, $\xi \lae \xi_i$, 
the maximum pressure contrast 
across the core alone will be $(P_c/P_i)_{\rm cr} \approx (P_c/P_s)_{BE}$. 
On the other hand, on the scale at which pressure gradients in the envelope  
are significant, $\xi \gg \xi_i$, the region within the core is 
essentially structureless; its only significance for the equilibrium is 
to set the interface values of various quantities.  
Thus, the expected maximum pressure contrast in the envelope alone will 
again approach the Bonnor-Ebert value.  Together, these results imply 
a {\it total} pressure contrast in the large-$\tau_i$ limit of
\bed
\left(\frac{P_c}{P_s}\right)_{CI,\rm cr} = \left(\frac{P_c}{P_i}
\right)_{\rm cr}
\left(\frac{P_i}{P_s}\right)_{\rm cr} \approx \left(\frac{P_c}{P_s}
\right)_{BE}^2 \simeq 197.
\eed
The critical stability curves of pressure contrast versus $r_i/r_s$ 
do indeed approach this asymptotic limit for $\tau_i \gg 1$ (Fig.\ 2$c$). 
Correspondingly, the total pressure contrast for a composite polytrope
consisting of $N$ components, each separated from the other by a large
value of $\tau_i$, approaches $(14.04)^N$.

\subsection{Radius and Mass}
\label{sec-iradm}

The dimensionless outer radius of the critical CIS can be much larger 
than that of the Bonnor-Ebert sphere; e.g.\ when $\tau_i = 2$ and 
$r_i/r_s = 0.1$, it extends to $\xi_{CI,{\rm cr}} = 15.7$, versus 
$\xi_{BE} = 6.45$.  Note that $\xi_{CI,{\rm cr}}$ is a function of the 
core-to-total radius ratio $r_i/r_s$.  As this ratio approaches zero 
(i.e., a pure envelope), one finds that $\xi_{CI,{\rm cr}} \goto \tau_i 
~\xi_{BE}$, or $\xi_{CI,{\rm cr}} = 12.9$ in the case considered here.  
Hence, the outer radius attains a maximum at some intermediate $r_i/r_s$. 

	Because the temperatures of the densest observed regions 
of molecular clouds consistently fall in the range 10-20 K, it is 
convenient to normalize
the mass with respect to this value rather than the surface value.
On the other hand, the pressure at the surface of a cloud
is dictated by the ambient conditions, and is
often better determined than the pressure in the cloud interior.   
We therefore
introduce a hybrid dimensionless mass based on the temperature
in the core and the pressure at the surface of the envelope:
\beq
m(r) \equiv \frac{M(r)}{\cc^4/(G^3 P_s)^{1/2}}.
\eeq
For fixed $\cc$ and $P_s$, $m(r)$ is directly proportional the
dimensional mass $M(r)$.
By contrast, the dimensionless mass
$\mu $ introduced above is normalized to the characteristic 
Jeans mass at $r$; the two are related by
\beq
\frac{\mu (r)}{m(r)}=\frac{\cc^4}{c^4(r)}
        \left[\frac{P(r)}{P_s}\right]^{1/2}=
        \frac{1}{\tau^2}\left[\frac{P(r)}{P_s}\right]^{1/2},
\label{eq-mum}
\eeq
where
\bed
\tau\equiv\frac{c^2}{\cc^2}.
\eed
For a CIS, $\tau=1$ in the core and $\tau=\tau_i$ in the envelope.
        
        The total dimensionless mass of a composite isothermal sphere 
is then
\beq
m_{CI} = \mc + m_{\rm env, I}, 
\label{eq-cimass}
\eeq
where
\beqa
\mc & \equiv & \frac{P_s^{1/2} G^{3/2}}{\cc^4} 
M_{\rm core} =
\left(4\pi \frac{P_c}{P_s}\right)^{-1/2} (\xi^2 \psi')_- 
\label{eq-cicore} \\
m_{\rm env,I} & \equiv & \frac{P_s^{1/2} G^{3/2}}{\cc^4} 
M_{\rm env,I} = 
\left(4\pi \frac{P_c}{P_s}\right)^{-1/2} \; \tau_i \; [ (\xi^2 
\psi')_s - (\xi^2 \psi')_+]. \non
\eeqa

        In the limit that the CIS is a pure core ($r_i/r_s \goto 1$, 
$\xi_- \goto \xi_s$), $m_{CI} = \mc$ must approach  
$m_{BI}$, the dimensionless mass of the bounded isothermal sphere,  
\beq
m_{BI}=\frac{P_s^{1/2}G^{3/2}}{c^4} M=
        \left(4\pi\,\frac{P_c}{P_s}\right)^{-1/2}(\xi^2\psi')_{BI,s}.
\label{eq-mbi}
\eeq
The last step, which follows from equations (\ref{eq-rc}) and
(\ref{eq-Mc}), shows that $m_{BI}$ indeed agrees with $\mc$
from equation (\ref{eq-cicore}) in this limit.  In the opposite limit 
in which the CIS is a pure envelope ($r_i/r_s \goto 0$, 
$\xi_+ \goto 0$), 
equation (\ref{eq-mum}) implies $m_{CI}\goto (c/\cc)^4m_{BI}
=\tau_i^2m_{BI}$.  The maximum mass of the CIS occurs for the
pure envelope (see Figure 3), and since $m_{BI}\leq m_{BE}$, we find
that the upper limit on the mass of a CIS is
\beq
m_{CI}\leq\tau_i^2 m_{BE}.  
\label{eq-mcilim}
\eeq

        The above expressions are also valid for the critically stable
CIS.  The critically stable pure core or envelope is the Bonnor-Ebert 
sphere with $\mu _{BE}=m_{BE}=1.1822$.  Note that because the
critical values of $P_i/P_s, \; \mc$, and $m_{\rm env,I}$ depend on
the fractional core radius $r_i/r_s, \;m_{CI,{\rm cr}}$ {\it is a function
of} $r_i/r_s$.  We plot $m_{CI,{\rm cr}}/m_{BE}$, 
in the isentropic case, for various values of $\tau_i$ in Figures 3$a$ 
and $b$.  The individual core and envelope components are also shown.   
Note that in general, as $\tau_i$ increases, the 
core becomes a lesser contributor to the overall mass of the cloud.  
As $\tau_i \goto \infty$, we recover the pure envelope case.
Finally, note that the above mass expressions are quite
general; they apply to both isentropic ($\gamma = \gamma_p$) and
non-isentropic ($\gamma \neq \gamma_p$) spheres.

\subsection{The Non-isentropic CIS}
\label{sec-niscis}

	Consider now the much larger class of non-isentropic, critical CISs; 
i.e.\ those with $\gamma \neq 1$.  We are most interested in the range 
$1 < \gamma < 4/3$, as discussed in {\S}\ref{sec-envp}.  Single-component 
isothermal 
spheres with $\gamma$ in this range were examined by Yabushita (1968), 
who showed that the dimensionless critical radius and pressure 
contrast of such spheres approached infinity as $\gamma \goto 32/25$.
The critical radius and pressure contrast remain infinite for
$32/25<\gamma<4/3$ (MH).

   These features of single-component spheres can largely be carried 
over to the CIS.  That is, CISs with $\gamc > 1$ and/or $\game > 1$ 
have pressure contrasts in excess of those found in the isentropic 
case, but smaller 
masses.  The maximum pressure contrasts (i.e., as a function of 
$r_i/r_s$) for CISs with $\tau_i = 2$ and $\tau_i = 3$ are plotted as 
contours in the $(\gamc, \game)$ plane in Figures 4$a$ and $b$.  
Note that as $\gamc$ and $\game$ approach the value 32/25, the 
various physical 
quantities diverge, and it proves difficult to integrate the equations 
all the way up to the singular point.  This is a feature of Fig.\ 4 
and several of our subsequent plots; however, we need not be concerned 
that any essential qualitative behavior is being missed.

     To outline the important qualitative changes in the non-isentropic CIS, 
let us restrict consideration to $\tau_i = 1$ and $1 < \gamma \lae 1.25$.  
Then the critical mass extrema always obtain in 
the limit of a pure core or envelope, depending on the relation between 
$\gamc$ and $\game$.  If $\gamc > 
\game$, then the minimum 
(maximum) mass is given by a pure core (envelope); and conversely if 
$\gamc < \game$.  Further, these extrema are given 
by the usual 
bounded isothermal sphere 
sequence beyond the mass peak (Fig.\ 1).  When $\tau_i > 1$ and/or 
$\gamc$ or $\game \gae 1.25$, the mass extrema no 
longer necessarily occur 
for a pure core or envelope: the values of $r_i/r_s$ corresponding to 
the mass peak depend on the exact values of $\tau_i, \;\gamc$, 
and $\game$.  As previously discussed, $\game > 1 
\imp m_{\rm env,I} < m_{BE}$, even in the pure envelope limit.  On the 
other hand, for $\tau_i > 1$, this effect competes with a relatively 
{\it larger} mass contribution from the envelope, due to its increased 
temperature ({\S}\ref{sec-iradm}).  This combination sometimes leads to  
a maximum in $m_{CI,{\rm cr}}\; (\leq m_{BE})$ at intermediate $r_i/r_s$, 
an effect manifested by
the slight asymmetry in the $(P_c/P_s)_{CI,{\rm cr}}^{\rm max}$ contours 
between corresponding pairs of $(\gamc, \game)$ in Fig.\ 4$a$.

\subsection{Summary---CIS}
\label{sec-cissum}

        Whereas the bounded isothermal sphere 
(usually assumed to
be isentropic, so that $\gamma=1$) has a maximum density and pressure
contrast between the center and the surface of a factor 14.04, 
composite isothermal spheres can have much larger drops: for two 
components with a temperature jump $\tau_i$, the pressure drop can 
be up to a factor $(14.04)^2=197$ (in the limit of large $\tau_i$), 
and the density drop a factor $\tau_i$ larger.  
Non-isentropic isothermal spheres, whether composite or not,
can have very large density and pressure drops provided
$\gamma$ is close to 4/3.
We attribute the increased stability of composite polytropes
and non-isentropic polytropes to the fact that they both
have an entropy that increases outwards.  The entropy is
proportional to $\ln(P/\rho^\gamma)$,
which increases outward as $ (\gamma-\gamma_p)
\ln(\rho_c/\rho)$ for a non-isentropic polytrope and
as $\gamma\ln\tau_i$ across a jump in a composite polytrope.

        The core mass of a CIS can never exceed its Bonnor-Ebert 
mass ($\mc\leq m_{BE}=1.18$), 
and the mass of the entire CIS can never 
exceed the Bonnor-Ebert mass of the envelope ($\mu _{CI}\leq 1.18$).  
As shown in Fig.\ 3 
and in equation (\ref{eq-mcilim}), however, the dimensionless mass of the CIS
can
exceed $m_{BE}$ whenever $\tau_i > 1$, since $m_{CI,{\rm cr}}\goto
\tau_i^2m_{BE}$ in the limit of a pure envelope.  
The dimensionless
mass of a non-isentropic CIS is less than that of an isentropic one 
(with the same $\tau_i$).

        In order to compare the mass of a CIS with that of an actual 
dense core or globule, we need to re-express equation (\ref{eq-cimass}) 
in dimensional form.  Scaling to the representative values $T_{\rm core} 
= 10$ K and $P_s/k = 2 \times 10^4$ cm$^{-3}$ K 
(Boulares \& Cox 1990; we have subtracted the cosmic ray pressure, since 
cosmic rays penetrate the molecular gas as well), one finds  
\beqa
M_{CI} & = & \frac{c_{\rm core}^4}{P_s^{1/2} G^{3/2}} m_{CI} = 
2.2 \;M_{\sun} \left(\frac{T_{\rm core}}{10\; {\rm K}}\right)^2 
\nonumber \\
&\times &\left(\frac{2 \times 10^4 \; {\rm cm}^{-3} {\rm
K}}{P_s/k}\right)^{1/2} m_{CI},
\label{eq-cimd}
\eeqa
where a mean molecular weight of $2.33\; m_H$ has been assumed.  The 
quantity $m_{CI}$, scaled by $m_{BE}$, can be read off from Fig.\ 3.
Assuming these representative values, the mass of the isentropic, 
critical CIS 
falls within the range $2.6\; M_{\sun} \leq M_{CI,{\rm cr}} \leq 2.6\; 
\tau_i^2 M_{\sun}$ (see eq.\ \ref{eq-mcilim}).  
The total molecular 
masses of dense cores and globules derived from ammonia and C$^{18}$O 
measurements range between a few $M_{\sun}$ and a few tens of $M_{\sun}$ 
({\S}\ref{sec-obs}), so critical CISs with $\tau_i \lae 5$ are in good 
agreement on this count.  

	Since observations show that PPCs and Bok globules
both have substantial nonthermal motions in their envelopes,
CISs are primarily of academic interest.  We turn now to
the more realistic case in which the envelope has a substantial
nonthermal pressure.

\section{THE COMPOSITE POLYTROPIC SPHERE}
\label{sec-polyr}

	In models having a polytropic envelope, we assume that the 
thermal velocity in the envelope is the same as that in the core, 
$\sigma_{\rm th,\ env} = \sigma_{\rm th,\ core}$.  The polytropic 
temperature, $T (r) \propto c^2 (r)$, may be interpreted as the sum 
of the kinetic temperature, $T_{\rm kin}$ (fixed at the core value, 
$T_{\rm core}$), and an ``effective'' temperature, $T_{\rm eff} (r) 
\propto (3 \sigma_w^2 + v_A^2)/2$ (see eq.\ \ref{eq-ss}).  
At the inner edge of the 
envelope, $T_{{\rm eff},+} = (\tau_i - 1) \;T_{\rm core}$, so that 
when $\tau_i > 1$, the jump in $T$ across the interface is assumed 
to be {\it entirely} due to a nonzero value of 
$T_{\rm eff} (r_i)$.\fn{In some cases, e.g., Bok globules exposed to 
the interstellar radiation 
field, the kinetic temperature may increase outward.  Then, an assumed 
$T_{\rm kin} (r)$ gives $\sigma_{\rm th} (r)$ and $T_{\rm eff} (r) = 
T(r) - T_{\rm kin} (r)$.}  We focus on the case of NIPs for the 
envelope ($n<-1,\ 0<\gamma_p<1$).
  
\subsection{Pressure and Density Structure}
 
Beginning again with the isentropic case, in Figure 5$a$ we plot, for 
various negative values of $n$ and for $\tau_i = 1$, the critical density 
and pressure contrasts of the composite polytropic sphere (CPS).
In Figs.\ 5$b$ and $c$, results are given for $\tau_i = 2$ and 
$\tau_i = 3$, 
respectively.  For $\tau_i = 1$, these curves approach the correct values 
for pure polytropes at $r_i/r_s = 0$ (Shu et al 1972)\fn{Note, however, 
the following error in Shu et al's Table 4.  While $(P_c/P_s)_{\rm cr} 
\goto 
1$ as $n \goto -1$, this is not true of $(\rho_c/\rho_s)_{\rm cr}$; rather, 
$(\rho_c/\rho_s)_{\rm cr} \goto 5.13$ in this limit (MH).} and the
Bonnor-Ebert value at $r_i/r_s = 1$.  For $\tau_i > 1$, the density curves 
approach $\tau_i$ {\it times} these same limiting values.  Observe that 
CPSs attain peak critical densities that are typically a few times 
larger than their pure polytropic counterparts for $\tau_i = 1$, and 
up to an order of magnitude (or more) larger for $\tau_i \gae 2$.  Also  
note that for $\tau_i > 1$ and $|n|$ sufficiently larger than 
1 (i.e., $\gamma_p\rightarrow 1$), 
$(P_c/P_s)_{\rm cr}$ can exceed $(P_c/P_s)_{BE}$ for a finite range of 
$r_i/r_s$, just as in the case of the CIS (see Fig.\ 2c).    

      Fig.\ 5 also shows that the limit $n \goto -1$ 
($\gamma_p\rightarrow 0$) is quite anomalous, in that the density 
contrast (but not the pressure contrast) greatly exceeds that of 
other CPSs over a large range of $r_i/r_s$.  
This can be understood as follows:  For small $r_i/r_s$, the CPS 
is mostly constant-pressure envelope, and is thus highly unstable. 
But as $r_i/r_s$ increases, stability is recovered due to the 
presence of the isothermal core.  More importantly, the high 
compressibility of 
the $n \simeq -1$ envelope has two distinct and complementary effects: 
first, the core is relatively insensitive to perturbations occurring 
at the cloud surface; and second, a relatively small pressure drop 
across the envelope corresponds to a large density drop.   
As in previous cases, these effects are only enhanced when $\tau_i > 1$.  

\subsection{Radius and Mass}

As in the case of the CIS, 
the dimensionless outer radius of the critical CPS has a maximum at an 
intermediate value of $r_i/r_s$.  This maximum is always greater than 
the Bonnor-Ebert critical radius, and therefore also greater than     
$\xi_{P,{\rm cr}}$, the critical radius of the pure polytrope 
(Shu et al 1972).  As $r_i/r_s \goto 0$ (i.e., a pure envelope), 
one finds that $\xi_{CP,{\rm cr}} \goto \tau_i^{(n+1)/2n}\; 
|n + 1|^{1/2}\; \xi_{P,{\rm cr}}$; e.g., $\xi_{CP,{\rm cr}} = 6.77$ 
when $n = -3$ and $\tau_i = 2$. 

    The mass of the CPS, in dimensionless form, is  
\bed
m_{CP} = \mc + m_{\rm env,P}, 
\eed
where $\mc$ is given by equation (\ref{eq-cicore}), 
\beqan
m_{\rm env,P} \equiv \frac{P_s^{1/2} G^{3/2}}{c_{\rm core}^4} 
M_{\rm env,P} & = &
\left(4\pi \frac{P_c}{P_s}\right)^{-1/2} \; \Delta \tau_s \; 
\frac{|n+1|}{\theta_s} \nonumber \\
& \times & [(\xi^2 \theta')_s - (\xi^2 \theta')_+],
\eeqan
and
$\tau_s\equiv\tau(r_s)=c_s^2/\cc^2$.
In the limit that the CPS is a pure core, one again obtains 
$m_{CP} \goto m_{BI}$, while in the pure envelope limit, 
\beq
m_{CP} = m_{\rm env,P} \goto \tau_s^2 m_P,  
\label{eq-menvl}
\eeq
where $m_P\equiv(P_s^{1/2}G^{3/2}/c_s^4)\; M$
is the polytropic analog of of $m_{BI}$ for a polytropic sphere. 
Note, once again, the factor of $\tau_s^2$ 
that enters since $m_{CP}$ is normalized to $\cc$.

        The critical mass of the CPS is plotted in the isentropic case 
for various envelope indices and $\tau_i = 1$ in Figure 6.  Note the 
asymptotic approach to the Bonnor-Ebert mass of the core at 
$r_i/r_s = 1$, and the different limiting 
values (according to $n$) at $r_i/r_s = 0$.  The latter values, 
which are for pure envelopes, differ from those of 
Shu et al (1972) due to the difference in non-dimensionalization.
Their values are normalized to the surface temperature so that
the critical mass is always {\it less} than the Bonnor-Ebert mass
(i.e.\ $\mu _{P,{\rm cr}} < \mu_{BE} = m_{BE}$ for NIPs),
whereas our values are normalized to the central temperature, and 
thus (for pure envelopes)
$m_{P,{\rm cr}}$ can exceed $m_{BE}$. 
The critical mass of a pure polytrope has a maximum of 
$m_{P,{\rm cr}} \simeq 1.4 \; m_{BE}$ at $n = -2.90$ ($\gamma_p=0.655$); 
this is in fact the maximum for {\it all} isentropic CPSs.  
CPSs with $-2.12 \leq n < -1$ have a 
maximum in $m_{CP,{\rm cr}}$ at intermediate $r_i/r_s$.  In the limit of a 
constant pressure polytrope, $n \goto -1$
$(\gamma_p\goto 0$), the critical mass shrinks to 
zero as the pressure gradient vanishes, while the mass for 
$|n| \gae 100$ 
is essentially constant at the Bonnor-Ebert value for all $r_i/r_s$.  
 
In Figures 7$a$ and $b$, we focus on the fiducial envelope index 
$n =-3$ ($\gamma_p=2/3$)
in order to examine the behavior of the critical mass as a function of 
$\tau_i$.  Note the similarity of these 
curves to those of the CIS, shown in Fig.\ 3.  The ``loops'' appearing 
in Fig.\ 7$a$ for $\tau_i \gae 3$, which mark the end of the envelope's 
dominant mass contribution, indicate that configurations with the same 
fractional core radius can have different critical masses.  
As in the case of the CIS ({\S}\ref{sec-isor}), this multi-valuedness 
disappears in a plot of $m_{CP,{\rm cr}}$ versus the fractional core 
mass, $M_i/M_s$ (Fig.\ 7$b$).  

\subsection{Velocity Dispersion}
\label{sec-vdisp}

	The observable component of the sound speed, $c(r)$, is the 
velocity dispersion, $\sigma (r)$, given by equation (\ref{eq-vel}).
The structure 
of the equilibrium CPS gives $P(r)$ and $\rho (r)$, so $c(r)$ is easily 
calculated from equation (\ref{eq-ss}).  
In order to gain a qualitative feel for $\sigma (r)$ and for the 
nonthermal velocity dispersion, $\sigma_{\rm nt}(r)=\sigma_w(r)$,
we set $v_A = 0$ and plot these quantities for an isentropic CPS with 
$n = -3,\; \tau_i = 1$, and various values of $r_i/r_s$ in Figure 8$a$.  
Also plotted 
are the same quantities for a pure $n = -3$ polytrope, with the central 
dispersion normalized to $\cc$.  Outside of the thermal core, 
both dispersions increase monotonically with radius.  Note how the 
slope of the nonthermal dispersion in the envelope increases with 
increasing core fraction.  Fig.\ 8$b$ shows how, in certain cases 
(here for $n = -1.5$, corresponding to $\gamma_p=1/3$), 
both $\sigma$ and its slope can exceed the pure 
polytropic values for a finite range of $r_i/r_s$ (here $0.2 \lae 
r_i/r_s \lae 0.55$).  This behavior will  
become increasingly important once we set out to model the observed 
increase of linewidth with size.  

In order to compare these model velocity dispersions with the observed 
molecular linewidths, one must first calculate the {\it projected} 
linewidth through the region of the cloud having $n > n_{\rm cr}$, 
where $n_{\rm cr}$ is the critical density for the excitation of a given 
molecular line (e.g.\ Genzel 1992).  We will discuss this procedure more 
fully, and present detailed modeling, in a subsequent paper. 

\subsection{The Non-isentropic CPS}
\label{sec-nicps}

	The CPS offers an improvement over the CIS, in 
that it is qualitatively consistent with the observed increase of 
velocity linewidth with size in molecular clouds.  But, as is evident  
from the critical masses plotted in Fig.\ 6, the isentropic CPS shares 
a common
problem with the CIS in that it cannot {\it quantitatively} reproduce 
the observed mass range of PPCs; in order to obtain 
masses and density contrasts that are as large as those inferred, one 
requires $\tau_i \gae 3$, a 
jump in $c^2$ for which there is neither theoretical nor observational 
justification.  As we have argued, insofar as molecular clouds can be 
described as polytropes, they are more likely to be non-isentropic in 
character.  It is therefore of some benefit to examine the quantitative 
properties of such polytropes.   

The theory of single-component, non-isentropic polytropes was reviewed 
and extended by MH.  They showed that, for a given $\gamma_p$ (or $n$), 
the value of $(\rho_c/\rho_s)_{\rm cr}$ increases 
as $\gamma$ increases above $\gamma_p$, and reaches infinity at a 
well-defined value $\gamma = \gamma_{\infty}$, given by 
\beq
\gamma_{\infty} = \frac{32 \gamma_p (2 - \gamma_p)}{(6 - \gamma_p)^2} 
= \frac{32 (n^2 - 1)}{(5n - 1)^2},
\label{eq-gaminf}
\eeq
corresponding to the family of singular polytropic configurations.  
Observe that in the isothermal case ($\gamma_p = 1, \; n = \pm \infty$), 
$\gamma_{\infty} = 32/25$, as found by Yabushita (1968).  
   
   Again focussing on the fiducial value $n = -3$, in Figures 9$a$ 
and $b$ we plot contours of maximum density contrast, 
$(\rho_c/\rho_s)_{\rm cr}^{\rm max}$, for $\tau_i = 1$ and $\tau_i = 2$.  
Recall that these maxima occur for different values of $r_i/r_s$, 
depending on the exact values of $\tau_i,\; \gamc$, and $\game$.  
The adiabatic index in the envelope, $\game$, ranges between the 
isentropic value, $\game = \gamma_{p,{\rm env}} = 2/3$, and 
$\game = \gamma_{\infty} = 1$, while $\gamc$ is confined to the 
corresponding range for the isothermal core ({\S}\ref{sec-niscis}).  
Note the qualitative similarity between Fig.\ 9$b$ and Figs.\ 4$a$ 
and $b$.

	   Since the CPS has one more free parameter than the CIS, 
it is interesting to examine properties of the critical state as $n$ 
varies.  In Figure 10$a$, we plot $(\rho_c/\rho_s)_{\rm cr}^{\rm max}$ and 
$(P_c/P_s)_{\rm cr}^{\rm max}$ versus $\gamma_{p,{\rm env}} = 1 + 1/n$ for 
$\tau_i = 1, 2, 3$, and 5.  Here, we fix $\gamc = 1.1$ and take 
$\game = \gamma_{p,{\rm env}} + 0.15$, in order to focus on the 
variation with $n$.  Because $\game = \gamma_{p,{\rm env}} + 0.15$ 
and equation (\ref{eq-gaminf}) have two points of intersection in the 
$(\game, \gamma_{p,{\rm env}})$ plane, there are two distinct values 
of $\gamma_{p,{\rm env}}$ where $(\rho_c/\rho_s)_{\rm cr}^{\rm max}$ and 
$(P_c/P_s)_{\rm cr}^{\rm max}$ become infinite: $\gamma_{p,{\rm env}} = 
0.22$ and 1.18.  The $\gamma_{p,{\rm env}}$ at which 
$(\rho_c/\rho_s)_{\rm cr}^{\rm max}$ is a minimum is roughly insensitive 
to $\tau_i$; it lies in the the range $0.75 \lae \gamma_{p,{\rm env}} 
\lae 0.90\; (-10 \lae n \lae -4)$ for $\tau_i \leq 3$.    

	As in the case of the non-isentropic CIS, critical masses well 
in excess of the Bonnor-Ebert mass of the core may obtain 
for the non-isentropic CPS, {\it even when} $\tau_i = 1$.
This divergence of the mass is associated with a divergence of 
$c_s^3/\rho_s^{1/2}$; if the properties of the surface are known, then 
of course there is no divergence ({\S}\ref{sec-sing}).  This behavior 
is illustrated in Fig.\ 10$b$, which shows the maximum critical 
mass (once again, obtained from the run of $m_{CP,{\rm cr}}$ versus 
$r_i/r_s$) for CPSs of different $\tau_i$, as a function of 
$\gamma_{p,{\rm env}}$ (solid curves).  As above, we fix $\gamc = 1.1$.  
The dotted curve is for isentropic pure polytropes, 
for which $m_{P,{\rm cr}}$ is always finite.  The dashed curves correspond 
to pure polytropes for the same values of $\tau_i$, that is, having central 
temperatures $T_c = \tau_i ~T_{\rm core}$.  Notice how the CPS 
mass is always equal to or greater than the pure polytropic mass.  Also, 
the former greatly exceeds the latter at low values of 
$\gamma_{p,{\rm env}}$, diverging more quickly than the pure polytrope 
for all values of $\tau_i$.  Conversely, at large $\gamma_{p,{\rm env}}$ 
the effect of the isothermal core is clearly seen, as 
$m_{CP,{\rm cr}}^{\rm max}$ 
remains finite; by contrast, $m_{P,{\rm cr}}$ rapidly approaches zero as 
$\gamma_{p,{\rm env}} \goto 6/5$ and $\gamma$ exceeds 4/3 (MH).  
The limiting mass as $\gamma_{p,{\rm env}} \goto 
6/5$ is slightly less than $m_{BE}$ simply because $\gamc > 
\gamma_{p,{\rm core}} = 1$ ({\S}\ref{sec-niscis}).
 
\subsection{Mean Quantities in the CIS and CPS}
\label{sec-mean}

The existence of large pressure contrasts in the CIS and CPS 
does not necessarily imply that their {\it mean} pressures are 
also larger than the Bonnor-Ebert value.  In fact, this is 
possible only 
for non-isentropic models.  The mean pressure is given by 
\bed
\ovl{P} = \lb c^2\rb \ovl{\rho},
\eed
where 
\bed
\lb c^2\rb = \frac{\int_0^{r_s} r^2 \rho c^2 dr}
{\int_0^{r_s} r^2 \rho dr} 
\eed
is the mass-average of the squared sound speed, and $\ovl{\rho} = 
M(r_s)/(4\pi r_s^3/3)$ 
is the mean density.  We have calculated the mean-to-surface pressure  
ratio, $\ovl{P}/P_s$, for a range of models, with the following results: 
(i) $\ovl{P}/P_s$ of the critical isentropic CIS or CPS is always less 
than or equal to the Bonnor-Ebert value, $(\ovl{P}/P_s)_{BE} = 2.43$.  
(ii) The non-isentropic CIS has $\ovl{P}/P_s > (\ovl{P}/P_s)_{BE}$ for 
all $r_i/r_s$, as shown in Figure 11$a$.  If $\gamc > \game$, then the 
maximum $\ovl{P}/P_s$ is obtained in the pure core limit, and vice-versa 
for $\game > \gamc$.  Since the upper bound is for a single component 
isothermal sphere, we can use the results of MH to set an upper limit of 
$\ovl{P}/P_s \leq 3.78$.
(iii) The non-isentropic CPS has $\ovl{P}/P_s$ less than the corresponding 
non-isentropic CIS.  When the core size is significant ($r_i/r_s \gae 0.2$, 
as shown in Fig.\ 11$b$ for $n=-3$), the pressure ratio can substantially 
exceed $(\ovl{P}/P_s)_{BE}$; otherwise, 
$\ovl{P}/P_s \simeq 
(\ovl{P}/P_s)_{NIP} < (\ovl{P}/P_s)_{BE}$.  The last result follows 
from the fact that $\ovl{P}/P_s$ is not greatly enhanced for 
non-isentropic versus isentropic NIPs, 
even as $\gamma \goto \gamma_{\infty}$.  As shown by MH, 
\bed
\left(\frac{\ovl{P}}{P_s}\right)_{NIP} \leq 1.26\; \frac{3 (2-\gamma_p)}
{(6-5\gamma_p)} = 1.26\; \frac{3 (n-1)}{(n-5)},
\eed  
so that $(\ovl{P}/P_s)_{\rm max}$ = 1.89
and 3.78 for $n=-3$ and $n=-\infty$, respectively.  These limits are 
consistent with the limiting values shown in Fig.\ 11$b$. 

\subsection{Summary---CPS}
\label{sec-cpssum}

        The results of this section lead to two general conclusions 
concerning the approximation of mean and fluctuating magnetic fields by 
isotropic pressure components (see also MH).  
First, 
both the maximum dimensionless mass $m_{CP}$ that can 
be supported in a given model and the maximum pressure contrast are
determined by 
the values of the polytropic {\it and} adiabatic indexes.
As $\gamma - \gamma_p \;(\gamma < 4/3)$ increases in 
either the core or the envelope, $(P_c/P_s)_{CP, \rm cr}$ 
becomes orders of magnitude larger than its isentropic 
value.  This behavior qualitatively mimics that found in equilibrium  
calculations of magnetized isothermal clouds, which feature stable 
pressure contrasts well in excess of the Bonnor-Ebert value for strong 
magnetic fields (Mouschovias 1976; Tomisaka et al 1988).  
However, the behavior of the ratio of the {\it mean} pressure 
to that at the surface
is quite different: we have extended the results of MH to
composite polytropes with negative-index envelopes
and shown that $\ovl{P}/P_s<4$, just as in the case of single
component NIPs.

	  Second, the restriction to 
isentropic NIPs, whether single-component or 
composite with $\tau_i\simeq 1$, puts severe limitations 
on the mass range of the resulting models.  Equilibrium
clouds of mass substantially larger than the Bonnor-Ebert mass
of the core do not exist under these restrictions.  On the other 
hand, in the more likely case of non-isentropic polytropes 
($\game > \gamma_{p,{\rm env}}$ due to the presence of 
a mean magnetic field and Alfv{\'e}n waves), both single-component 
NIPs and composite CPSs can have 
much larger critical masses.  

   The distinction between isentropic and non-isentropic polytropes 
is best seen by putting the critical mass in dimensional 
form. The mass of a composite polytrope is
\beqa
M_{CP} & = & \frac{c_{\rm core}^4}{P_s^{1/2} G^{3/2}} m_{CP} = 
2.2 \;M_{\sun} \left(\frac{T_{\rm core}}{10\; {\rm K}}\right)^2 
\nonumber \\
&\times &\left(\frac{2 \times 10^4 \; {\rm cm}^{-3} {\rm
K}}{P_s/k}\right)^{1/2} m_{CP},
\eeqa
Thus $2.6 \; M_{\sun} \leq M_{CP} \leq 3.6 \; \tau_s^2 
~M_{\sun}$, assuming these representative values (see eq \ref{eq-menvl}).  
The larger upper limit, compared to the CIS (\ref{eq-cimd}), is  
due to the increased mass for pure $n = -2.9$ polytropes 
when expressed in terms of the central temperature.  For a given 
surface temperature and pressure, $M_{CP,{\rm cr}}\leq 3.6\;\tau_s^2 \; 
M_{\sun}$ depends only weakly on whether the polytropic envelope is 
isentropic or not.  However, the core temperature is rarely less
than about 10 K; for a fixed core temperature and surface
pressure, the maximum critical mass scales as
$\tau_s^2\propto (P_c/P_s)^{-2(1-1/\gamma_p)}$, which is limited
for isentropic NIPs, but can be unbounded for non-isentropic ones.
This fact has important consequences for the observed structure of 
embedded dense cores, a topic we take up in the following section.  

        Although the composite polytropes discussed in this paper 
represent the simplest possible improvement on pure polytropic models, 
they already suggest a number of interesting implications for star 
formation generally.  We discuss a few of these below.   

\section{DISCUSSION}

\subsection{Near-Critical Cores with Massive Envelopes}

An interesting consequence of our results for the non-isentropic CPS 
(\S \ref{sec-nicps}) is the existence of equilibria featuring massive 
envelopes surrounding approximately Bonnor-Ebert mass cores, a property 
reminiscent of the observations reviewed in \S 1.2.  Since the derived 
mass ratio of $^{13}$CO envelopes to NH$_3$ cores is of order 
10--100, we choose an example CPS which illustrates this mass range.

     The core, envelope, and total dimensionless masses of a $n=-3$ CPS
with $\gamc = 1.2,\; \game = 0.94$ are displayed as a function of 
the core fraction $r_i/r_s$ in Figure 12$a$.  The critical mass 
of the core alone, $m_{\rm core,cr}$, has also been plotted, in the 
same non-dimensionalization.  For a fixed central sound speed 
$c_{\rm core}$ and outer surface pressure $P_s$, the pressure on the 
core, $P_i$, increases as the core fraction shrinks; hence, 
$m_{\rm core,cr}\propto M_{\rm core,\;cr}\propto P_i^{-1/2}$ 
decreases with decreasing $r_i/r_s$.  Fig.\ 12$a$ 
has a very interesting property: the mass of the core remains very 
near (and indeed, somewhat exceeds\fn{Recall that $m > m_{\rm cr}$ 
for non-isentropic states with $(P_c/P_s)_{BE} < 
P_c/P_i < (P_c/P_i)_{\rm cr}$ (Fig.\ 1).}) its critical mass for  
$r_i/r_s > 1.3 \times 10^{-2}$, despite the fact that the 
corresponding mass of 
the envelope varies by more than an order of magnitude over the same 
range.  If observed in a high-density tracer, such a core would 
appear to have a virial parameter of nearly the Bonnor-Ebert value 
($\alpha=2.06$; Bertoldi \& McKee 1992), as is observed.  
In terms of our dimensionless variables, the virial parameter 
becomes (see \S \ref{sec-virial}) 
\bed
\alpha = \frac{5 \lb\sigma^2\rb \ovl{R}}{GM} = \left(4\pi 
\frac{P_c}{P_s}\right)^{1/2} \frac{5 \;\lb \sigma^2\rb\;
\ovl{\xi}}{m\sigma_{\rm th}^2}.
\eed
Taking $\lb\sigma^2\rb=\sigma_{\rm th}^2,\; \ovl{\xi} = \xi_i$, and 
$m = \mc$, we find $\alpha_{\rm core} \simeq 2$ for all $r_i/r_s 
> 1.3 \times 10^{-2}$, as shown by the upper curve in Fig.\ 12$b$.  
It is of interest to compare $\alpha_{\rm core}$ with the virial 
parameter for the entire cloud, $\alpha_{\rm tot}$.  We estimated 
the mean velocity dispersion of the latter using equation 
(\ref{eq-vel}), and by assuming $\lb\sigma^2\rb \gg \sigma_{\rm th}^2$
(valid for $r_i\ll r_s$)
and an Alfv\'en Mach number of 1, i.e. $v_A^2 = 3\lb\sigma^2\rb$ 
(Myers \& Goodman 1988; Crutcher 1999).  This gives $\lb\sigma^2
\rb \simeq c^2/3$.  Strictly speaking, the latter should  
change with $r_i/r_s$, but this estimate suffices to get a rough 
idea of the magnitude of $\alpha_{\rm tot}$.  Finally, taking 
$\ovl{\xi} = \xi_s$ and $m = m_{CP}$, we find that $\alpha_{\rm tot}$ 
varies between 0.73 and 1.5 as $r_i/r_s$ ranges from 0 to 1 (Fig.\ 12$b$). 
Values of $\alpha_{\rm tot} < 1$ simply reflect the limited validity of 
our estimated $\lb\sigma^2\rb$---as $r_i\goto r_s$, the cloud becomes 
thermally dominated with $\lb\sigma^2\rb\goto c^2$, not $c^2/3$ as we 
have assumed.

Insofar as the derived virial parameters of many clumps on sub-GMC 
scales are much greater than unity (\S \ref{sec-virial}), we conclude 
(with Bertoldi \& McKee 1992), that such clumps are pressure-confined. 
Sub-critical CPSs having $M_{\rm env} = 10-100 \; M_{\rm core}$ 
are easily constructed.  In our view, these results present a reasonable 
explanation for why it is that PPCs, which are nearly critical by virial
theorem estimates (\S \ref{sec-virial}), can exist within clumps and GMCs 
that are far more massive 
but nonetheless stable or close to critical.  
This constitutes an explicit realization of the formalism developed 
by Bonazzola et al (1987), who 
proposed that a turbulent cloud could be stable on large scales and 
unstable on small ones 
(note that our models are stable by construction,
but would become unstable if the core mass were increased
further for $r_i/r_s > 1.3 \times 10^{-2}$).
A more detailed comparison of this aspect of the models with existing 
data is, however, beyond the scope of this paper.

\subsection{The Formation of Very Low-Mass Stars}

The enhanced pressure contrasts attainable in the composite models 
may also have implications for forming very low-mass stars such as 
brown dwarfs.  A simple argument for an isolated core forming a 
single star serves to illustrate this point.  In order to form a star 
of mass $M_*$, it is necessary to collect together at least a 
Bonnor-Ebert mass, 
$M_{BE} = 2.6\; (T/10\; K)^2\; [P_s/(2 \times 10^4\; k_B)$ 
K cm$^{-3}]^{-1/2}\; M_{\sun}$, of gas.  
Observed cores in GMCs have surface pressures ranging from $P_s/k_B 
\sim 4 \times 10^4$ (Taurus, Cepheus) to $3 \times 10^5$ K cm$^3$ 
(Orion) (Solomon et al 1987; Bertoldi \& McKee 1992), so it is 
easy to form stars with $M_* \simeq 0.7 - 1.8\; M_{\sun}$.  But to 
create a star of much smaller mass (brown dwarfs have $M_* \lae 0.08 
M_{\sun}$), it is necessary to either reduce the temperature below 10 K 
(which is difficult due to radiative trapping-- de Jong et al 1980), 
or to increase $P_s/k_B$ well beyond the observed range.  Inserting 
$M_J \lae 0.08\; M_{\sun}$ into the above, one finds $P_s/k_B \gae 1.3 
\times 10^8$ K cm$^3$, nearly three orders of magnitude above the observed 
pressures.  In a CPS with $\mc = 0.08\; M_{\sun}$, however, it is the 
pressure on the core-envelope {\it interface}, $P_i$, that counts.  Here 
one is not limited by the Bonnor-Ebert value $P_i/P_s = 14.04$, since 
the pressure drop through a non-isentropic envelope can be much larger 
(Fig.\ 10$a$).  
Thus, non-isentropic CPSs lead to the possibility of
a structure near the brink of instability with a core
small enough to make a brown dwarf.  Whether it is possible
for the core to undergo collapse without having a significant
mass from the envelope fall on it, thereby increasing its mass,
cannot be determined from hydrostatic calculations such as ours.

\subsection{Protostellar Collapse}

One of the more robust predictions of studies of cloud collapse is 
the mass accretion rate onto the nascent central object. The 
well-studied SIS, for example, features a time-independent accretion 
rate, $\dot{M}_{SIS} = 0.975\; \sigma_{\rm th}^3/G$ (Shu 1977).  
Observational estimates 
of $\dot{M}$ have been inferred for many Class 0 and Class I infrared 
sources, from the peak of their spectral energy distributions in the 
far-infrared (Hartmann 1998).  These fall within the range $1.6 \times 
10^{-6} M_{\sun}$ yr$^{-1}$ (Class I) $\lae \dot{M} \lae 10^{-4} 
M_{\sun}$ yr$^{-1}$ (Class 0).  For a typical $\sigma_{\rm th} \sim 0.2$ 
km s$^{-1}$, the SIS is in accord with the lower limit of this range, but 
is inconsistent with the high $\dot{M}$'s of the Class 0 objects.  This 
again raises the question of the relevance of the SIS for the earliest 
stages of protostellar evolution (Foster \& Chevalier 1993;
Safier et al 1997). 

	We now attempt to predict how the collapse 
of a composite polytrope will differ from that of previous models.  
Since we have modeled the core of a CPS with an isothermal
sphere, the early behavior is likely to be similar to that of 
the Bonnor-Ebert sphere, which was investigated most recently
by Foster \& Chevalier (1993).  They found that $\dot{M}$ is largest 
at the onset 
of collapse, then decreases, eventually approaching $\dot{M}_{SIS}$.
The late-time behavior is more uncertain, and depends on the 
characteristics of the envelope.  If the envelope can be described
by a NIP, then two recent studies suggest that the accretion
rate will increase at late times:
McLaughlin \& Pudritz (1997) considered expansion wave solutions
in singular polytropic spheres and found that $\dot{M} 
\propto t^{-3/n} \; (n < -1)$; i.e., the accretion rate is small 
at early times, then increases monotonically.
Henriksen, Andr{\'e}, \& Bontemps (1997) studied the collapse of 
a pressureless model (so that there is no expansion wave)
with a piecewise continuous density profile: 
$\rho \propto r^0$ in the core and $\rho \propto r^{-2n/(n-1)}$ 
(corresponding to a singular polytropic sphere) in 
the envelope.  Their results show that after an initial burst of 
accretion ($\dot{M}$ is formally infinite in their model, due 
to the pressureless, flat inner region), models with $n<-1$ 
pass through a minimum in $\dot{M} (> \dot{M}_{SIS})$, before increasing 
monotonically at large $t$ as $t^{-3/n}$,
in agreement with the results of McLaughlin \& Pudritz.  
On the other hand, if the envelope is supported by
a static magnetic field so that the late-time accretion is regulated
by ambipolar diffusion, the accretion rate is similar to that
of an SIS (Safier et al 1997).  More definitive results for the
collapse of a CPS await detailed numerical calculations.

\section{SUMMARY}

In this paper, we have developed a polytropic theory of molecular 
clouds that improves on previous work in several ways.  
The consideration of spatially separate core and envelope regions 
is in better accord with the known velocity structure of star-forming 
cloud cores.  Such an approach allows a thermally-dominated core to 
coexist with a nonthermally-dominated envelope in a single, 
self-gravitating equilibrium structure. 
In the formalism developed here, the core and the envelope
are each represented by a single pressure component with properties
chosen to approximate the combined effects of thermal pressure
and static magnetic fields in the core (assumed to be
described by a polytropic index $\gamma_p=1$), and 
these plus turbulent motions in the envelope.
The results for isothermal envelopes are summarized in 
{\S}\ref{sec-cissum}, and those for polytropic envelopes in 
{\S}\ref{sec-cpssum}.  Just as composite polytropic 
models of stars were created to reflect changing physical conditions 
with radial scale, cloud models can now do the same.  Our models 
are non-singular, a feature in better concord with observations of the 
densest pre-protostellar cores
than the singular hypothesis.  This agreement carries over 
to the quantitative comparison of both the density and velocity 
dispersion profiles with those observed, as we present in 
detail in a subsequent paper. 

	The increase of temperature near the surface and
of line width with size had led previous workers to
consider negative-index polytropes (NIPs) as models
of interstellar clouds (Shu et al 1972; Maloney 1988).
In general, these models were based on the assumption that
the gas is isentropic---i.e., that the polytropic index
$\gamma_p$ is the same as the adiabatic index $\gamma$.
The maximum pressure and density drops in such NIPs are always 
less than that of a Bonnor-Ebert sphere, however, 
suggesting that they are 
inadequate as models for molecular clouds.
Isentropic composite polytropes of reasonable core fractions 
already provide pressure and density enhancements well in excess of 
those of the isothermal sphere.  The critical mass and radius of an 
isentropic composite polytropic sphere (CPS) can also exceed the 
Bonnor-Ebert mass of the core in cases where the kinetic temperature 
of the envelope is higher than that of the core.  

However, the pressure components in molecular gas,
particularly magnetic fields and turbulence (which
we model with Alfv{\'e}n waves), are {\it not} isentropic (MH).
Non-isentropic CPSs 
permit pressure and density drops in stable structures that
are as large as those 
observed, and in addition
permit masses that can be large compared with the
Bonnor-Ebert mass of the core, even
without a temperature difference between core and envelope.  
However, as found by MH for individual NIPs,
the ratio of the {\it mean} pressure to the surface pressure
for composite polytropes with negative-index
envelopes must be less than a factor $\sim 4$, which appears
to be a general feature of hydrostatic models for self-gravitating
clouds with outwardly increasing temperatures.
The greater flexibility of the CPS allows one to transcend the 
limitations of many previous polytropic models, which were unable 
to reproduce even the bulk physical properties of PPCs, not to 
mention the detailed spatial structure of these quantities. 

\vskip 0.3cm
\noi
The authors would like to thank Dean McLaughlin for useful discussions.  
This research is supported in part by a NASA grant to the Center for 
Star Formation Studies.  The research of CC was supported in part 
by an NSERC Postdoctoral Fellowship. The research of CFM is
supported in part by a grant from the NSF (AST95-30480),
a Guggenheim Fellowship, and a grant from the Sloan
Foundation to the Institute for Advanced Study.  CC thanks 
the Astronomy Group at the University of Western Ontario for 
their hospitality while this work was completed. 
CFM wishes to express his appreciation to John Bahcall of
the Institute for Advanced Study and Edith Falgarone of
the Ecole Normale Superieur for their hospitality.

\vskip 0.5cm
\begin{center}
{\bf APPENDIX
\vskip 0.2cm
ON THE STABILITY CRITERION FOR POLYTROPIC MODELS OF MOLECULAR CLOUDS} 
\end{center}

As discussed in {\S}\ref{sec-prev}, the stability analysis of bounded
polytropes has been pursued in the literature according to two very 
different assumptions about the response of the cloud to small 
perturbations.  Some authors have assumed that the central temperature 
remains constant during the perturbation, whereas we (in agreement
with MH) have assumed that the entropy remains constant.  For an 
isothermal gas, the two assumptions are equivalent; but in general 
they are not.  The issue is not one of mere formalism: whether a given 
model is stable or not is highly sensitive to the choice.  

The constant central temperature condition was justified by Maloney 
(1988) on the basis of the ``\ldots general uniformity of gas temperatures 
observed in molecular clouds over a large range of spatial and density 
scales \ldots''  
As Turner et al (1992) pointed out, however, it is the value of $T_c$ for a 
{\it given} cloud {\it during} a perturbation that is relevant to the issue 
of stability.  Still asserting the constancy of $T_c$, however, 
the latter authors argued that it 
``\ldots is determined by the cloud's extinction and the local UV 
[field], and variations in [$T_c$] due to (small) variations in 
[the boundary pressure] may reasonably be considered negligible.''  

        The issue becomes more involved when we consider that these authors 
view the increasing outward temperature of NIPs as being partly or entirely 
{\it nonthermal} in origin.  In the models of both Maloney and McLaughlin 
\& Pudritz (1996), e.g., the kinetic temperature is spatially constant 
(and supposed equal to $T_c$), but superposed on this is an isotropic, 
turbulent velocity field $\delta {\bf v}$, whose {\it effective} 
temperature, $T_{\rm eff} \propto \lb\delta {\bf v}\rb^2$, behaves like 
that of a NIP.  This is, 
in fact, in accord with recent observations of Bok globules showing that 
the gas temperature throughout a given globule is remarkably uniform, 
even though the nonthermal linewidth can be up to twice the thermal 
value (Lemme et al 1996).  The origin of the turbulent motions producing 
the nonthermal linewidths is the object of much speculation (e.g.\ 
Alfv\'en waves, cloud-cloud motions, 
protostellar outflows, etc.), but it is certainly different than
the mechanisms that determine the thermal motions in the gas.

In a polytropic model, the surface temperature is related
to the central temperature by
\beq
\frac{T_s}{T_c}=\left(\frac{P_c}{P_s}\right)^{1/(n+1)}
	= \left(\frac{P_c}{P_s}\right)^{(1-\gamma_p)/\gamma_p}.
\eeq
The central region is often sufficiently shielded that the temperature 
there is determined by cosmic ray heating.  Thus, 
while it is plausible that $T_c$ remains constant during the perturbation, 
there is no justification for assuming that $T_s$, 
which is represented in the turbulent models by 
the velocity dispersion at the surface of the cloud, 
will change just so as to satisfy this relation with $T_c$=const.
Indeed, there is no known relation between
the physical processes determining the kinetic temperature 
and the nonthermal velocity dispersion 
{\it anywhere in the cloud.}  Furthermore, as pointed out by MH, 
both Maloney and McLaughlin \& Pudritz considered isentropic polytropes.
Such polytropes cool upon compression if $\gamma_p<1$, as they
assumed.  In order to maintain a constant central temperature, energy
would have to be supplied, and no plausible energy source has been 
identified.  In the language of \S \ref{sec-locad} and MH, 
the isothermal perturbations considered by these authors are neither 
locally nor globally adiabatic, since heat is not simply redistributed 
but rather is injected by some unknown agent.

In our analysis, we avoid these contradictions and follow most previous 
authors by taking the perturbations to be locally adiabatic (\S 
\ref{sec-locad}): no heat enters or leaves the system (cf.\ Shu et al.\ 
1972; Viala \& Horedt 1974; Stahler 1983).  However, we depart from 
these treatments by allowing for {\it non-isentropic} pressure 
components, whose adiabatic index $\gamma$, which dictates the 
stability properties of the gas, differs from the polytropic index 
$\gamma_p$ [or $n = 1/(1-\gamma_p)$], which determines the equilibrium 
structure.  For NIPs in particular, taking $\gamma > \gamma_p$ moves 
the critical point of equilibrium to larger dimensionless radii, and 
more centrally condensed states (Fig.\ 1), than in the isentropic 
($\gamma = \gamma_p$) case.  These newly-accessible equilibria are 
identical to those {\it assumed} to exist by the above authors, but 
are not derived from the questionable assumption of unnamed energy 
sources.

\newpage
\bfig
\plotone{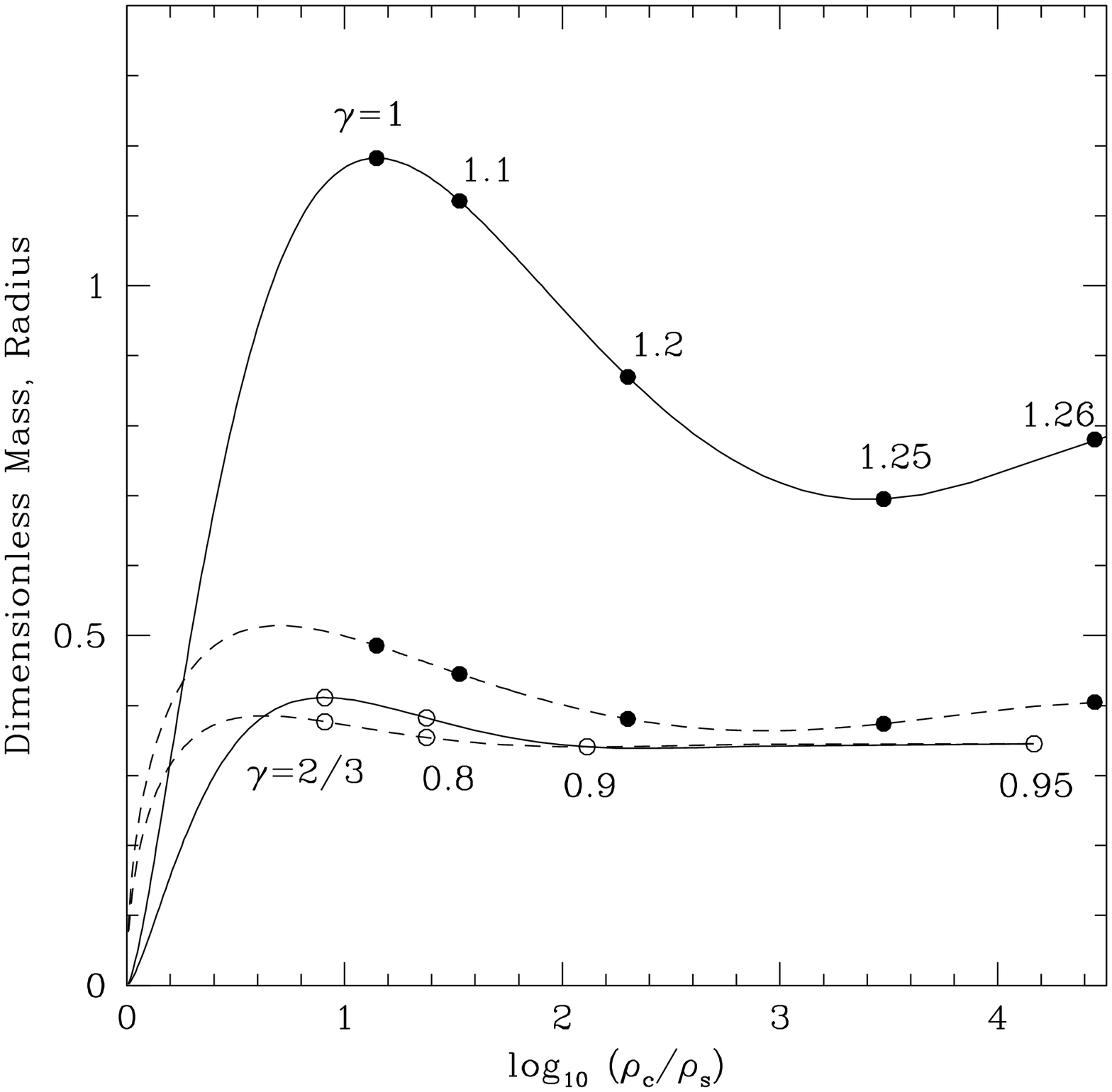}
\caption{Equilibrium sequences of the isothermal sphere (curves 
with filled circles) and an $n=-3$ polytrope (curves with open
circles).  Solid lines show the dimensionless mass; dashed lines, 
the dimensionless radius.  Dots indicate critical points at 
particular values of $\gamma$, as labeled. \label{fig1}} 
\efig

\bfig
\plottwo{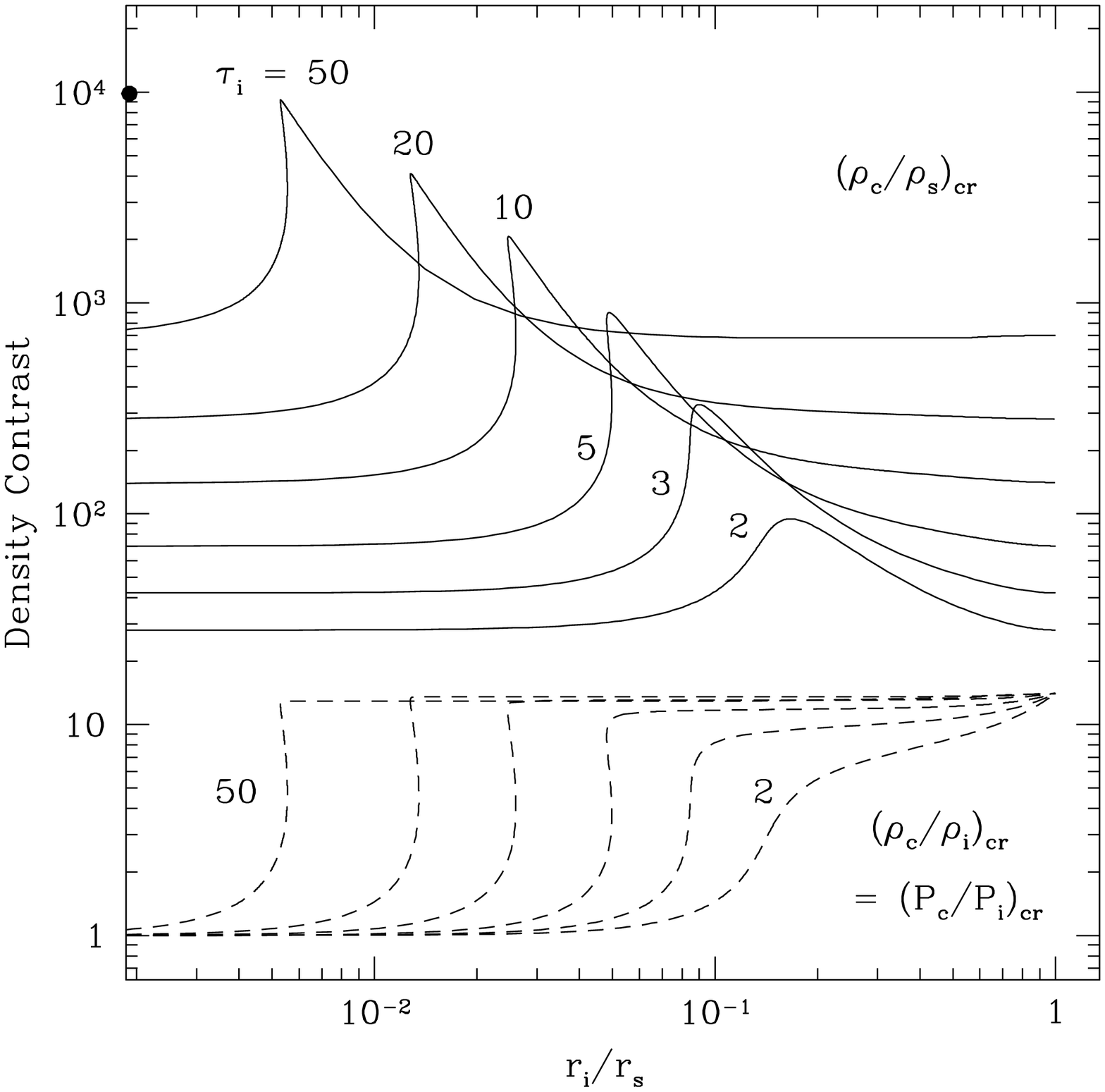}{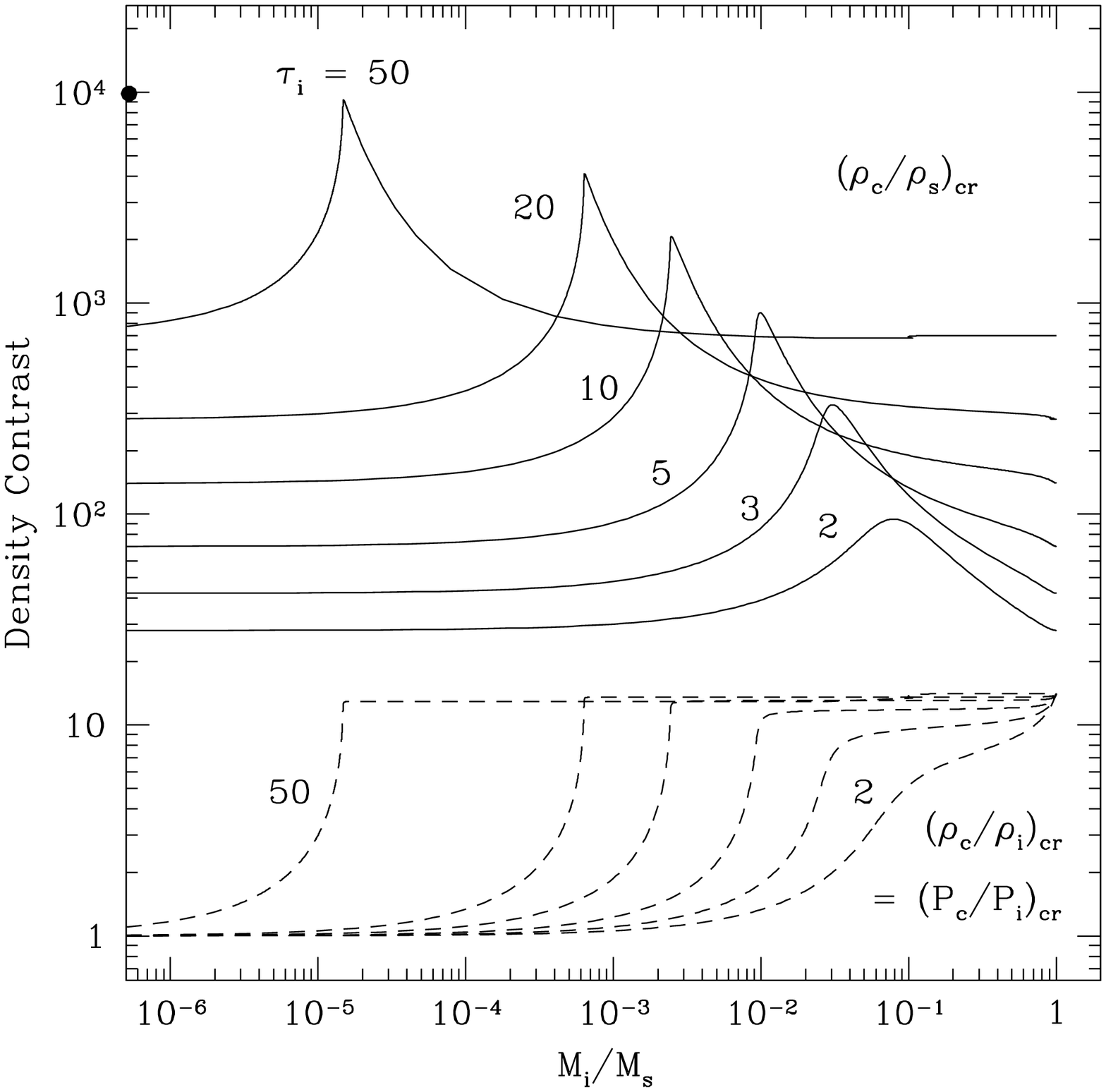}
\epsscale{.5}
\plotone{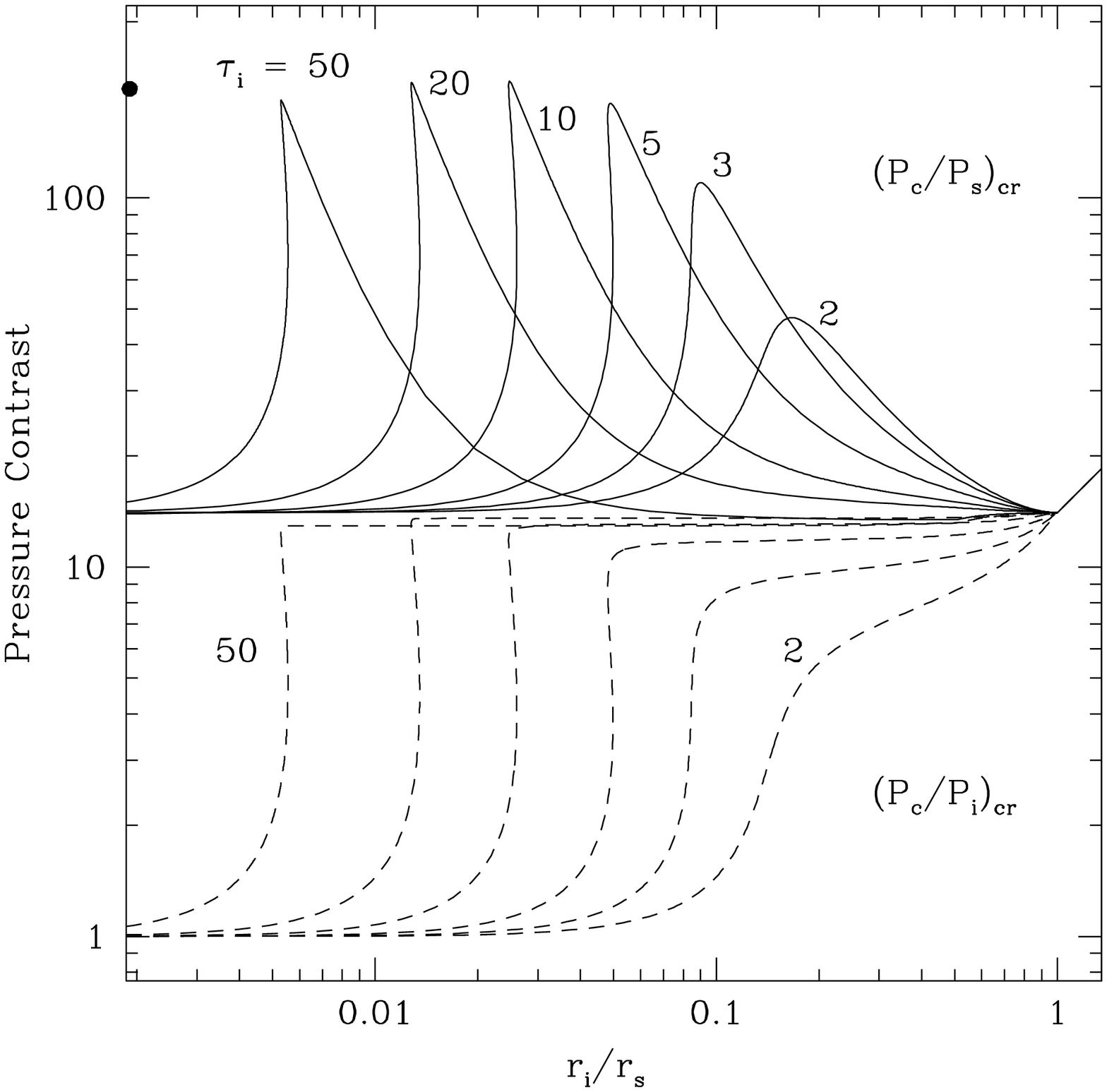}
\caption{Density and pressure contrasts of the critically stable, 
isentropic CIS. 
(a) Critical density contrast as a function of fractional core radius, 
$r_i/r_s$.  Each point on a curve represents a single, critically stable, 
equilibrium model.  Solid curves are center-to-surface contrast, dashed 
curves center-to-core boundary contrast.  Curves are labeled by values 
of $\tau_i$.  The filled circle on the ordinate indicates a density contrast 
of $\tau_i (\rho_c/\rho_s)_{BE}^2 = 50(14.04)^2 = 9855$. 
(b) Critical density contrast as a function of fractional core mass, using 
same notation as in (a). 
(c) Critical pressure contrast as a function of fractional core radius, 
using same notation as in (a).  The filled circle on the ordinate indicates 
a pressure contrast of $(P_c/P_s)_{BE}^2 = (14.04)^2 = 197$. \label{fig2}} 
\efig

\bfig
\plottwo{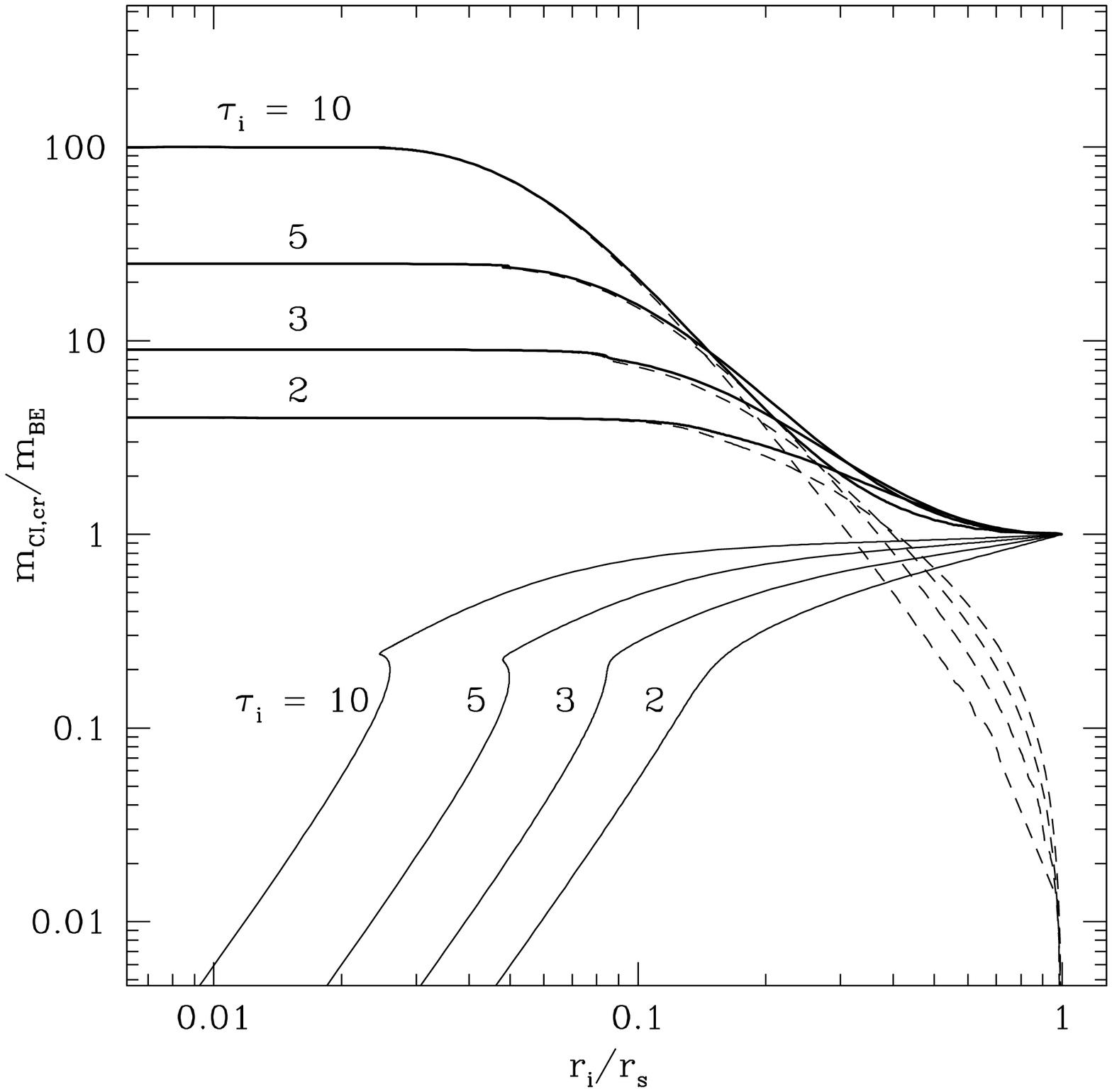}{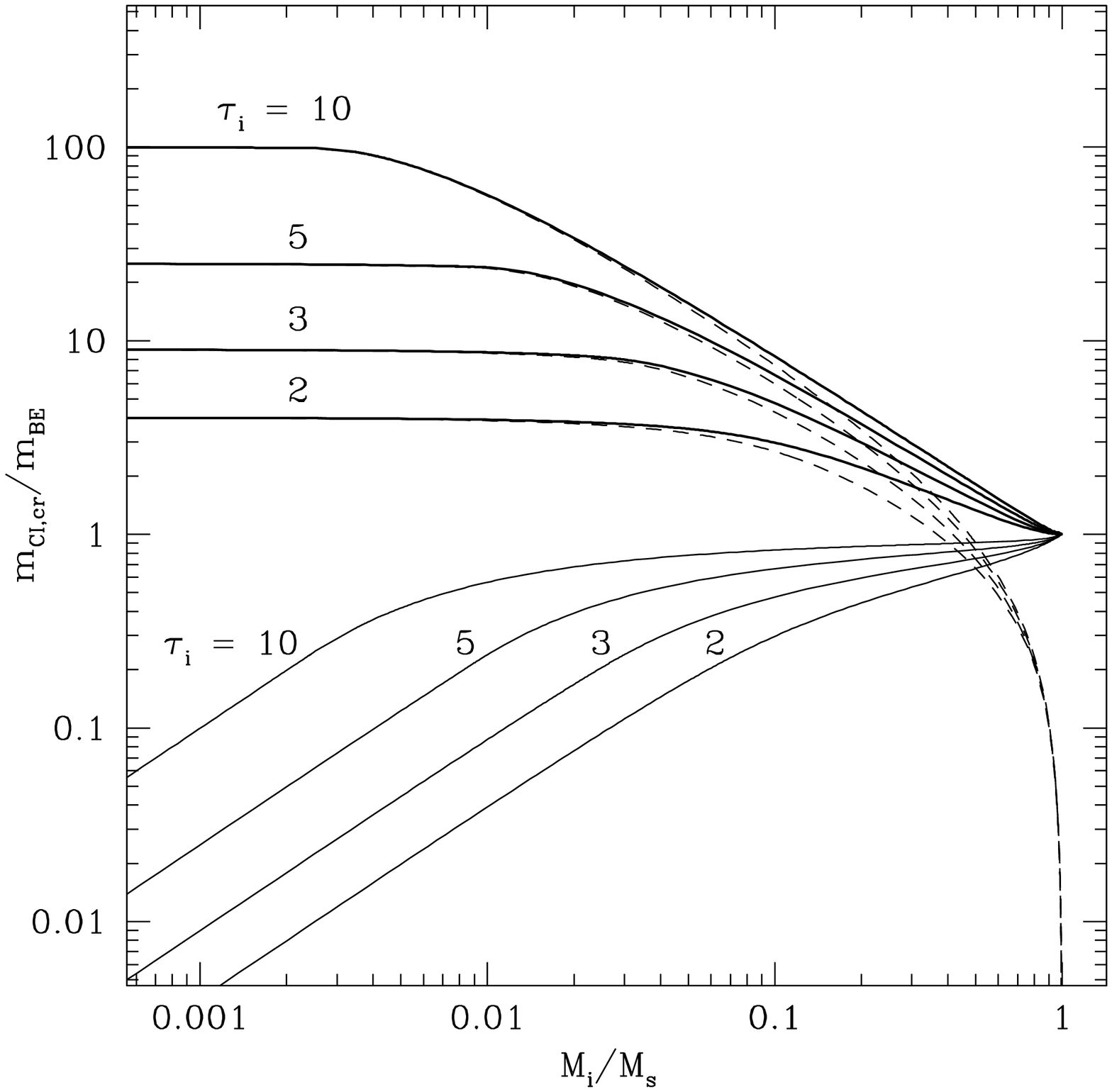}
\caption{Dimensionless mass of the critically stable, isentropic CIS, 
normalized to the Bonnor-Ebert mass, $m_{BE}=1.182$. (a) Critical 
mass as a function of fractional core radius, $r_i/r_s$. 
Solid curves show the core mass alone; dashed curves, the envelope mass; 
and the heavy solid curve the total of the two.  Curves are labeled by 
values of $\tau_i$. 
(b) Critical mass as a function of fractional core mass, using same 
notation as in (a). \label{fig3}} 
\efig
 
\bfig
\plottwo{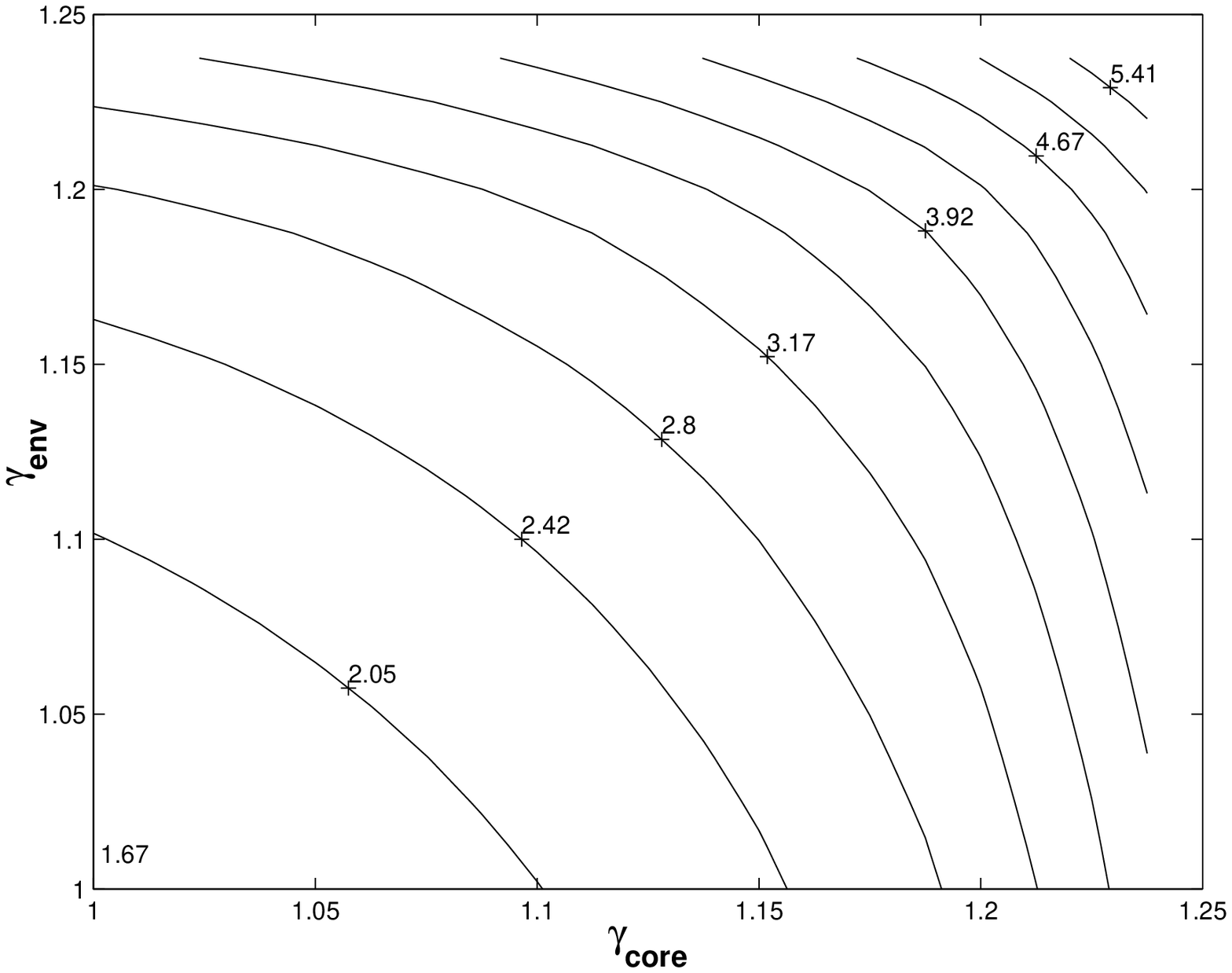}{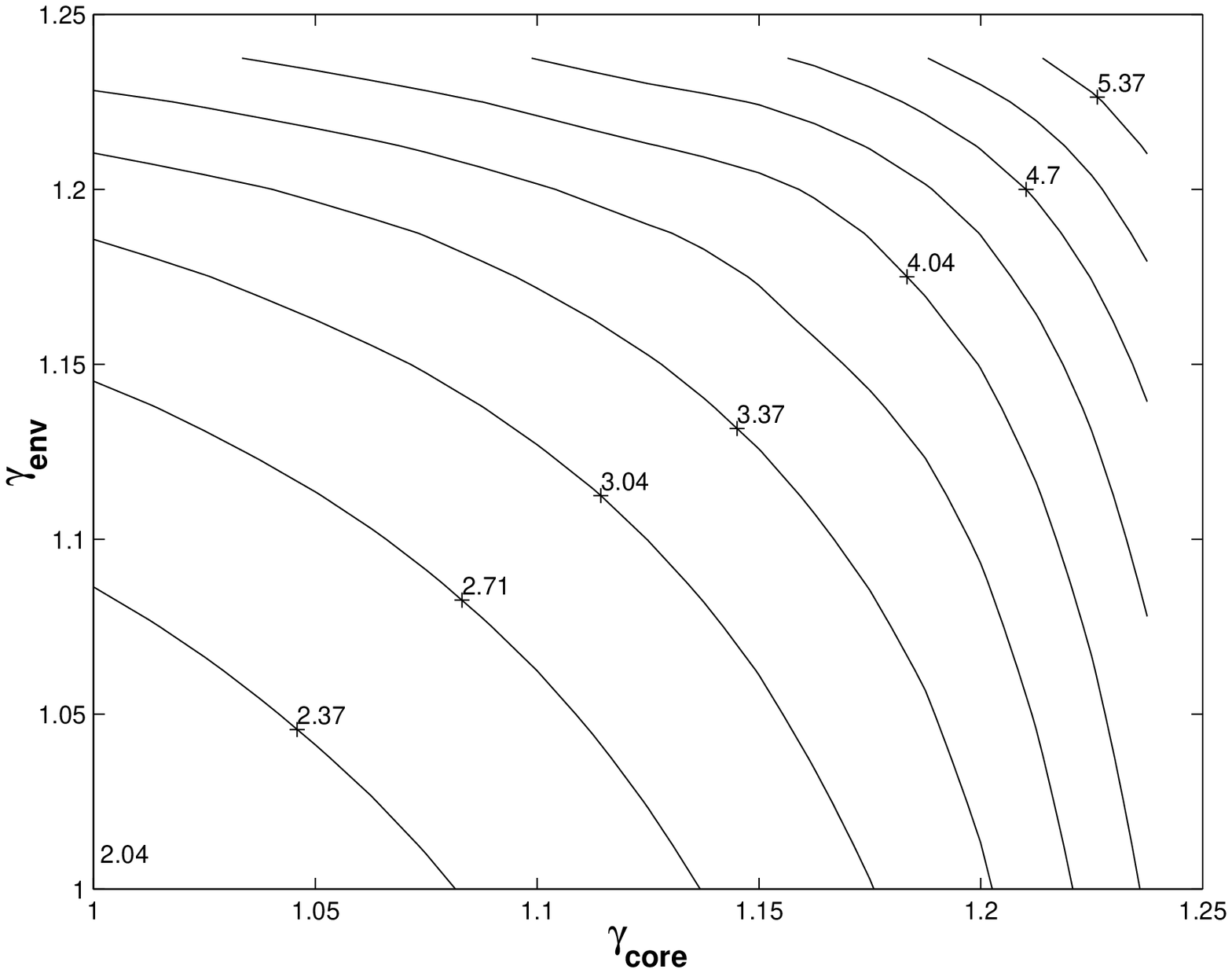}
\caption{Maximum pressure contrast of the critically stable, 
non-isentropic CIS. 
Contours of equal log$_{10} (P_c/P_s)_{\rm cr}^{\rm max}$ are plotted as a 
function of the core and envelope adiabatic indices, $\gamc$ 
and $\game$, respectively, 
for (a) $\tau_i = 2$; (b) $\tau_i = 3$. \label{fig4}}   
\efig
 
\bfig
\plottwo{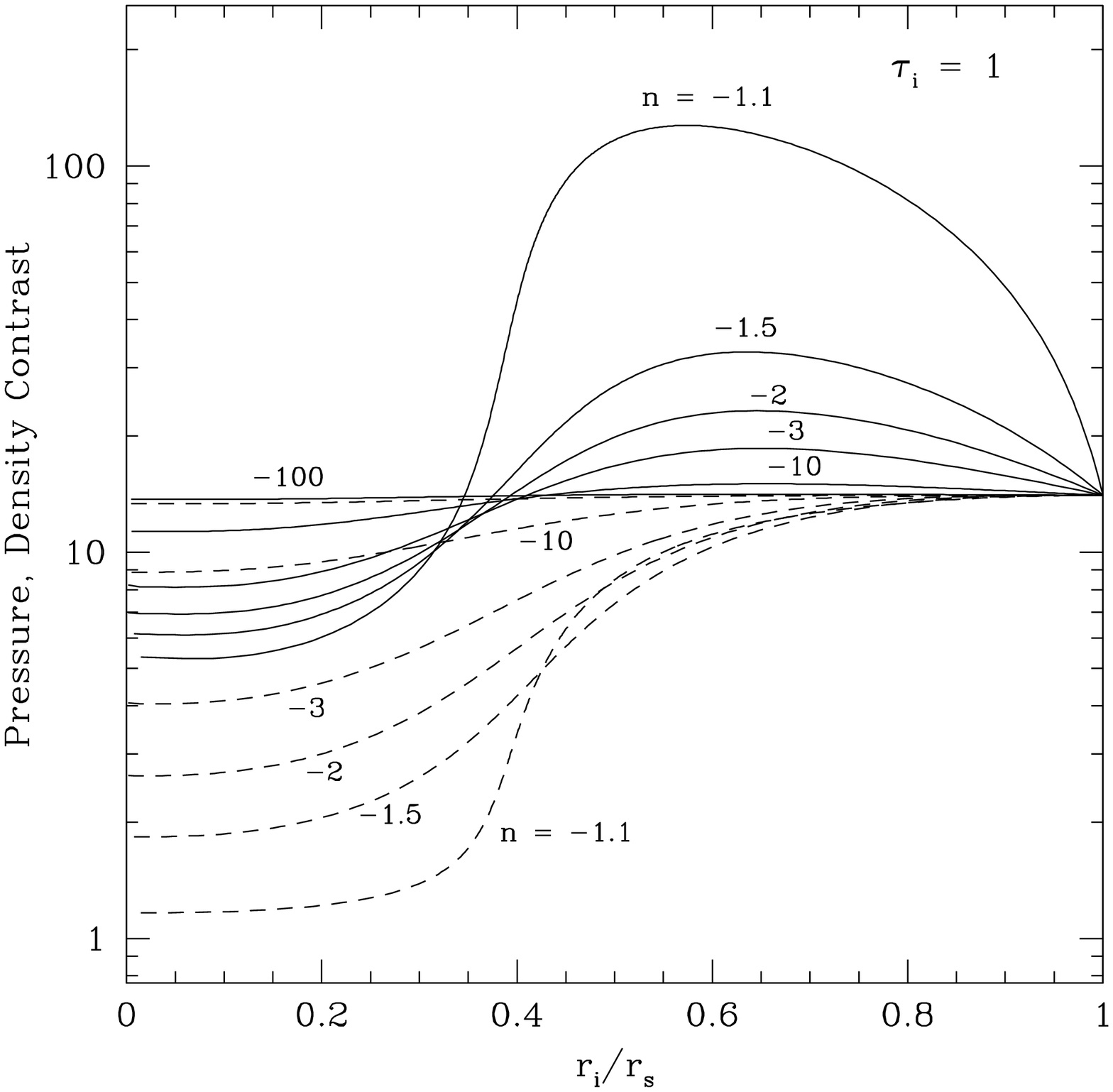}{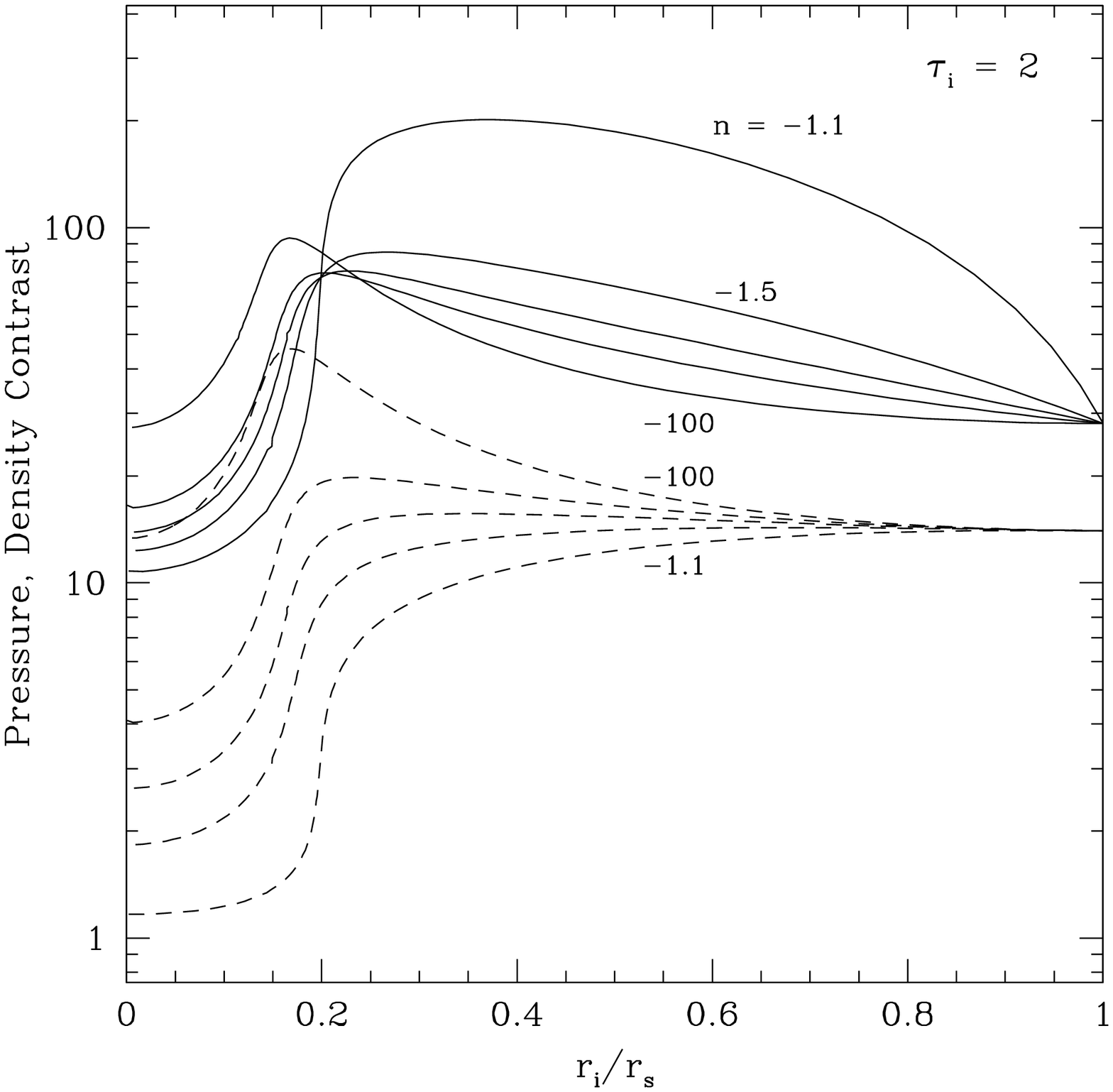}
\epsscale{.5}
\plotone{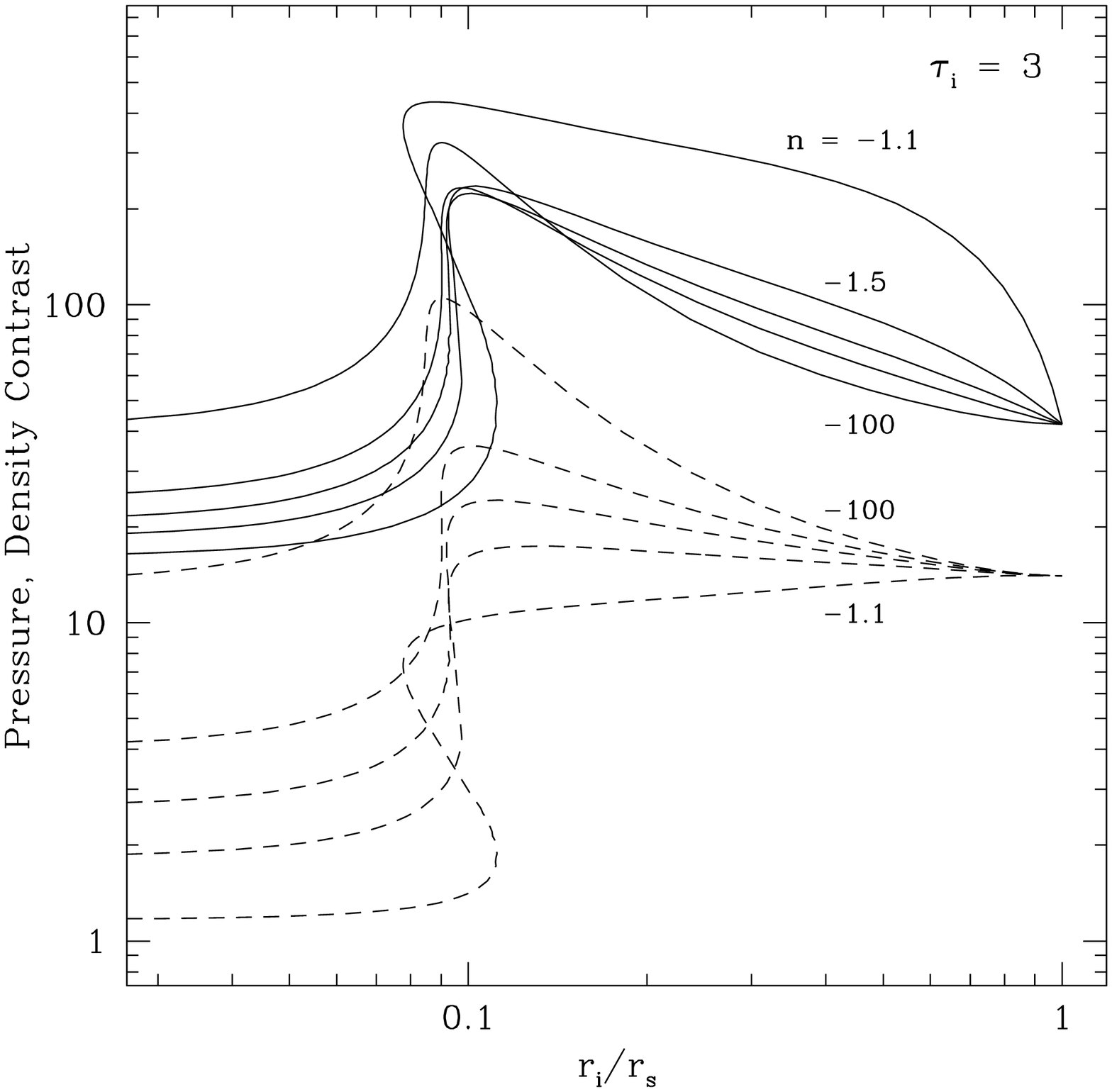}
\caption{Pressure and density contrasts of the critically stable, 
isentropic CPS, in the cases (a) $\tau_i = 1$; (b) $\tau_i = 2$; and 
(c) $\tau_i = 3$.  Solid curves show the center-to-surface density 
contrast; dashed curves, the center-to-surface pressure contrast. 
Both are plotted as a function of the fractional core radius.  
Curves are labeled by values of $n$. \label{fig5}} 
\efig

\bfig
\plotone{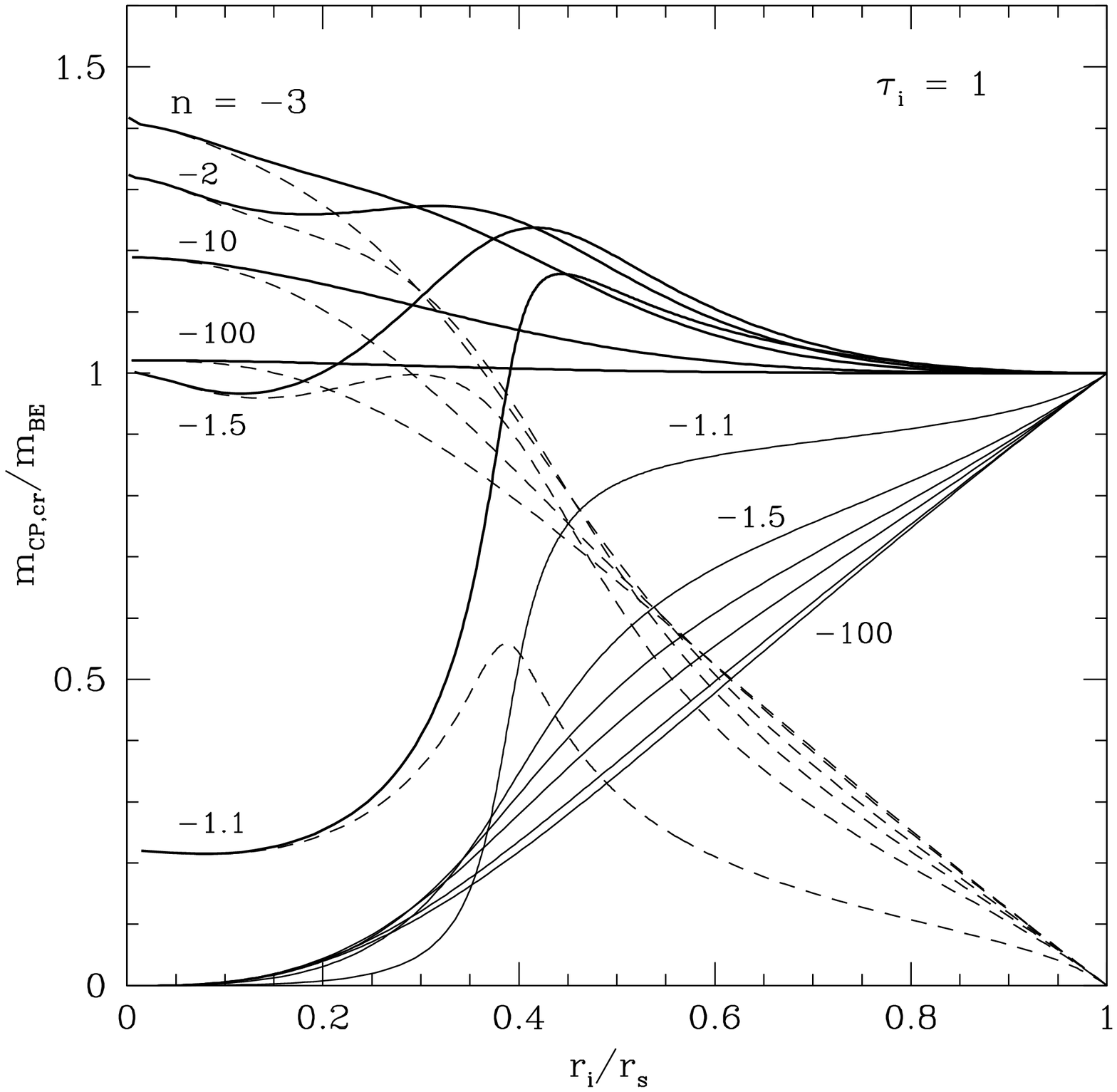}
\caption{
Dimensionless mass of the critically stable, isentropic CPS with 
$\tau_i = 1$, as a function of fractional core radius.  Solid curves 
indicate the core mass, dashed curves the envelope mass, and heavy 
solid curves the sum of the two.  Curves are labeled by values of $n$. 
\label{fig6}} 
\efig

\bfig
\plottwo{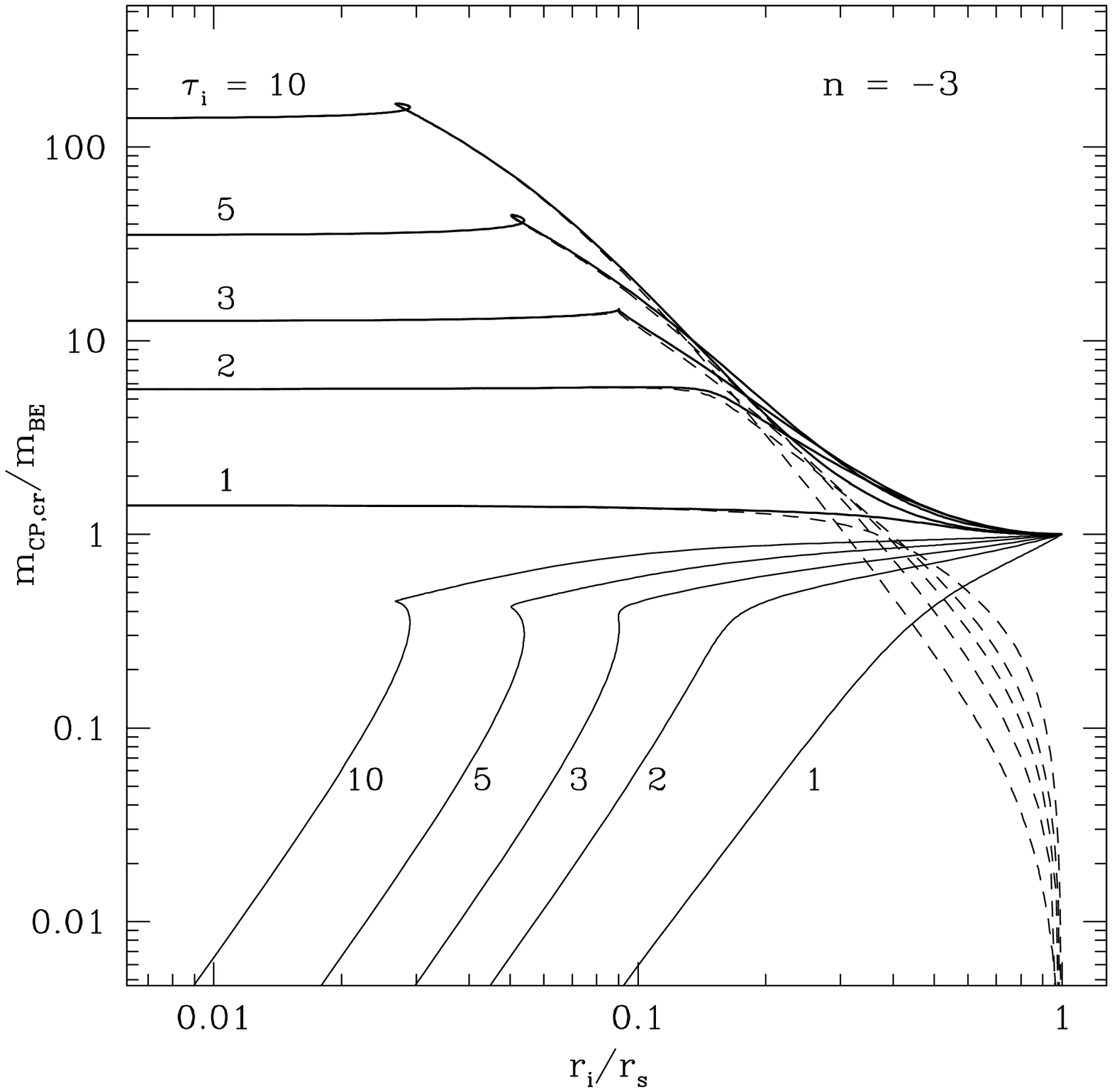}{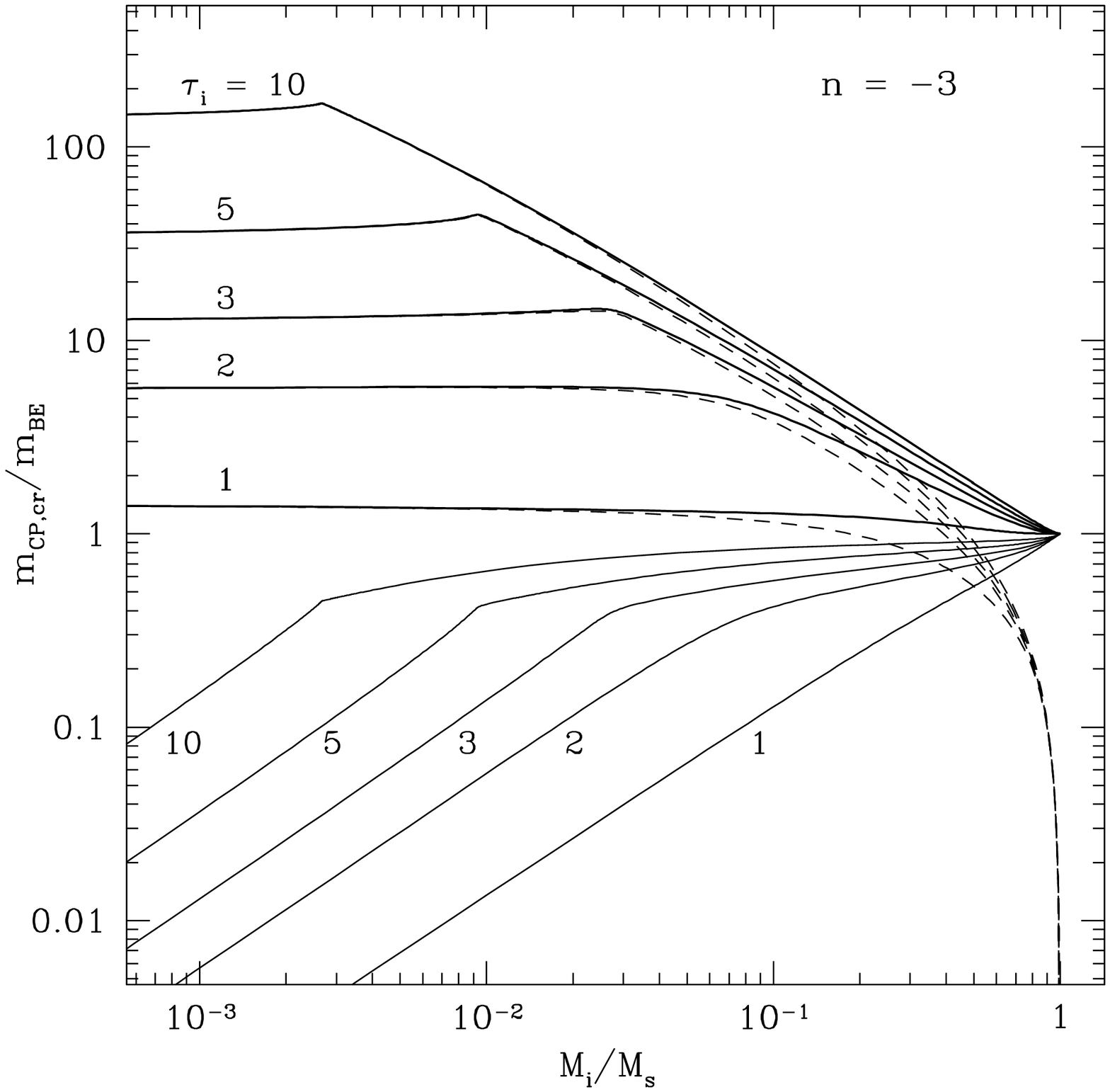}
\caption{
Dimensionless mass of the critically stable, isentropic CPS. 
(a) Critical mass, normalized to the Bonnor-Ebert mass $m_{BE}$,
as a function of fractional core radius, $r_i/r_s$.  
Solid curves show the core mass alone; dashed curves, the envelope mass; 
and heavy solid curves the sum of the two.  Curves are labeled by values 
of $\tau_i$. 
(b) Critical mass as a function of fractional core mass, using same 
notation as in (a). \label{fig7}}
\efig
 
\bfig
\plottwo{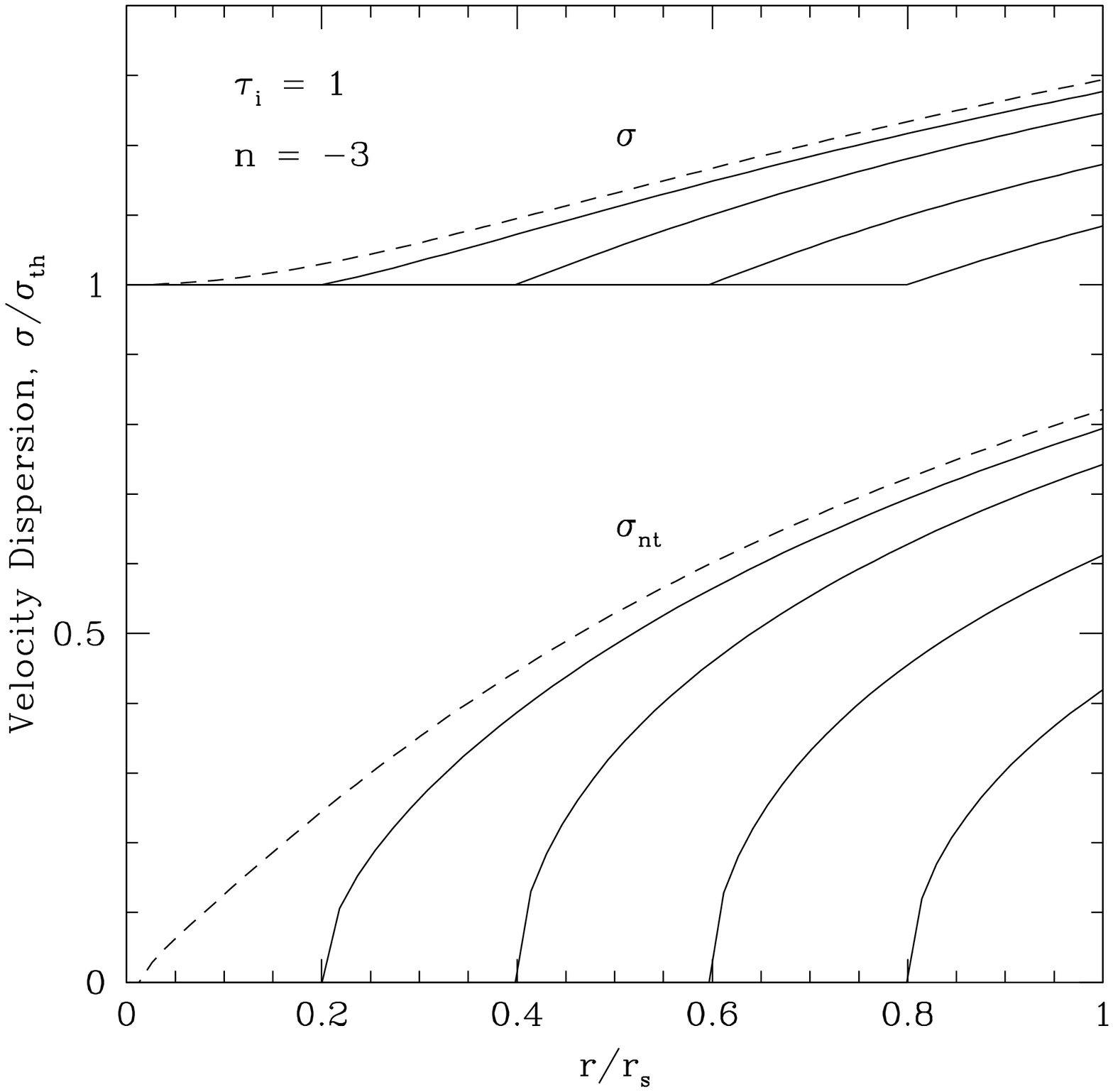}{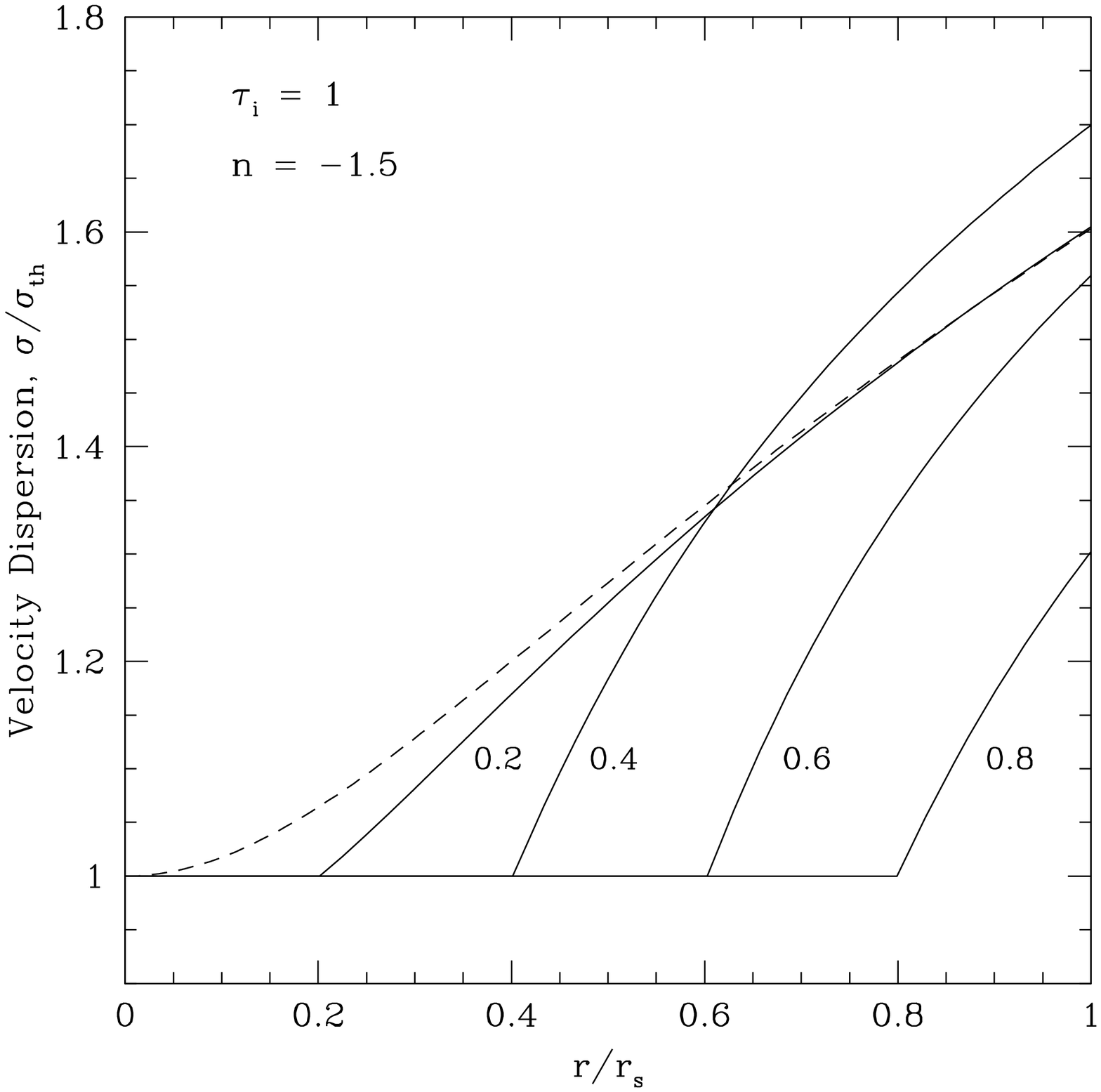}
\caption{
Velocity dispersion in the critically stable, isentropic CPS, assuming 
$v_A = 0$. 
(a) Total (top curves) and nonthermal (bottom curves) velocity dispersions 
for CPSs with $\tau_i = 1,\; n = -3$, and various values of $r_i/r_s$.  From 
left to right: $r_i/r_s = 0.2,\; 0.4,\; 0.6$, and 0.8.  Also shown are the 
corresponding dispersion for a pure $n = -3$ polytrope (dashed curves). 
(b) Similar to (a), but for $n = -1.5$.  Only the total dispersion is 
plotted. \label{fig8}}
\efig

\bfig
\plottwo{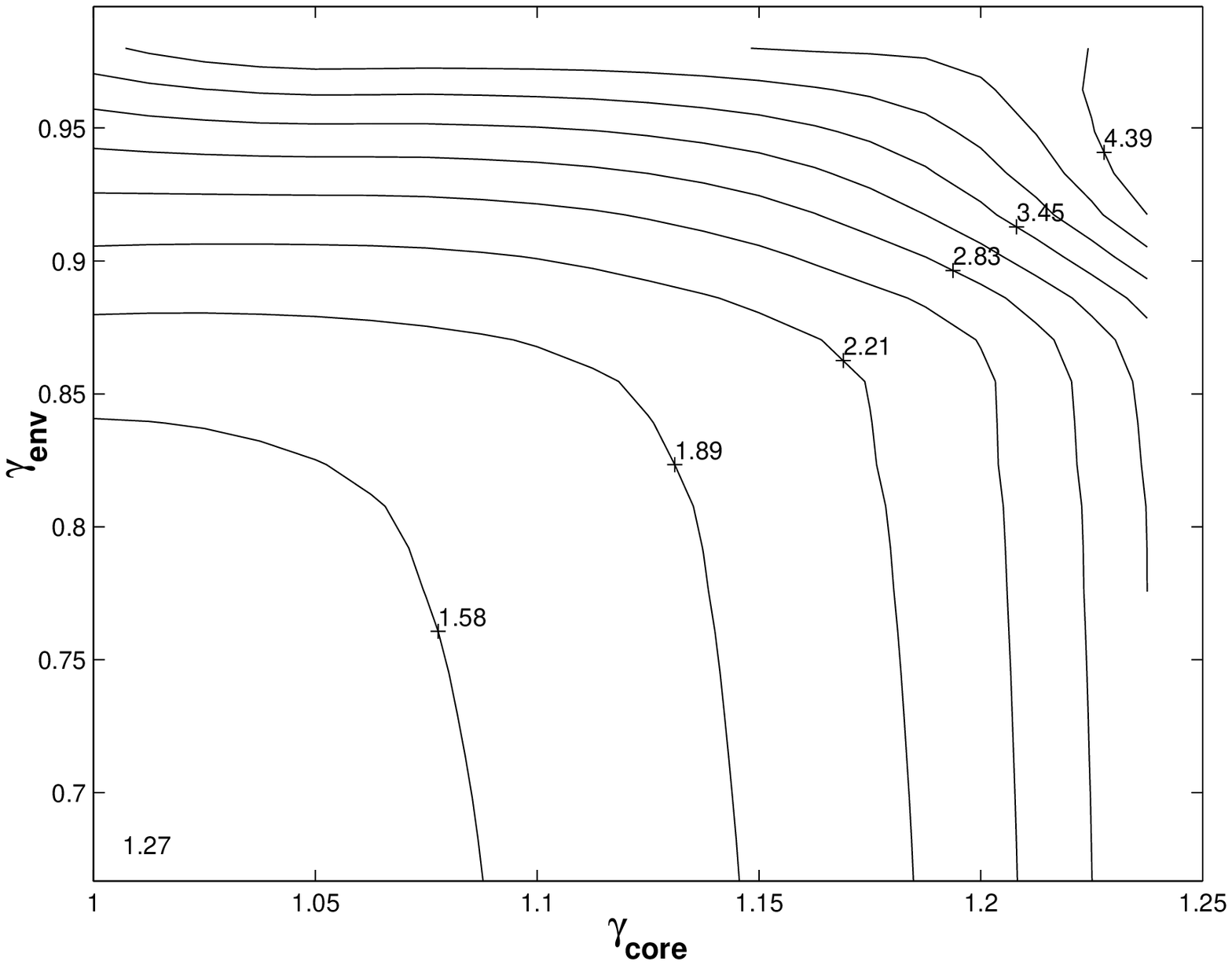}{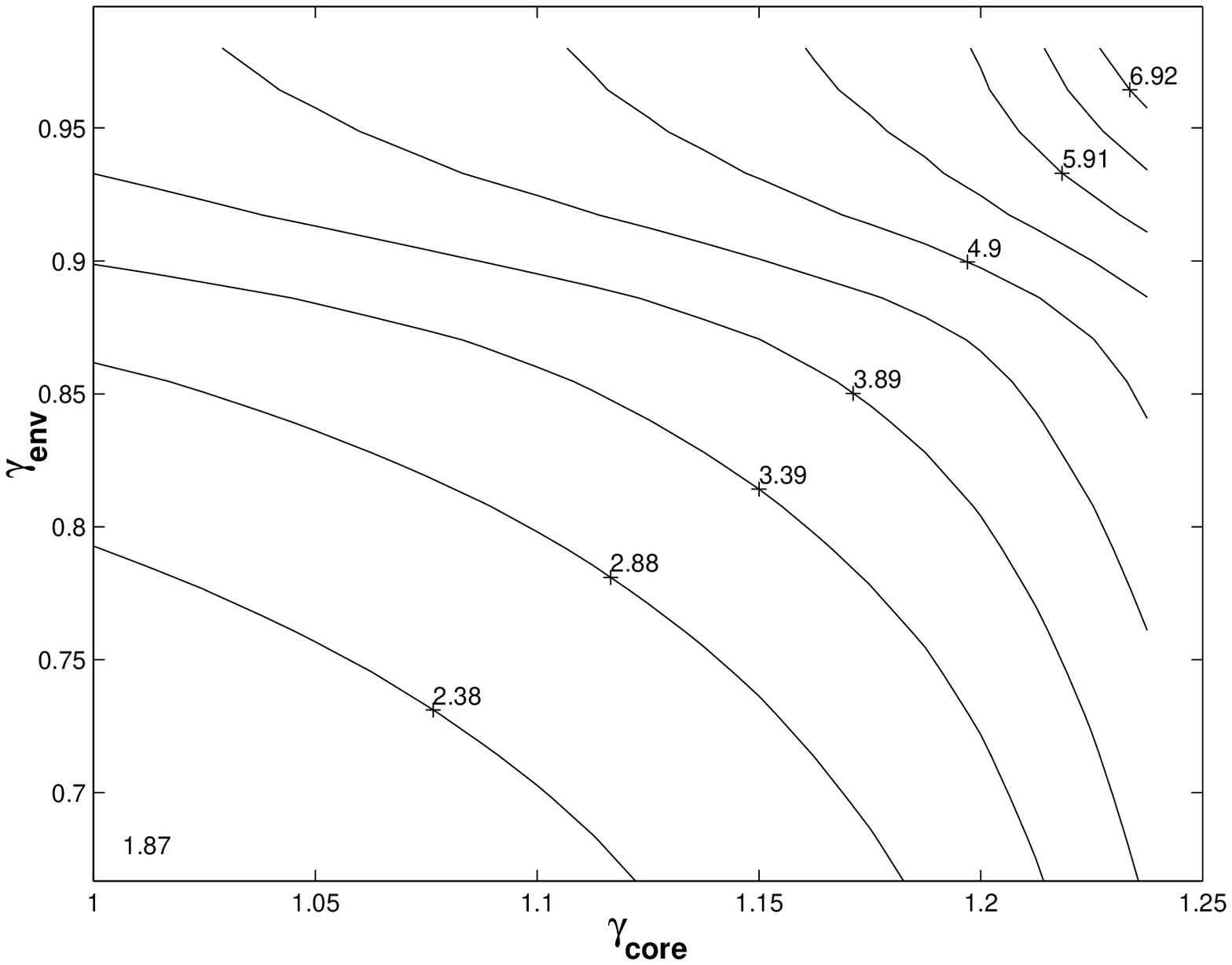}
\caption{
Maximum density contrast of the critically stable, $n = -3$ non-isentropic 
CPS.  Contours of equal log$_{10} (\rho_c/\rho_s)_{\rm cr}^{\rm max}$ are 
plotted as a function of the core 
and envelope adiabatic indices, $\gamc$ and $\game$, 
respectively, for (a) $\tau_i = 1$; (b) $\tau_i = 2$. \label{fig9}}  
\efig

\bfig
\plottwo{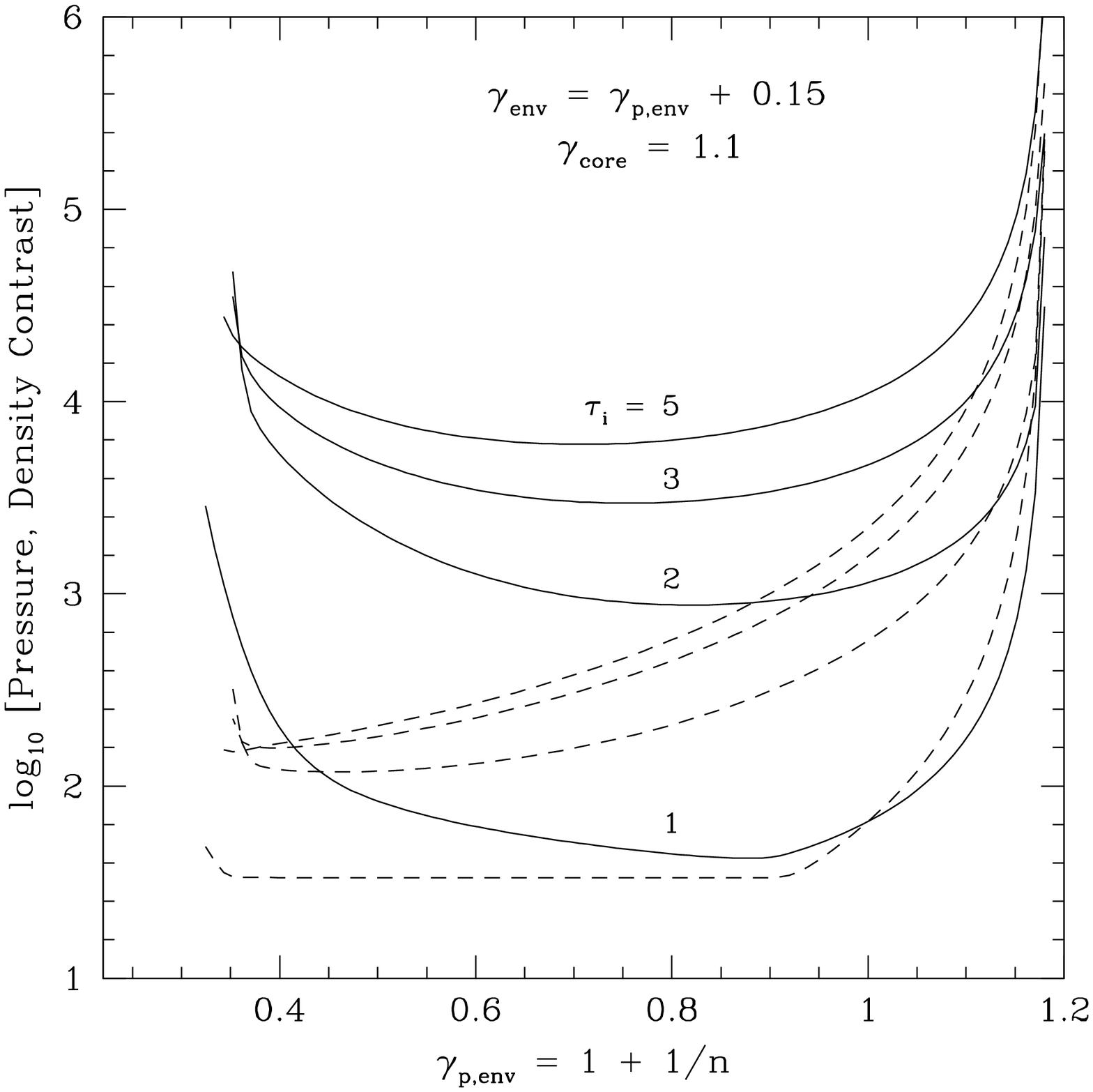}{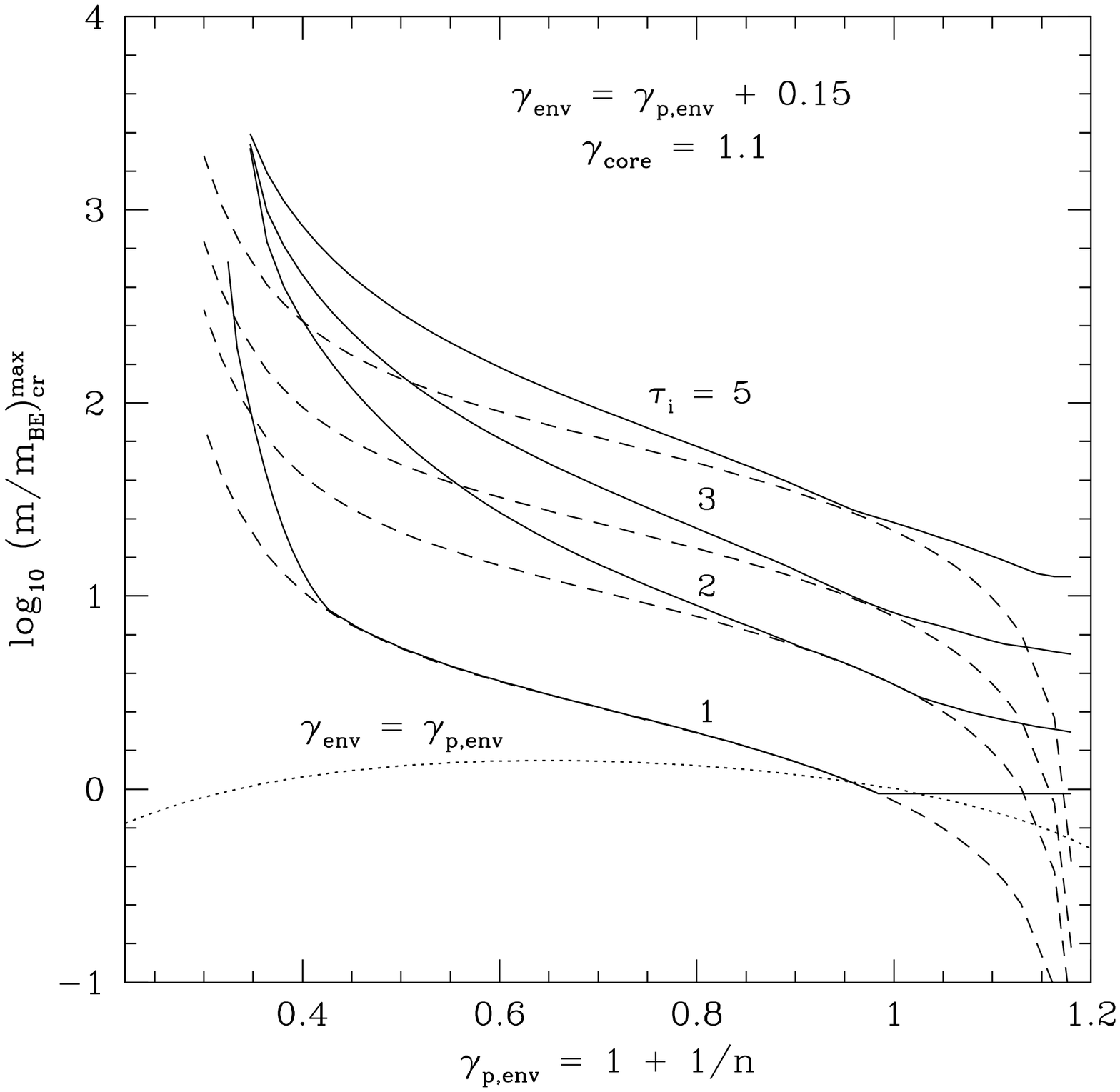}
\caption{
Properties of non-isentropic CPSs with $\gamc = 1.1$ and 
$\game = \gamma_{p,{\rm env}} \;+\; 0.15$.  
(a) Maximum critical pressure (dashed curves) and density (solid curves) 
contrasts as a function of $\gamma_{p,{\rm env}}$.  Curves are labeled by 
$\tau_i$. 
(b) Maximum critical masses as a function of $\gamma_{p,{\rm env}}$ 
(solid curves). 
Also shown are the critical masses of pure polytropes with 
$T_{\rm core} = \tau_i ~T_{\rm core} (\tau_i = 1)$ (dashed curves).  
The dotted line shows the mass of isentropic polytropes. \label{fig10}} 
\efig

\bfig
\plottwo{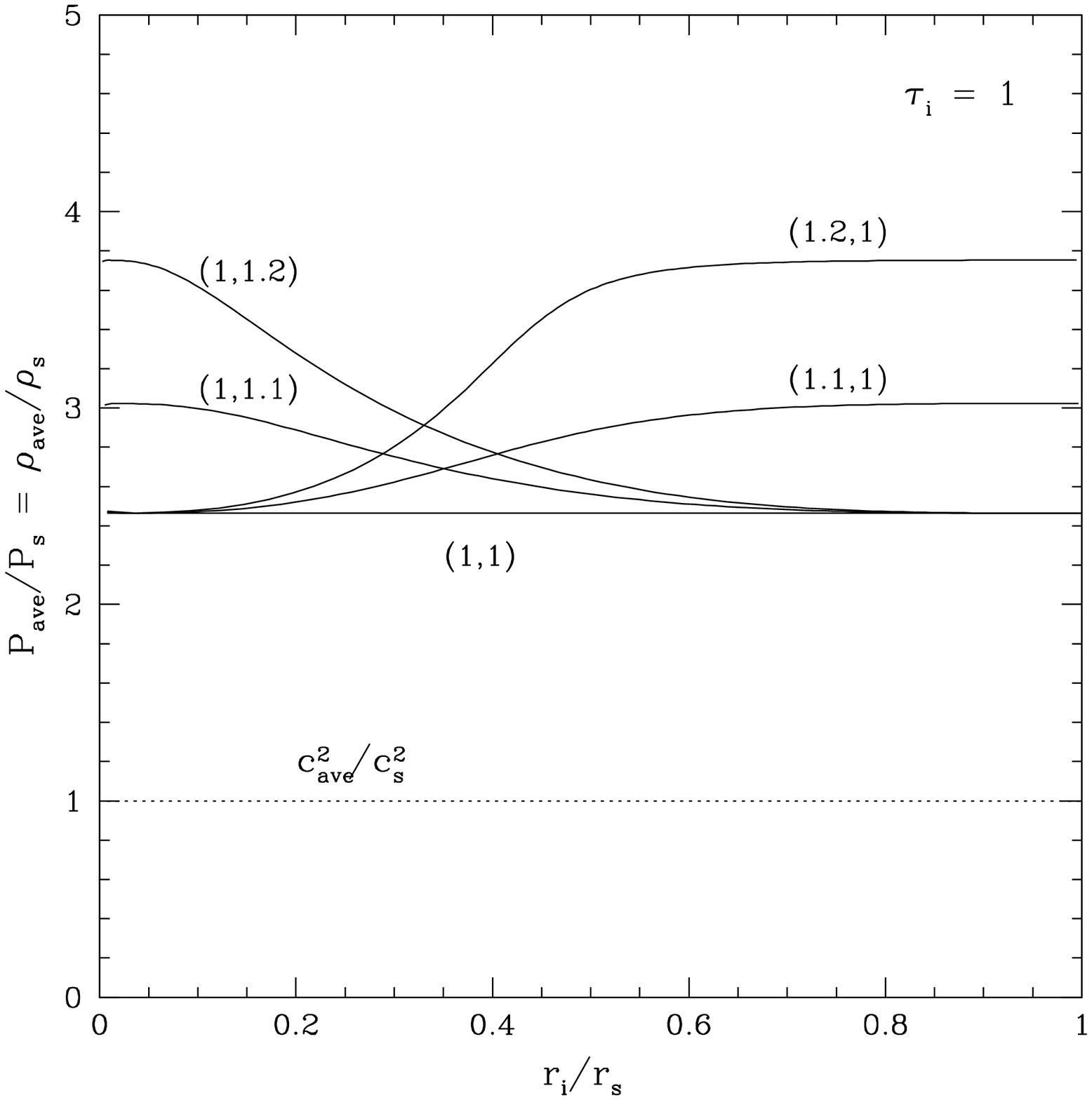}{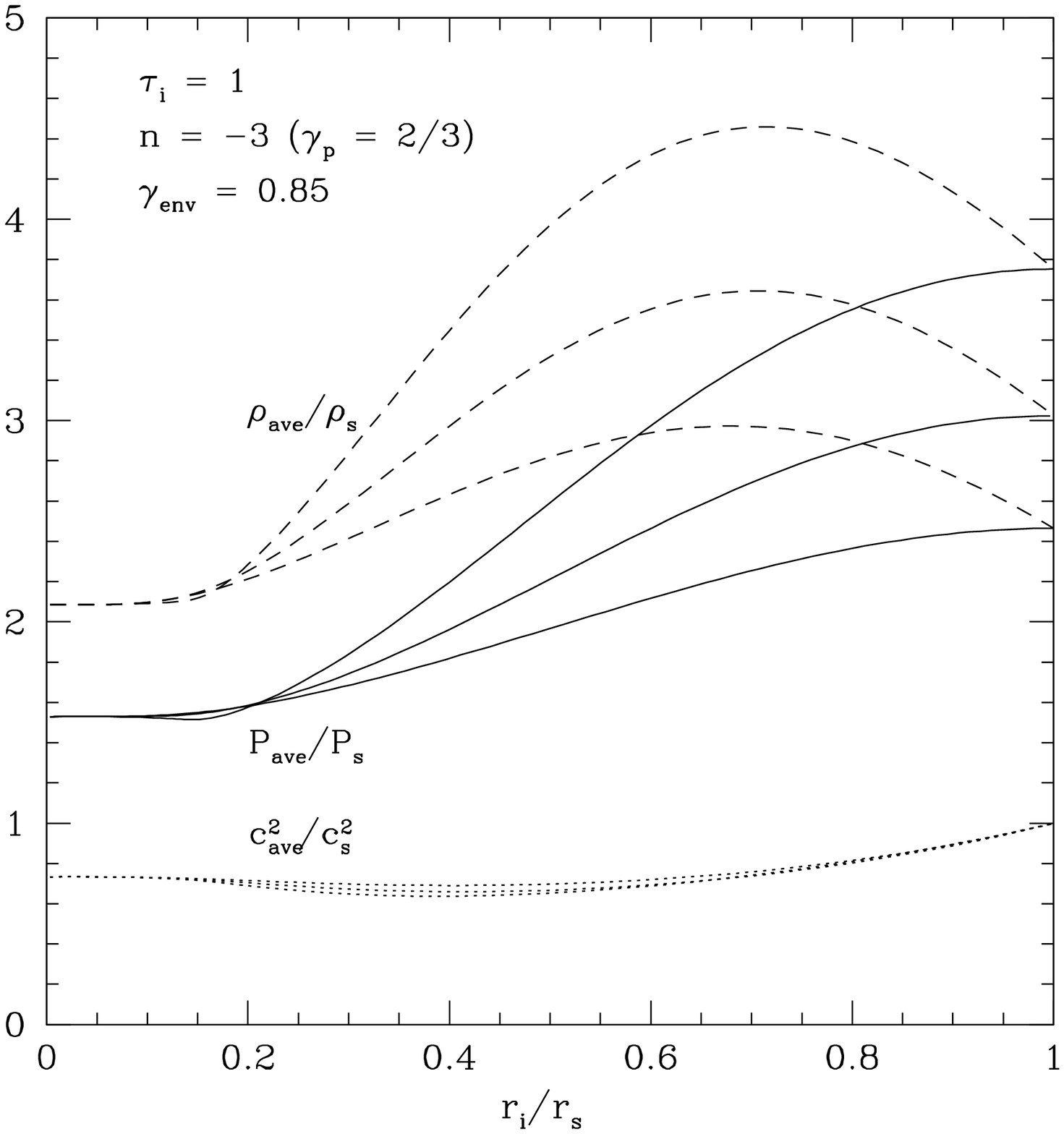}
\caption{
(a) Mean pressure contrasts (solid curves) as a function of fractional 
core radius in the $\tau_i = 1$, 
non-isentropic CIS.  Each curve is labeled by its corresponding 
value of $(\gamc,\game)$.  The dotted line shows the mean 
sound speed contrast, which is equal to unity since $\tau_i = 1$.  
(b) Mean pressure contrasts (solid curves) in the $\tau_i = 1$, 
non-isentropic CPS with $n=-3$ and $\game = 0.85$.  Mean density 
contrasts (dashed curves) and mean sound speed contrasts (dotted 
lines) are also shown.  Within each set of curves, the corresponding 
value of $\gamc$ is, from bottom to top, 1, 1.1, and 1.2. \label{fig11}}  
\efig

\bfig
\plottwo{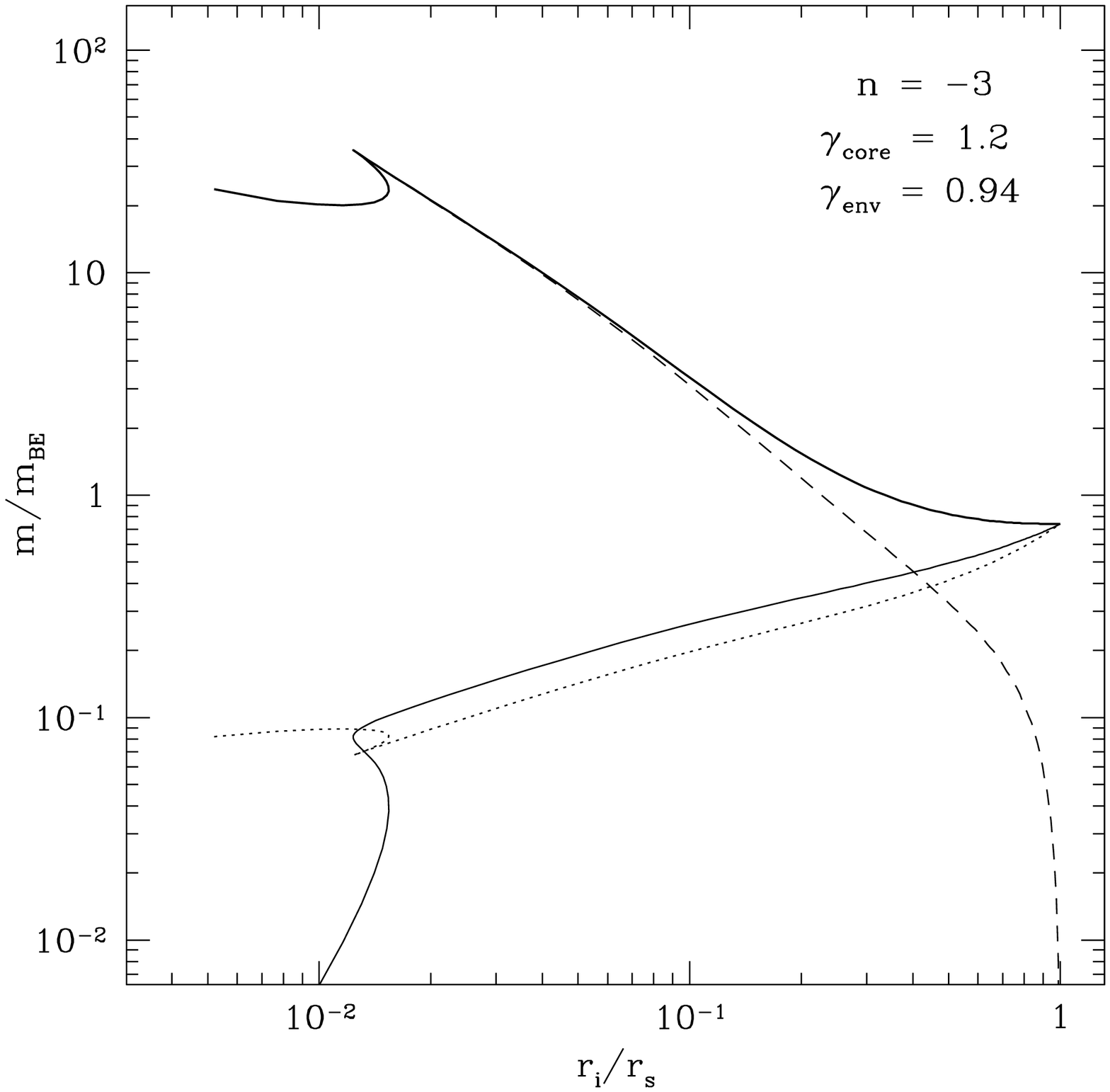}{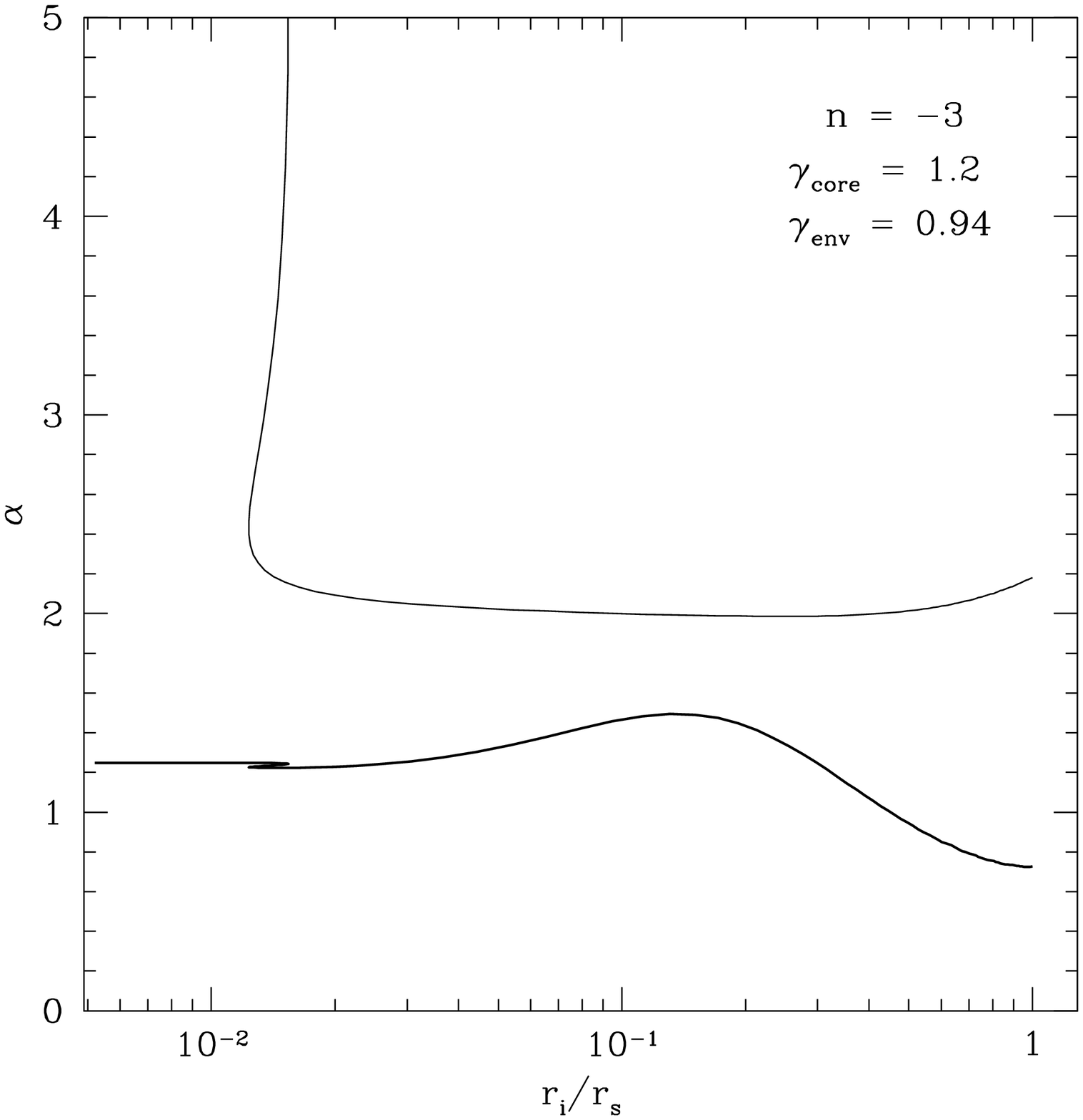}
\caption{
(a) Dimensionless masses of a critically stable, non-isentropic CPS 
with $\tau_i = 1$, as a function of the fractional core radius.  
The solid curve shows the core mass alone; 
the dashed curve, the envelope mass; and the heavy solid curve the total 
of the two.  The dotted curve is the critical mass of the core alone. 
(b) Virial parameters for the core alone (solid curve) and the entire 
cloud (heavy solid curve).  See text for details. \label{fig12}}  
\efig

\end{document}